\documentclass{aa}  

\usepackage{adjustbox}
\usepackage{threeparttable}
\usepackage{graphicx,url,twoopt,natbib}       
\usepackage{hyperref}   
\usepackage{longtable}
\usepackage{pifont}

\hypersetup{
  colorlinks=true,   
  urlcolor=blue,     
  linkcolor=blue,    
  citecolor=blue}

\usepackage{txfonts}
\usepackage{amsfonts}
\usepackage{caption}
\usepackage{subcaption} 
\usepackage{array}

\newcommand{\kms}{\mbox{km\,s$^{-1}$}}

\newcommand{\mloss}{\mbox{$\dot{M}$}}

\newcommand{\my}{\mbox{$M_{\odot}$~yr$^{-1}$}}

\newcommand{\ta}{\mbox{$T^*_{\rm A}$}}  
\newcommand{\tmb}{\mbox{$T_{\rm MB}$}}
\newcommand{\eff}{\mbox{$\eta_{\rm eff}$}} 
\newcommand{\iram}{\mbox{IRAM-30\,m}}
\newcommand{\farc}{\mbox{$.\!\!^{\prime\prime}$}}

\newcommand{\vexp}{\mbox{$V_{\rm exp}$}}

\newcommand{\tex}{\mbox{$T_{\rm ex}$}}

\newcommand{\rmax}{\mbox{$R_{\rm max}$}}
\newcommand{\rinner}{\mbox{$R_{\rm inner}$}}
\newcommand{\tinner}{\mbox{$T_{\rm inner}$}}

\newcommand{\sio}{SiO}
\newcommand{\hcn}{HCN}

\newcommand{\hnc}{HNC}
\newcommand{\hctresn}{HC$_{3}$N}

\newcommand{\hcoplus}{HCO$^{+}$}
\newcommand{\hocplus}{HOC$^{+}$}
\newcommand{\hh}{H$_{2}$}
\newcommand{\hhplus}{H$_{2}^{+}$}
\newcommand{\hhhplus}{H$_{3}$$^{+}$}

\newcommand{\coplus}{CO$^{+}$}
\newcommand{\sicdos}{SiC$_{2}$}
\newcommand{\sis}{SiS}

\newcommand{\cplus}{C$^{+}$}

\newcommand{\hcnhplus}{HCNH$^{+}$}

\newcommand{\cch}{C$_{2}$H}
\newcommand{\cn}{CN}

\newcommand{\ndoshplus}{N$_{2}$H$^{+}$}

\newcommand{\doceuno}{$\rm^{12}$CO\,($J$=1-0)}
\newcommand{\docedos}{$\rm^{12}$CO\,($J$=2-1)}

\newcommand{\hcouno}{$\rm HCO^{+}$\,($J$=1-0)}
\newcommand{\hcodos}{$\rm HCO^{+}$\,($J$=3-2)}
\newcommand{\hcnuno}{HCN\,($J$=1-0)}
\newcommand{\hcndos}{HCN\,($J$=3-2)}
\newcommand{\hncuno}{HNC\,($J$=1-0)}
\newcommand{\hnctres}{HNC\,($J$=3-2)}
\newcommand{\siouno}{SiO\,($J$=2-1)}
\newcommand{\siodos}{SiO\,($J$=6-5)}
\newcommand{\hctresnuno}{HC$_{3}$N\,($J$=10-9)}
\newcommand{\hctresndos}{HC$_{3}$N\,($J$=29-28)}
\newcommand{\sicdosuno}{SiC$_{2}$\,(11$_{4,7}$-10$_{4,6}$)}
\newcommand{\sicdosdos}{SiC$_{2}$\,(11$_{4,8}$-10$_{4,7}$)}
\newcommand{\sicdostres}{SiC$_{2}$\,(12$_{0,12}$-11$_{0,11}$)}
\newcommand{\sisquince}{SiS\,($J$=15-14)}
\newcommand{\water}{\mbox{H$\rm_2$O}}

\def\snu#1{\ifmmode {S_\nu\,\propto\,\nu^{#1}}
          \else \hbox{$S_\nu$\,$\propto$\,$\nu^{#1}$}\fi}
\def\cm#1{\ifmmode {\,{\rm cm^{-#1}}}                  
        \else \hbox{$\,${\rm cm$^{\rm -#1}$}}\fi}
\def\raw {\ifmmode\rightarrow\else$\rightarrow$\fi}
\def\ex#1{\ifmmode {\cdot 10^{#1}}         
        \else \hbox{{$\times 10^{\rm #1}$}}\fi}

\renewcommand{\descriptionlabel}[1]{\hspace{\labelsep}\textbf{#1}}

\begin{document}

   \title{Exploring circumstellar chemistry in X-ray emitting AGB stars}

   \author{J. Alonso-Hernández \inst{1, 2} \and C. Sánchez Contreras \inst{1} \and M. Agúndez \inst{3} \and R. Sahai \inst{4} \and J. P. Fonfría \inst{1,3,5} \and L. Velilla-Prieto\inst{3} \and G. Quintana-Lacaci\inst{3}  \and J. Cernicharo\inst{3} }

   \institute{Centro de Astrobiologia (CAB), CSIC-INTA. Postal address: ESAC, Camino
Bajo del Castillo s/n, 28692, Villanueva de la Ca\~nada, Madrid, Spain \\ 
\email{jalonso@cab.inta-csic.es}  
\and Escuela de Doctorado UAM, Centro de Estudios de Posgrado, Universidad Autónoma de Madrid, E-28049 Madrid, Spain   \and Department of Molecular Astrophysics, Instituto de Física Fundamental (IFF-CSIC), C/ Serrano
121, 28006 Madrid, Spain \and Jet Propulsion Laboratory, California Institute of Technology, Pasadena, CA 91109, USA \and Observatorio Astronómico Nacional (IGN), Alfonso XII No 3, 28014 Madrid, Spain
\\}

   \date{Received 20 January 2025 / Accepted 25 April 2025}

  \abstract
   {}
   {Our goal is to characterise the chemistry and physical conditions of the circumstellar envelopes (CSEs) of Asymptotic Giant Branch (AGB) binary candidate stars with UV-excess and X-ray emission. In particular, our aim is to identify the effects of the internal X-ray emission in the abundance of certain key molecules.}
   {We observed the 86.0-94.0 and 260.0-272.5 GHz spectral ranges and searched for rotational transitions of the X-ray sensitive molecule \hcoplus\ in four AGB stars. Two detected in both UV and X-ray emission, and the other two detected only in UV. We derived the CSEs's physical parameters from previous CO observations and determined the molecular abundances of the detected species using radiative transfer models. We developed chemical kinetics models that account for the effects of internal X-ray emission (as well as UV radiation) and compared our predictions with observations.} 
  {We report the detection of \hcoplus\ in the X-ray emitting C-rich AGB T\,Dra, while it remains undetected in the spectra of the other three sources. In T\,Dra we also detected \sio, \hcn, \hnc, \hctresn, \sicdos, \cch, and \sis. For the other targets, only HCN and SiO were detected. The high fractional abundance of \hcoplus\ derived for T\,Dra ($[1.5-3.0]\times 10^{-8}$) is in good agreement with the predictions from our chemical kinetics models including the effects of internal X-ray emission, and one order of magnitude higher than the values expected for C-rich AGB stars. Additionally, we identified abundance enhancements for \hnc\ and \hctresn\ alongside a depletion of CO in the innermost regions of T\,Dra's envelope.} 
  {An internal X-ray source can significantly alter molecular abundances in AGB CSEs, and enhance \hcoplus, \ndoshplus, \hnc, and \hctresn\ while depleting parent species like CO. The UV radiation has a weaker effect unless the envelope is optically thin or porous.}

   
   \maketitle

\section{Introduction}\label{intro}

Asymptotic Giant Branch (AGB) stars are very efficient factories of dust and molecular gas, which are expelled to the interstellar medium (ISM) via slow (\vexp $\sim$ 5-15 \kms) dusty winds with typical gas mass-loss rates of \mloss $ \simeq 10^{-7}-10^{-4} \my$ \citep[e.g.][]{Hofner_2018}. These winds result in the stars being embedded in dense circumstellar envelopes (CSEs) with a very rich chemistry. 

The molecular composition of the CSEs around AGB stars has been intensively explored during the past several decades. AGB stars are chemically classified according to their elemental $\rm C/O$ ratio, and C-rich AGB stars are chemically richer than their O-rich counterparts. The source that contains the highest number of detected molecular species and that has the most complete chemical description is the C-rich IRC+10216 \citep[see e.g.][]{agundez_2006, cordiner_2009, Cernicharo_2010, cherchneff_2012, agundez_2017}. Examples of O-rich AGB stars with rich chemistries are IK\,Tau \citep[the O-rich AGB star with most detailed chemical description, see e.g.][]{Kim_2010, Decin_2010b, De_Beck_2013, Velilla-Prieto_2017}, and OH\,231.8+4.2 \citep[the best example of fast-shocks induced chemistry, see][]{Morris_1987, sanchez-contreras_2000,sanchez-contreras_2015, Velilla-Prieto_2015}. A recent overview of the state of circumstellar chemistry in both molecular gas and dust was provided by \cite{agundez_2020}. 

\begin{table*}[h!]

\renewcommand{\arraystretch}{1.2}
\small
\centering

\caption{Astronomical parameters and UV/X-ray fluxes of the sample.} 
\label{tab:astronomy}

\begin{adjustbox}{max width=\textwidth}
\begin{threeparttable}[b]

\begin{tabular}{l >{\centering\arraybackslash}p{1.80cm} >{\centering\arraybackslash}p{2.00cm} >{\centering\arraybackslash}p{2.00cm} >{\centering\arraybackslash}p{2.00cm} >{\centering\arraybackslash}p{1.50cm} >{\centering\arraybackslash}p{1.50cm} >{\centering\arraybackslash}p{1.50cm} >{\centering\arraybackslash}p{1.50cm}}
\hline\hline 
Source & RA  & DEC & parallax & D  & Chem & NUV  & FUV & X-ray \\
       & (hh:mm:ss) & (dd:mm:ss) & (mas) & (pc) &   & ($\rm \mu Jy$) & ($\rm \mu Jy$) & ($\mathrm{erg \,cm^{-2}\, s^{-1}}$)\\
\hline 

T Dra & 17 56 23 & +58 13 07   & 1.02$\pm$0.08 & 980$\pm$80 &  C &  30  &  8  &  3.6e-13 $^{(a)}$\\
EY Hya & 08 46 21 & +01 37 56   & 2.34$\pm$0.07 & 427$\pm$7 & O &  110  &  80   &   4.9e-13 $^{(a)}$\\
VY Uma & 10 45 04 & +67 24 41   & 2.38$\pm$0.08 & 420$\pm$14 & C &  140  &  8  &  $<$4.8e-13 $^{(b)}$\\
V Eri & 04 04 19 & -15 43 30   & 3.35$\pm$0.16 & 298$\pm$14 &  O & 100  &  70  &  $<$1.2e-14 $^{(b)}$\\

\hline
\end{tabular} 
\begin{tablenotes}
\item \normalsize \textbf{Notes.} Column (1): Name, Col. (2): Right ascension, Col (3): Declination, Col (4): Gaia DR3 parallax, Col (5): Gaia DR3 distance, Col. (6): Chemical type, Col (7): time-averaged GALEX NUV flux, Col (8): time-averaged GALEX FUV flux, Col (9): X-ray flux: $^{(a)}$ are X-ray fluxes obtained in the energy range (0.3-10.0 keV) from \cite{Ortiz_2021} and $^{(b)}$ are upper limits (1$\sigma$) X-ray fluxes obtained in the energy range (0.2-12.0 keV) from The RapidXMM upper limit server \citep{Ruiz_2022}. 
\end{tablenotes}

\end{threeparttable}
\end{adjustbox}

\renewcommand{\arraystretch}{1.0}

\end{table*}

Some years ago, a population of AGB stars exhibiting unexpected UV excesses was discovered \citep[hereafter uvAGBs; see][]{Sahai_2008}. These stars are in striking contrast to the majority of AGB stars, which show no detectable UV emission. Furthermore, a few uvAGBs were also found to have X-ray emission \citep[hereafter X-AGBs; see][]{Ramstedt_2012, Sahai_2015, Ortiz_2021, Sahai_2022, Schmitt_2024, Guerrero_2024}. The origin of this internal high-energy emission, is still under debate \citep[see][]{Sahai_2011, Montez_2017}. However, for X-AGBs, the presence of active accretion processes is the most likely explanation. 

Recently, \cite{Alonso-Hernandez_2024} studied a sample of uvAGBs (some of which are also X-AGBs) and reported an anticorrelation between the CO line intensity and the observed UV emission, a trend that most likely reflects enhanced CO photodissociation in the central regions of their CSEs. Another theoretically predicted effect of internal UV emission is a drastic change in the abundances of some key molecules, in both O-rich and C-rich CSEs 
\citep[see][]{Van_de_Sande_2018, Van_de_Sande_2022, Van_de_Sande_2023}. However, the impact of internal X-ray emission on the chemistry of X-AGBs remains largely unexplored, both theoretically and observationally. We expect these effects are quite significant due to the larger photodissociation and photoionisation capabilities of X-rays compared to UV, making them a promising driver of chemical enrichment in AGB CSEs, especially in the formation of certain key molecules such as \hcoplus.

In standard\footnote{In the context of this paper, we refer to 'standard' AGB stars as those without UV excess, X-ray emission, or any other observational characteristics that distinguish them as outliers from the general trends of the AGB class.} AGB stars, \hcoplus\,is expected to form in the outer regions of the envelopes due to chemical reactions induced by ionisation from cosmic rays and the UV interstellar radiation field \citep[ISRF;][]{agundez_2006, Li_2016}. 
The detection of \hcoplus\,in standard AGB stars remained elusive during decades of observational searches, although there are now firm detections in a few AGB CSEs, most of them O-rich. Even in these cases, the \hcoplus\ emission is generally weak, as this ion is typically found in low abundances in these outer regions \citep[\hcoplus/\hh$\sim$5$\times$$10^{-8}$ for O-rich AGB stars and $\sim$5$\times$$10^{-9}$ for the C-rich AGB star IRC+10216, see][and references therein]{Pulliam_2011}.

In this paper, we introduce a new perspective on photoionisation and photodissociation-induced chemistry by using HCO$^{+}$ as a sensitive probe to study the effects of X-rays on circumstellar chemistry. While X-ray irradiation has been studied before in a variety of astrophysical systems, such as young stellar objects and X-ray binaries \citep[e.g.][]{Wolfire_2022}, its impact on AGB CSEs remains largely unexplored. In protoplanetary disks, for example, the abundances of key molecular ions, particularly HCO$^{+}$, are strongly influenced by X-rays from the central star, with variations that appear to correlate with X-ray stellar flares \citep{Cleeves_2017, Waggoner_2023}. Recently, in the planetary nebulae NGC\,7027, significant differences have been observed in the spatial distributions of CO$^+$ and HCO$^{+}$, and these variations are probably linked to UV or X-ray-induced chemical reactions \citep{Bublitz_2023}. However, the physical and chemical conditions in AGB CSEs—such as density, temperature, and chemical gradients—differ substantially from those in the environments studied so far. This makes the study of X-ray-driven chemistry in AGB CSEs a unique contribution, with the potential to broaden understanding of how X-rays influence circumstellar environments. Our observational study focuses on the search for \hcoplus\, emission in X-AGBs. It is complemented by chemical kinetics models simulating X-ray irradiated AGB CSEs, thus providing new insights into the molecular abundances and processes altered by X-ray irradiation.

This paper is organised as follows. In Sect.\,\ref{sample} we introduce our sample. A description of the observations is given in Sect.\,\ref{obs}. The detected lines and their measured parameters are given in Sect.\,\ref{spectra}. The data analysis, which includes radiative transfer modelling of the line emission, is described in Sect.\,\ref{anal}. The chemical kinetics modelling and comparison of empirically derived versus predicted abundances are presented in Sect.\,\ref{chem}. The discussion and interpretation of the main results are given in Sect.\,\ref{dis}. Finally, the conclusions are summarised in Sect.\,\ref{summ}.

\section{The sample}\label{sample}

In this work, we carried out a exploratory chemistry study over a small sample of four AGB stars. Two of these are both UV-excess and X-ray emitting AGB stars (T\,Dra and EY\,Hya), while the other two are UV-excess AGB stars without X-ray emission (V\,Eri and VY\,UMa). Since all X-AGBs also show ultraviolet excesses \citep[see e.g.][]{Sahai_2015}, this sample was designed to compare and empirically isolate the effects of X-rays from those of UV radiation in the circumstellar chemistry. We included one C-rich and one O-rich AGB star in each subsample to compare the effects of X-ray emission on different types of circumstellar chemistry.

Our targets were selected from the sample of uvAGBs with CO detections presented in \cite{Alonso-Hernandez_2024}, where a preliminary analysis led to the estimate of their mass-loss rates (\mloss $\approx$10$^{-7}$-10$^{-6}$\,\my) and other envelope parameters. The targets are listed in Table~\ref{tab:astronomy} together with their equatorial coordinates, Gaia DR3 \citep{GAIA} parallaxes and distances ($D$), chemistry type, GALEX NUV and FUV fluxes, and X-ray fluxes. We considered Gaia distances to perform a uniform analysis over our sample and to make a direct comparison with the preliminary estimations from \cite{Alonso-Hernandez_2024}, which were also based on Gaia distances. Even though Gaia distance estimations are not especially reliable for AGB stars \citep[see][]{El-Badry_2021, Andriantsaralaza_2022}, we noted that the estimated Gaia distances for our sample are in relatively good agreement (within a 50\% deviation) with those found in the literature and obtained from alternative methods, such as based on luminosity assumptions \citep[250 pc for V\,Eri, 330 pc for VY\,Uma, and 300 pc for EY\,Hya from][respectively]{Kahane_1994, shoier_2001, Olofsson_2002}, period-luminosity relationships \citep[610 pc for T\,Dra from][]{shoier_2001}, and Gaia parallaxes corrected by comparison with Very Long Baseline Interferometry (VLBI) parallax measurements \citep[444 pc for VY\,Uma and 901 pc for T\,Dra from][]{Andriantsaralaza_2022}.

\section{Observations}\label{obs}

\subsection{IRAM\,30}

The observations presented in this paper were carried out with the \iram\ radiotelescope (Pico Veleta, Granada, Spain) using the spectral line Eight MIxer Receiver \citep[EMIR,][]{Carter_2012}. Observations were taken in two observational campaigns in 2020, in which all the targets were observed, and 2024, in which only T\,Dra was observed. 

We simultaneously observed the lines presented in Appendix~\ref{line_table} with E090 and E230 receivers operated in dual sideband (2SB) mode and dual (H+V) polarisation.  Observations were performed with four Fast Fourier Spectrometers (FTS) backends, which provide a spectral resolution of $\sim$200\,kHz ($\sim$0.2 and 0.7\,\kms\ at 1 and 3\,mm, respectively). EMIR 2SB mode offers a spectral coverage $\sim$8\,GHz per band and per polarisation, the selected spectral coverage was $\sim$86$-$94 and $\sim$260$-$268 GHz for the 2020 observations and $\sim$86$-$94 and $\sim$264.5$-$272.5 GHz for the 2024 observations.

Observations were performed in wobbler-switching mode with a wobbler throw of 120\arcsec\ and frequencies of 0.5\,Hz. Calibration scans on the standard two load system were taken every $\sim$10$-$15 min. Pointing and focus were checked regularly on nearby continuum sources. After pointing corrections, the typical pointing accuracy was $\sim$2\arcsec$-$5\arcsec. On-source integration times were in the range 5$-$9\,h (with a mean value of 7\,h) per target for the 2020 observations and 13\,h for T\,Dra in 2024. 

The observations were reduced using CLASS\footnote{CLASS is a world-wide software used to process, reduce and analyse spectroscopic single-dish observations developed and maintained by the Institut de Radioastronomie Millimétrique (IRAM) and distributed with the GILDAS software, see {\tt \url{http://www.iram.fr/IRAMFR/GILDAS}}} following standard procedures (baseline fitting and subtraction, averaging of scans after excluding high-noise ones, etc.). We present the spectra in main beam-temperature (\tmb) scale, which was obtained from the antenna temperature (\ta) applying the well-known relation \tmb=\ta/\eff, where \eff\ is the main beam efficiency: \eff\ $\sim$0.86 at 90\,GHz and \eff\ $\sim$0.57 at 270\,GHz. The point source sensitivity (S/\ta, i.e.\,the K-to-Jy conversion factor) is $\sim$5.8 and $\sim$8.8 at 90 and 270\,GHz, respectively. The half power beam width (HPBW) of the \iram\ antenna is 27\farc3 at 3mm and 9\farc1 at 1mm\footnote{The \iram\ efficiencies and beam widths can be found at {\tt \url{https://publicwiki.iram.es/Iram30mEfficiencies.}}}.

The rms noise of the spectra for a velocity resolution of 0.7\,\kms\ is in the range 5.3$-$13\,mK (with a median value of 10\,mK) at 1mm and 3.3$-$5.1\,mK (with a median value of 3.9\,mK) at 3mm. The uncertainty of the relative calibration of our observations was estimated by observing IRC+10216, a standard source with some intense rotational lines emission that is commonly used as a line calibrator, in the same spectral bands at the beginning of every observational run. We estimate total line flux uncertainties of $\sim$20$-$25\% at 1mm and of $\sim$10$-$15\%\ at 3mm.  

\subsection{Archival UV photometry and X-ray fluxes}
UV and X-ray fluxes from archival data for our targets are given in Table~\ref{tab:astronomy}. The UV fluxes were obtained from the GALEX MAST archive \citep{Conti_2011}, where we time-averaged fluxes from multiple epochs. X-ray fluxes were taken by the compilation made by \cite{Ortiz_2021} for X-AGBs. For sources without an X-ray detection, we obtained upper limits from The RapidXMM server \citep{Ruiz_2022}. The time-resolved GALEX fluxes and additional related details can be found in \cite{Alonso-Hernandez_2024}.

\section{Observational results}\label{spectra}

\begin{figure}[h!]
     \centering
     \begin{subfigure}[b]{\linewidth}
         \centering
         \includegraphics[width=\linewidth]{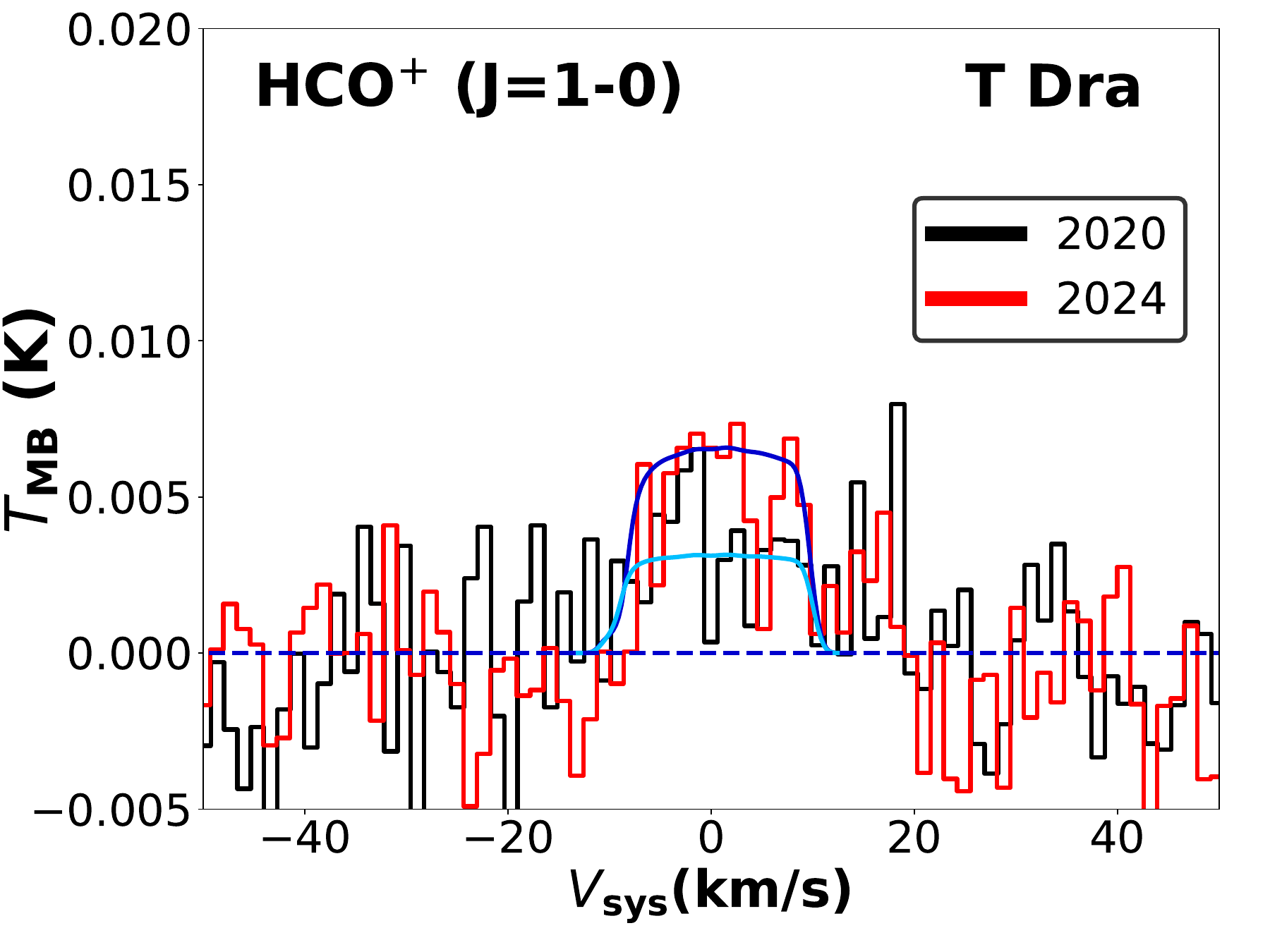}
     \end{subfigure}
     \begin{subfigure}[b]{\linewidth}
         \centering
         \includegraphics[width=\linewidth]{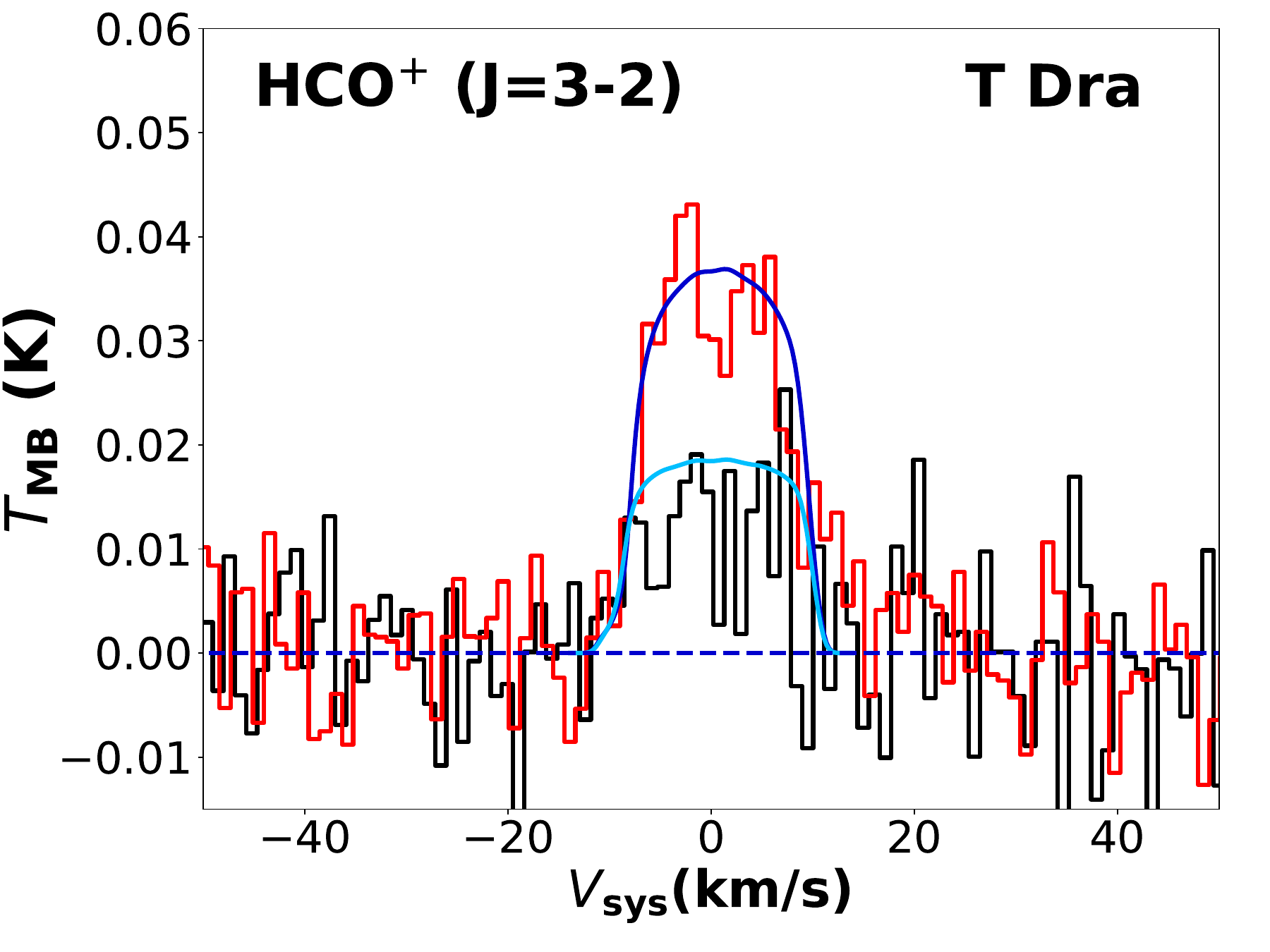}
     \end{subfigure}
        \caption{Spectra of \hcoplus\,in the C-rich X-AGB T\,Dra. Observed spectra are shown in black for 2020 observations and in red for 2024 observations (velocity resolution is $\delta v$=1.4\,\kms). Synthetic spectra are shown in dark blue for 2020 and in light blue for 2024 (details on and results from the radiative transfer model are given in Sect.~\ref{anal}).}
        \label{fig:HCO+_T-DRA_spectra}
\end{figure}

\begin{figure*}[h!]
     \centering
     \begin{subfigure}[b]{0.33\linewidth}
         \centering
         \includegraphics[width=\linewidth]{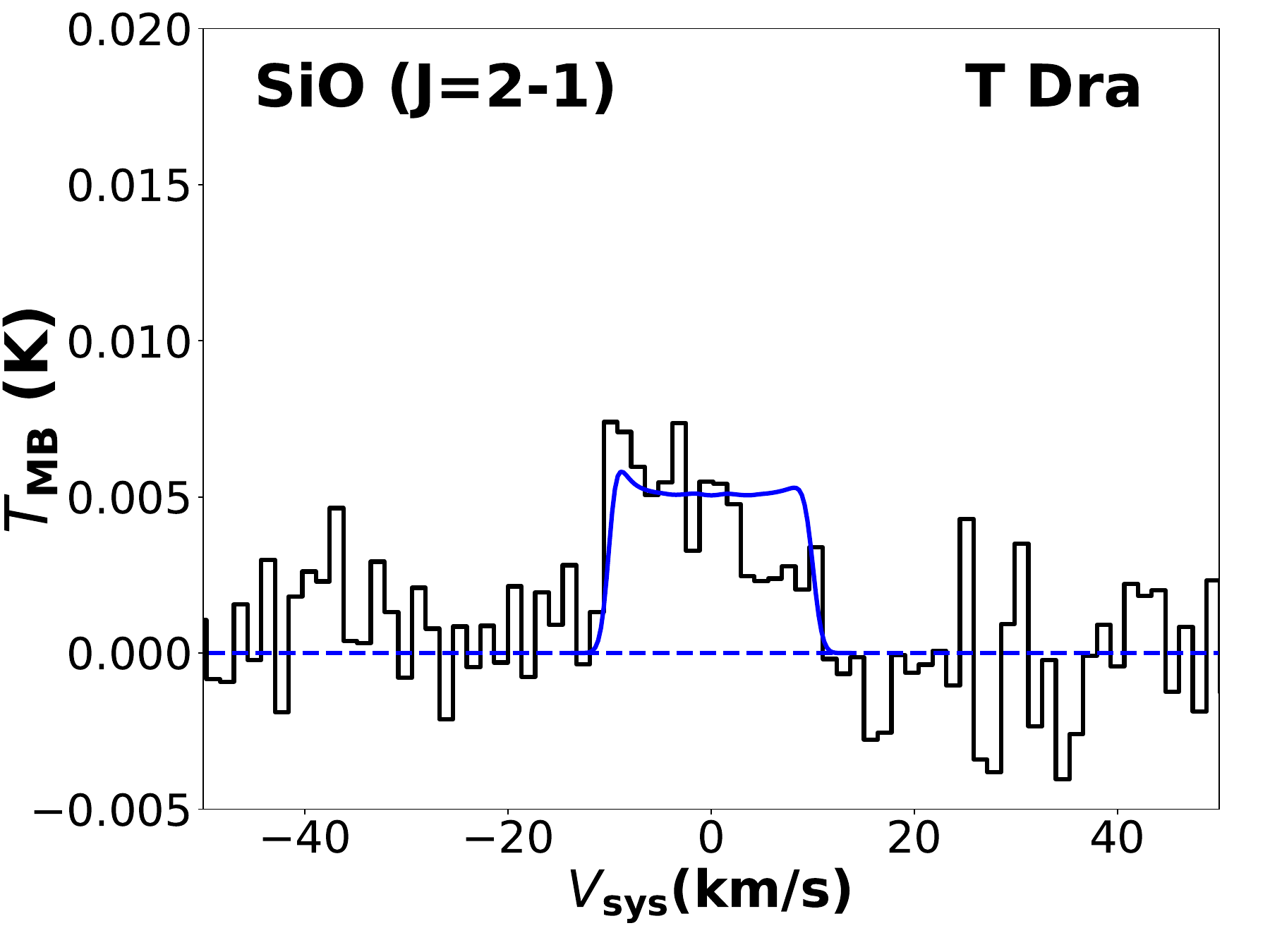}
     \end{subfigure}
     \begin{subfigure}[b]{0.33\linewidth}
         \centering
         \includegraphics[width=\linewidth]{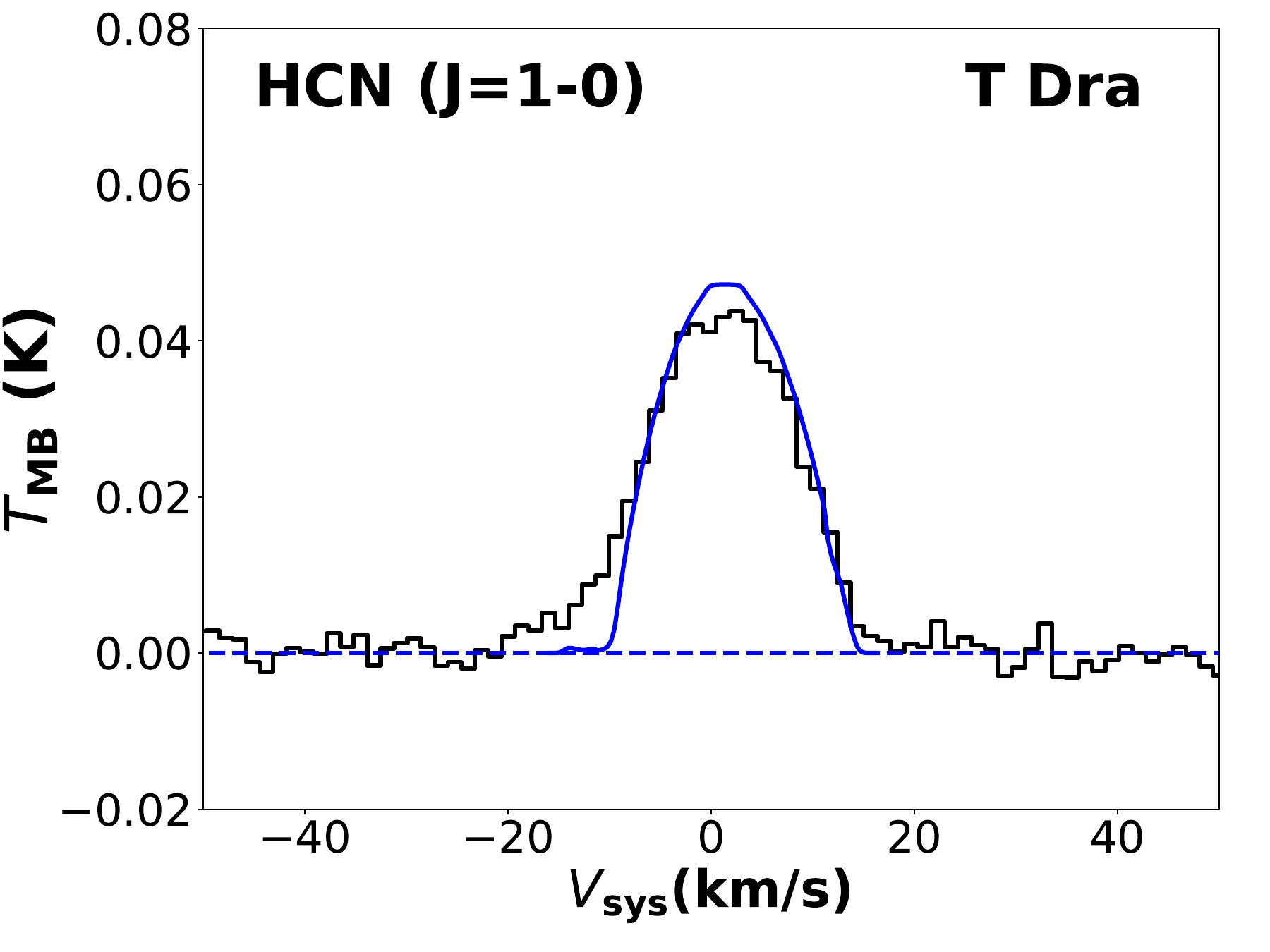}
     \end{subfigure}
     \begin{subfigure}[b]{0.33\linewidth}
         \centering
         \includegraphics[width=\linewidth]{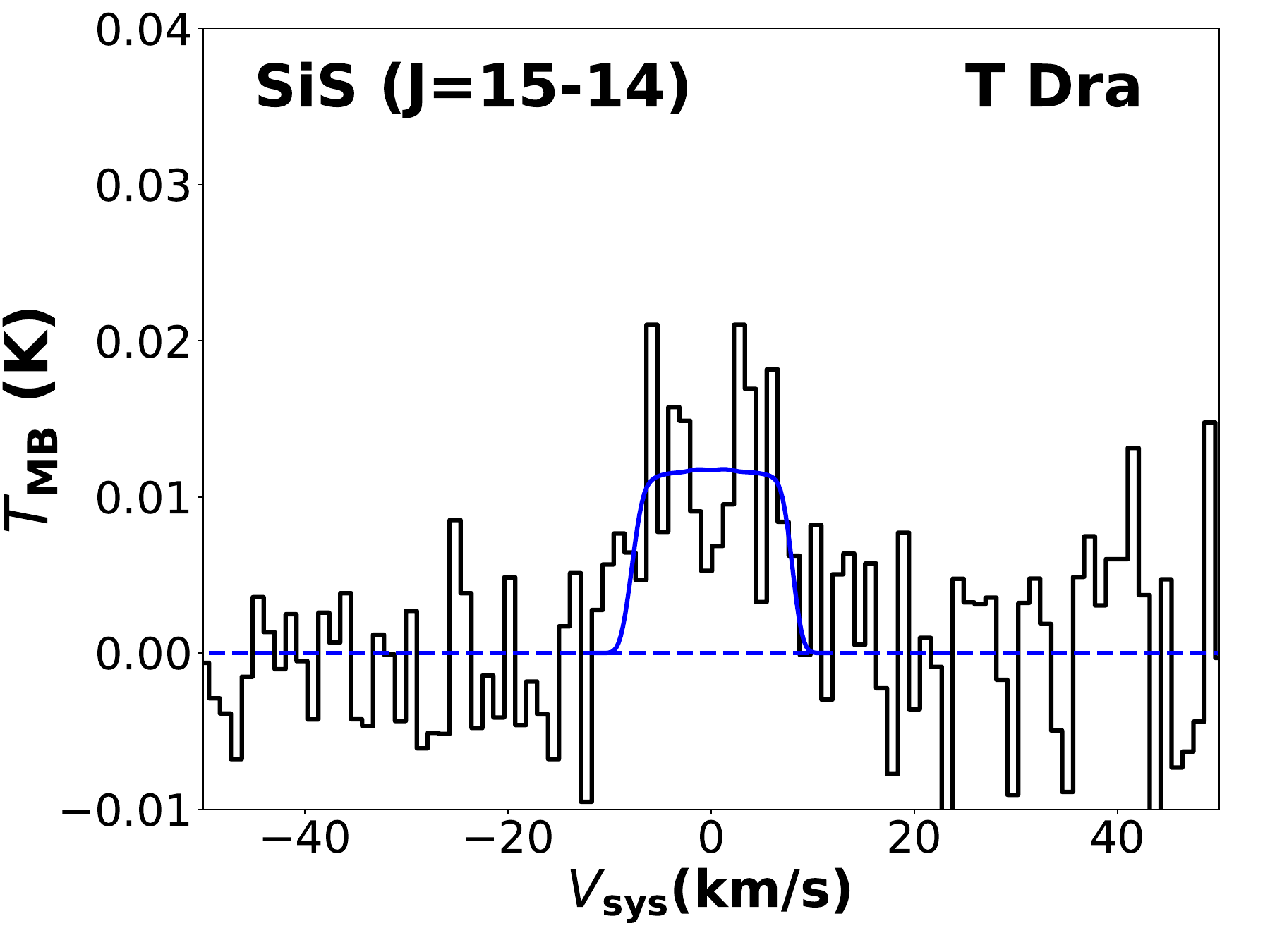}
     \end{subfigure}

     \begin{subfigure}[b]{0.33\linewidth}
         \centering
         \includegraphics[width=\linewidth]{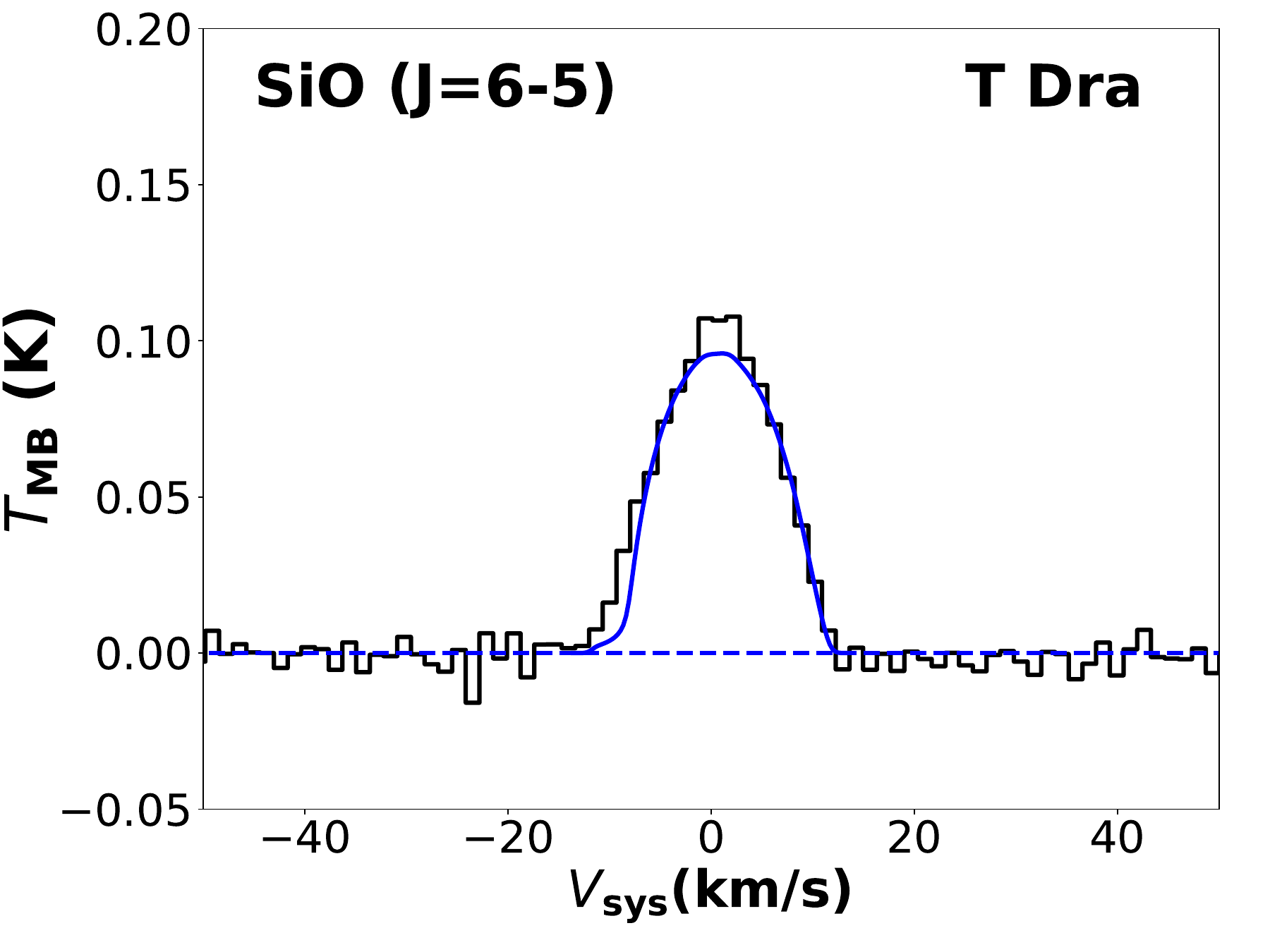}
     \end{subfigure}
     \begin{subfigure}[b]{0.33\linewidth}
         \centering
         \includegraphics[width=\linewidth]{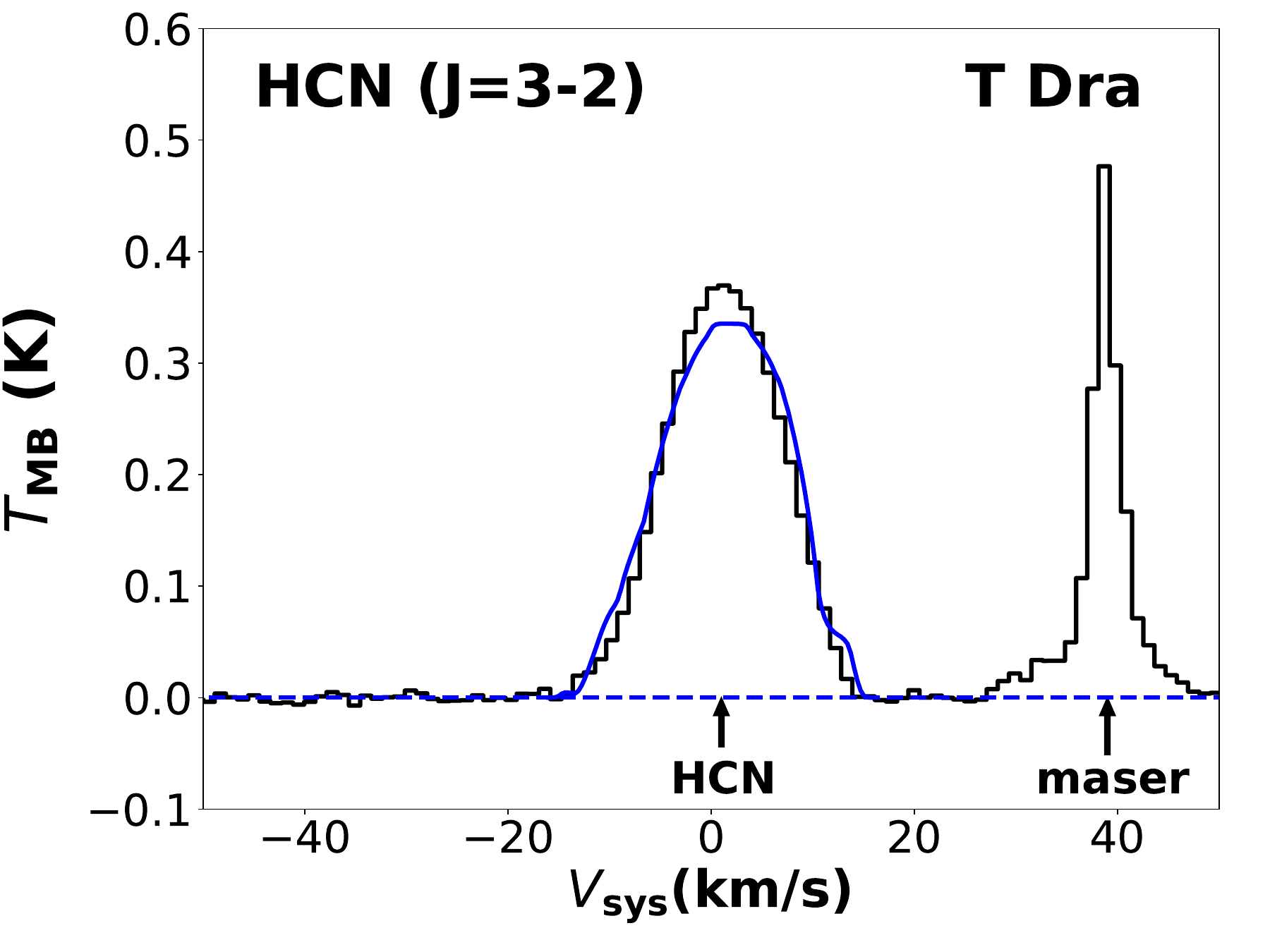}
     \end{subfigure}
     \begin{subfigure}[b]{0.33\linewidth}
         \centering
         \includegraphics[width=\linewidth]{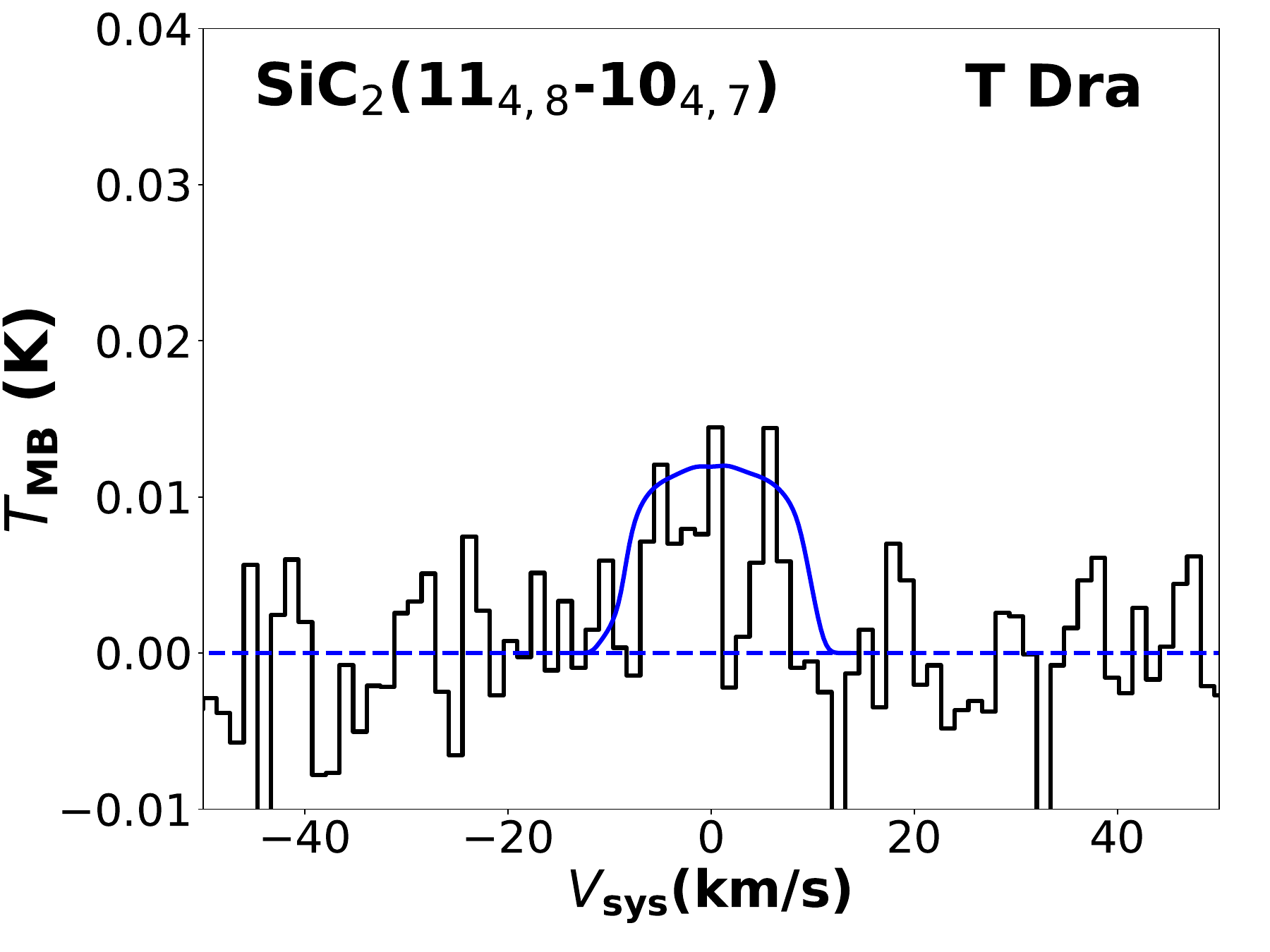}
     \end{subfigure}

     \begin{subfigure}[b]{0.33\linewidth}
         \centering
         \includegraphics[width=\linewidth]{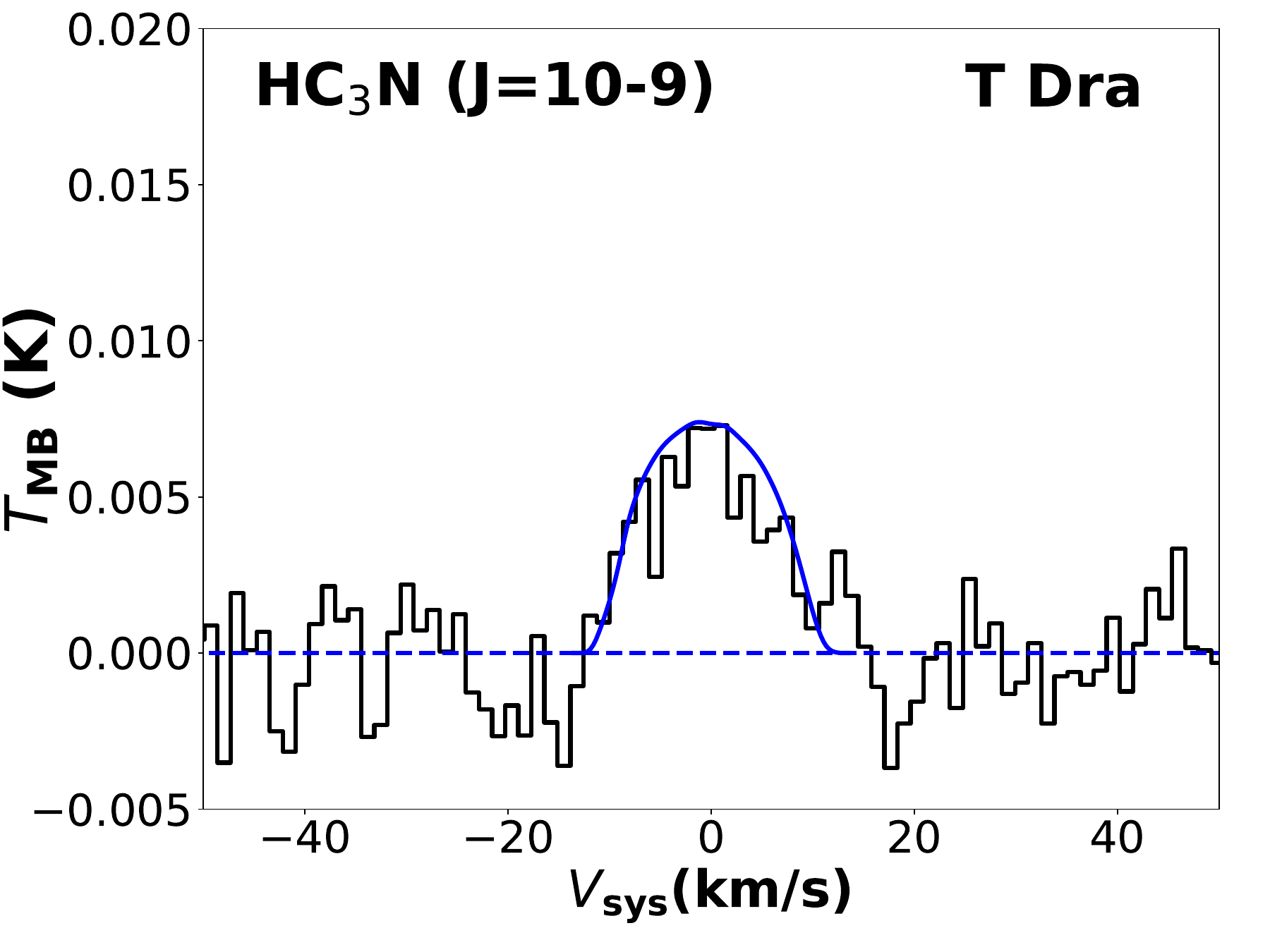}
     \end{subfigure}
     \begin{subfigure}[b]{0.33\linewidth}
         \centering
         \includegraphics[width=\linewidth]{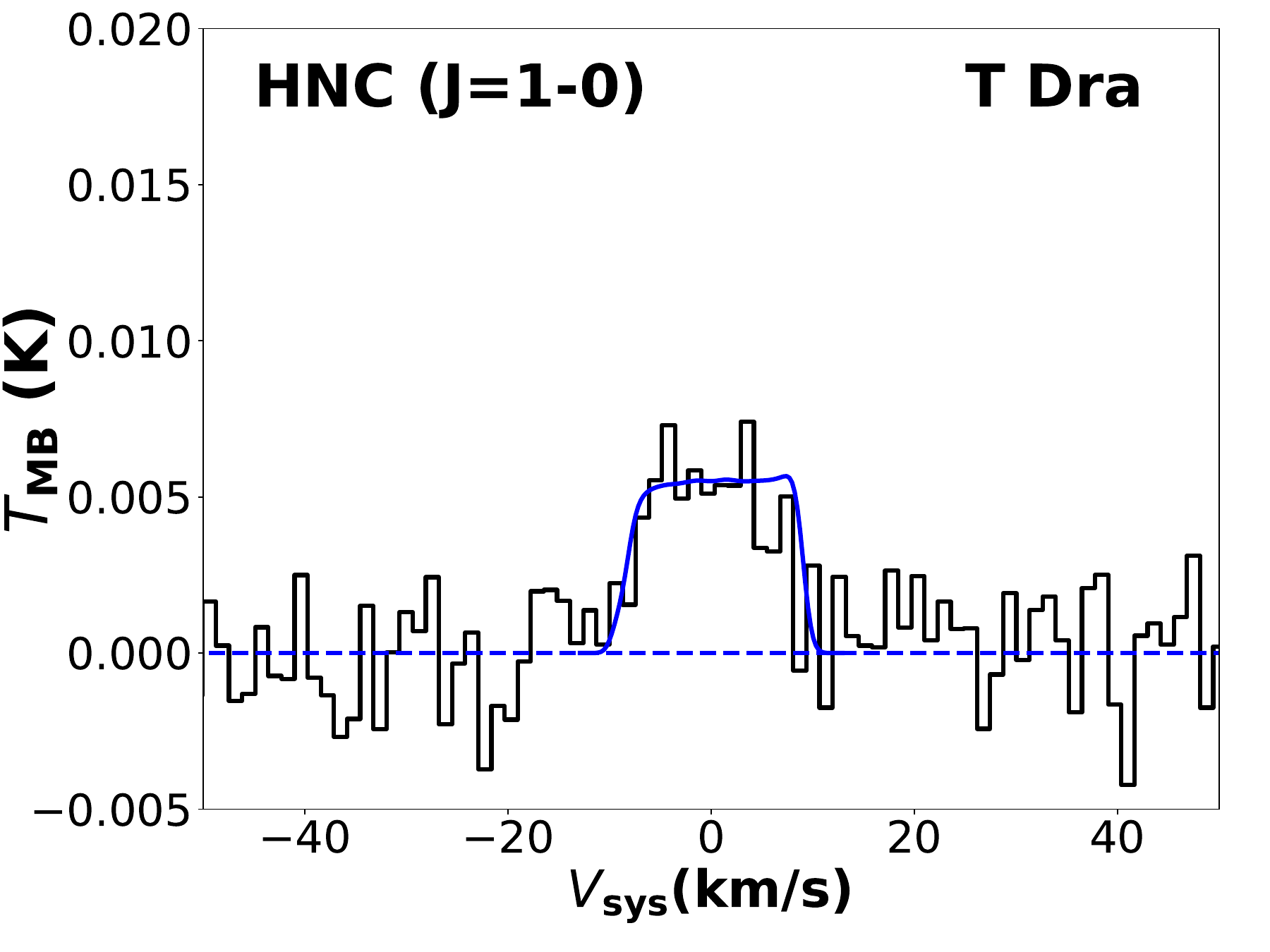}
     \end{subfigure}
     \begin{subfigure}[b]{0.33\linewidth}
         \centering
         \includegraphics[width=\linewidth]{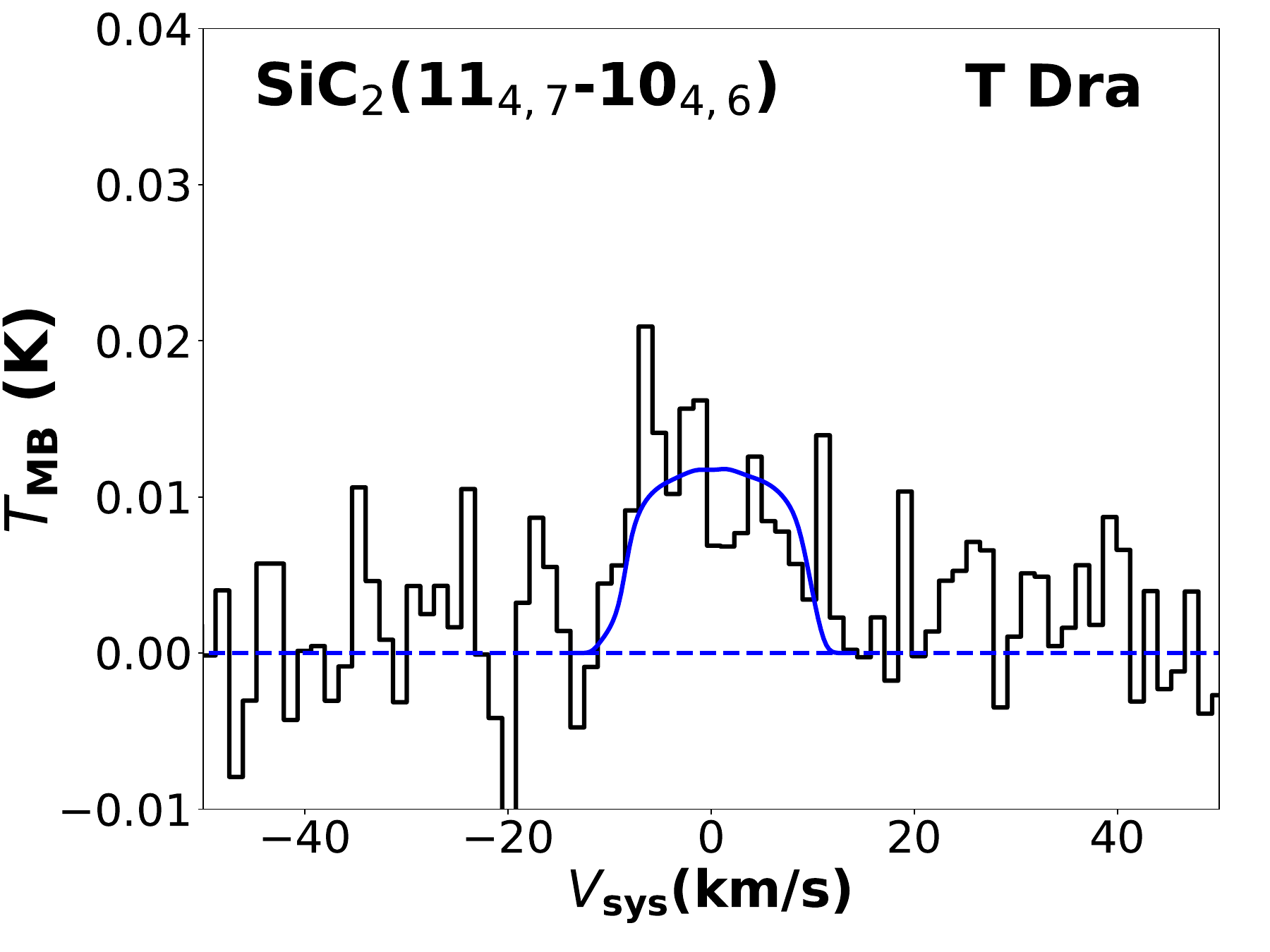}
     \end{subfigure}  

     \begin{subfigure}[b]{0.33\linewidth}
         \centering
         \includegraphics[width=\linewidth]{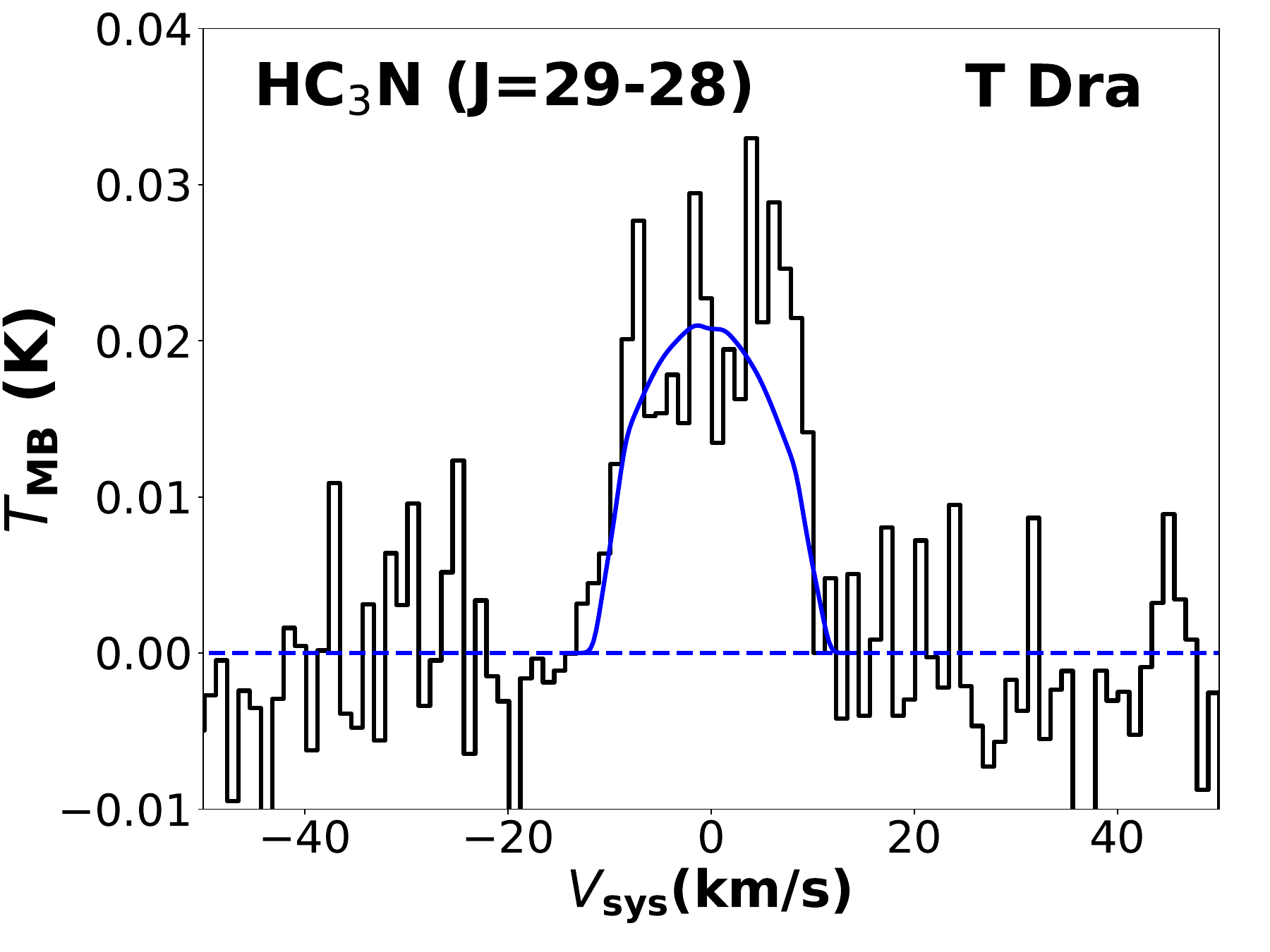}
     \end{subfigure}
     \begin{subfigure}[b]{0.33\linewidth}
         \centering
         \includegraphics[width=\linewidth]{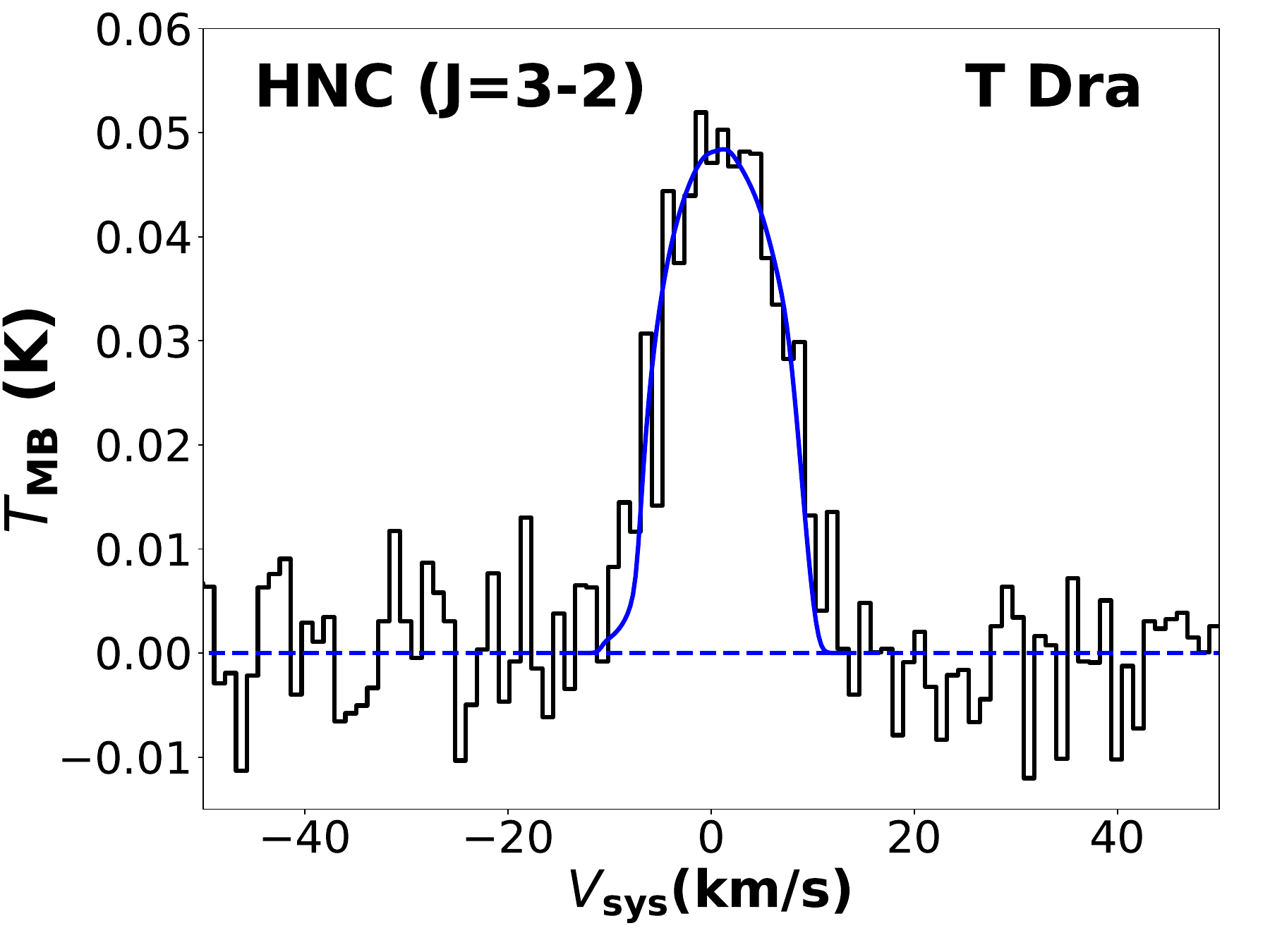}
     \end{subfigure}
     \begin{subfigure}[b]{0.33\linewidth}
         \centering
         \includegraphics[width=\linewidth]{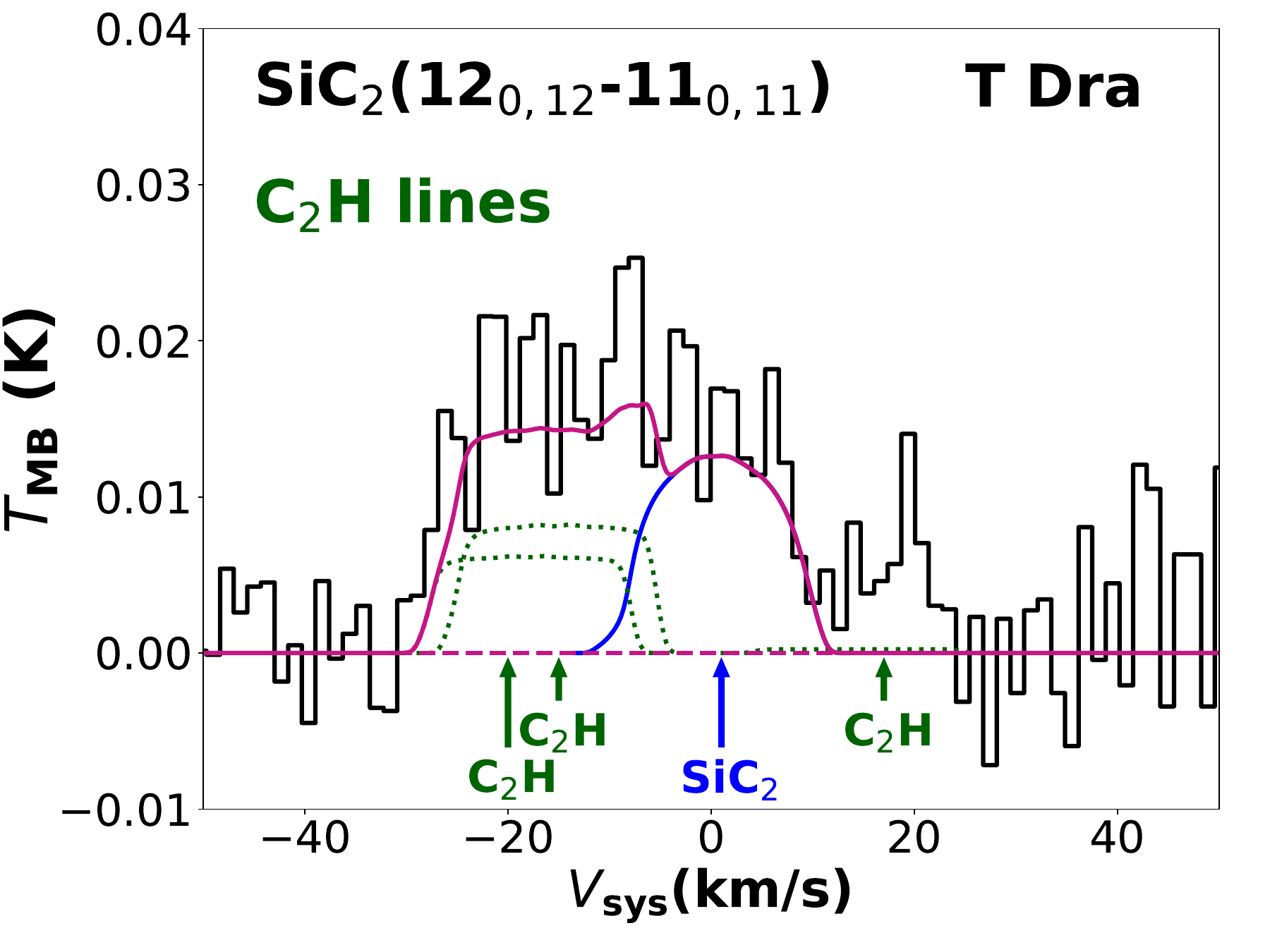}
     \end{subfigure}
        \caption{Spectra of detected lines (other than \hcoplus\, and CO) in the C-rich X-AGB T\,Dra. Observed spectra are shown in black (velocity resolution is $\delta v$=1.4\,\kms). Synthetic spectra are shown in blue. In the case of \sicdostres, the blended \cch\, lines are shown in green and the sum of the emission lines in purple.} 
        \label{fig:T-DRA_spectra}
\end{figure*}

\begin{figure*}[h!]
     \centering

     \begin{subfigure}[b]{0.33\linewidth}
         \centering
         \includegraphics[width=\linewidth]{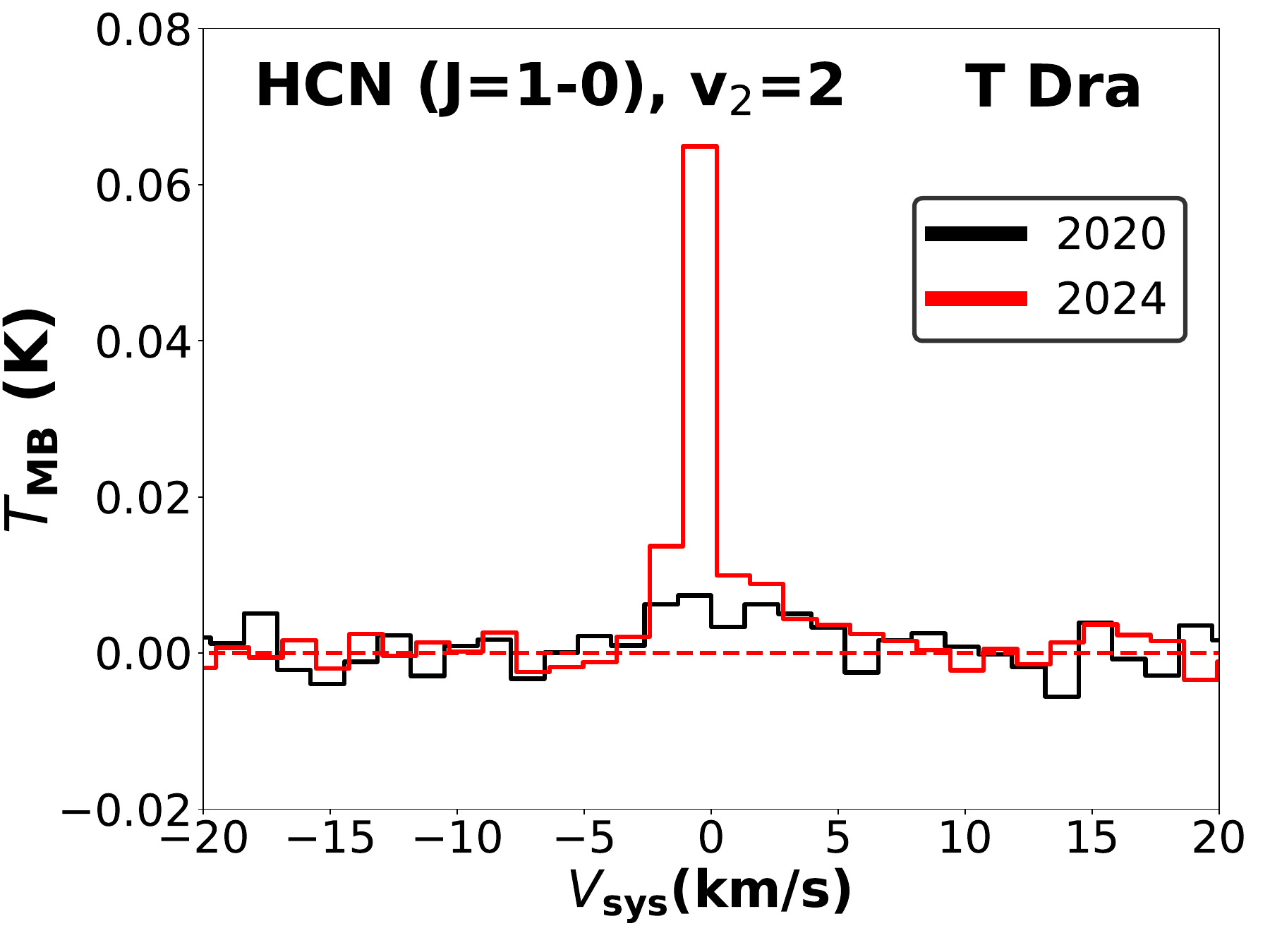}
     \end{subfigure}
     \begin{subfigure}[b]{0.33\linewidth}
         \centering
         \includegraphics[width=\linewidth]{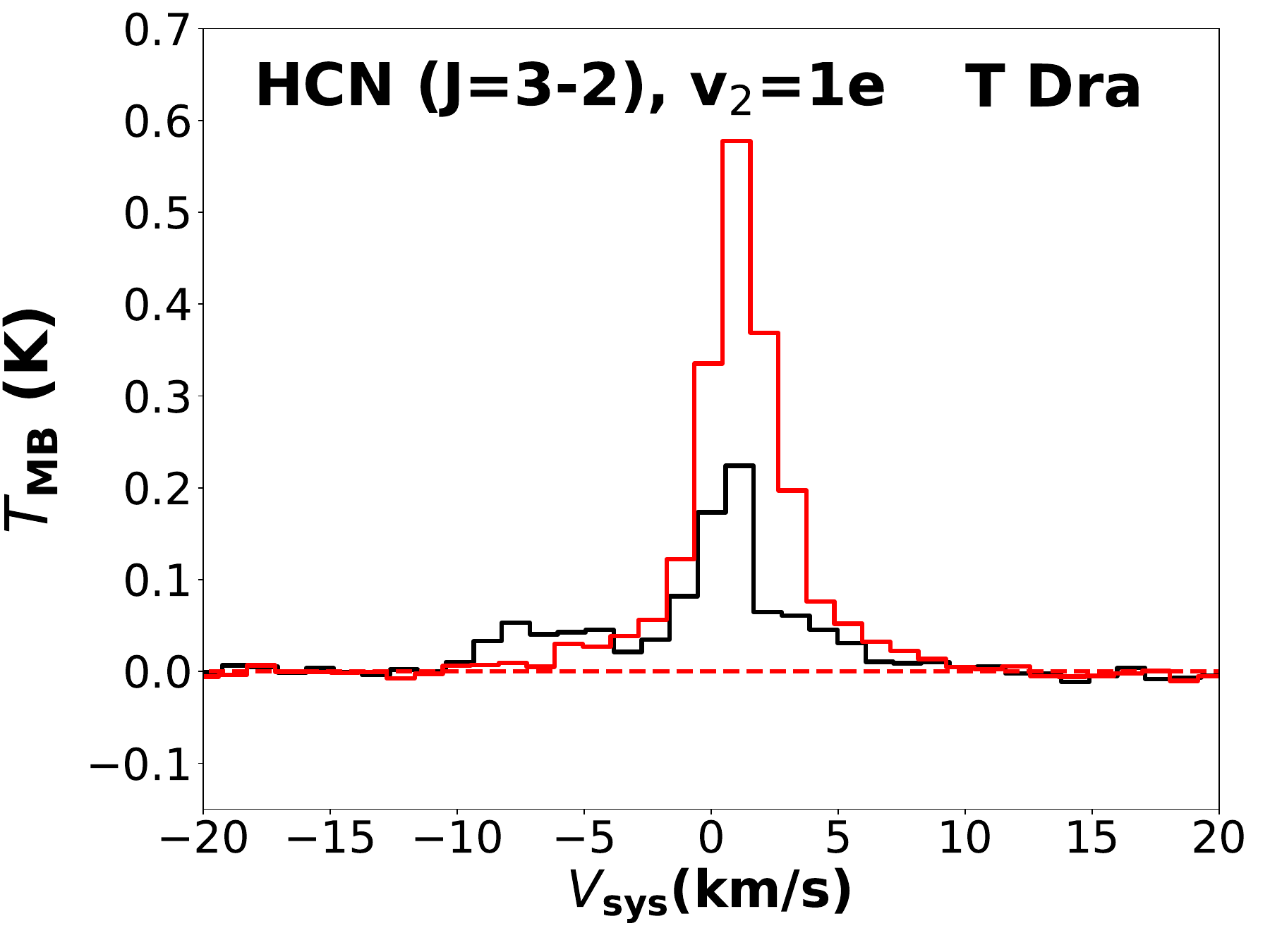}
     \end{subfigure}
     \begin{subfigure}[b]{0.33\linewidth}
         \centering
         \includegraphics[width=\linewidth]{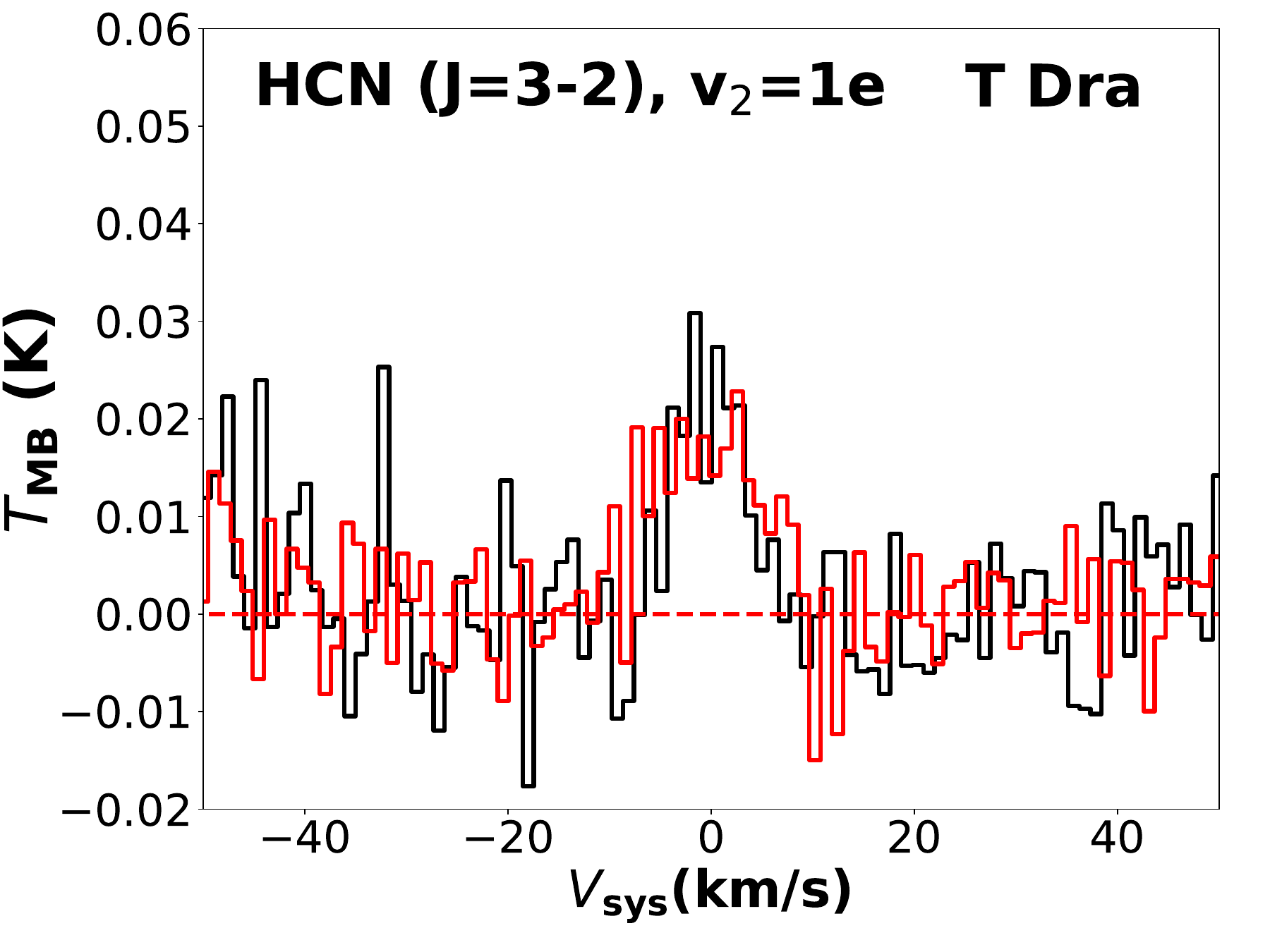}
     \end{subfigure}
        \caption{Spectra of \hcn\, vibrationally excited lines in T~Dra. Observed spectra are shown in black for 2020 observations and in red for 2024 observations (velocity resolution is $\delta v$=1.4\,\kms). } 
        \label{fig:T-DRA_spectra_vib}
\end{figure*}

\begin{figure*}[h!]
     \centering
     \begin{subfigure}[b]{0.33\linewidth}
         \centering
         \includegraphics[width=\linewidth]{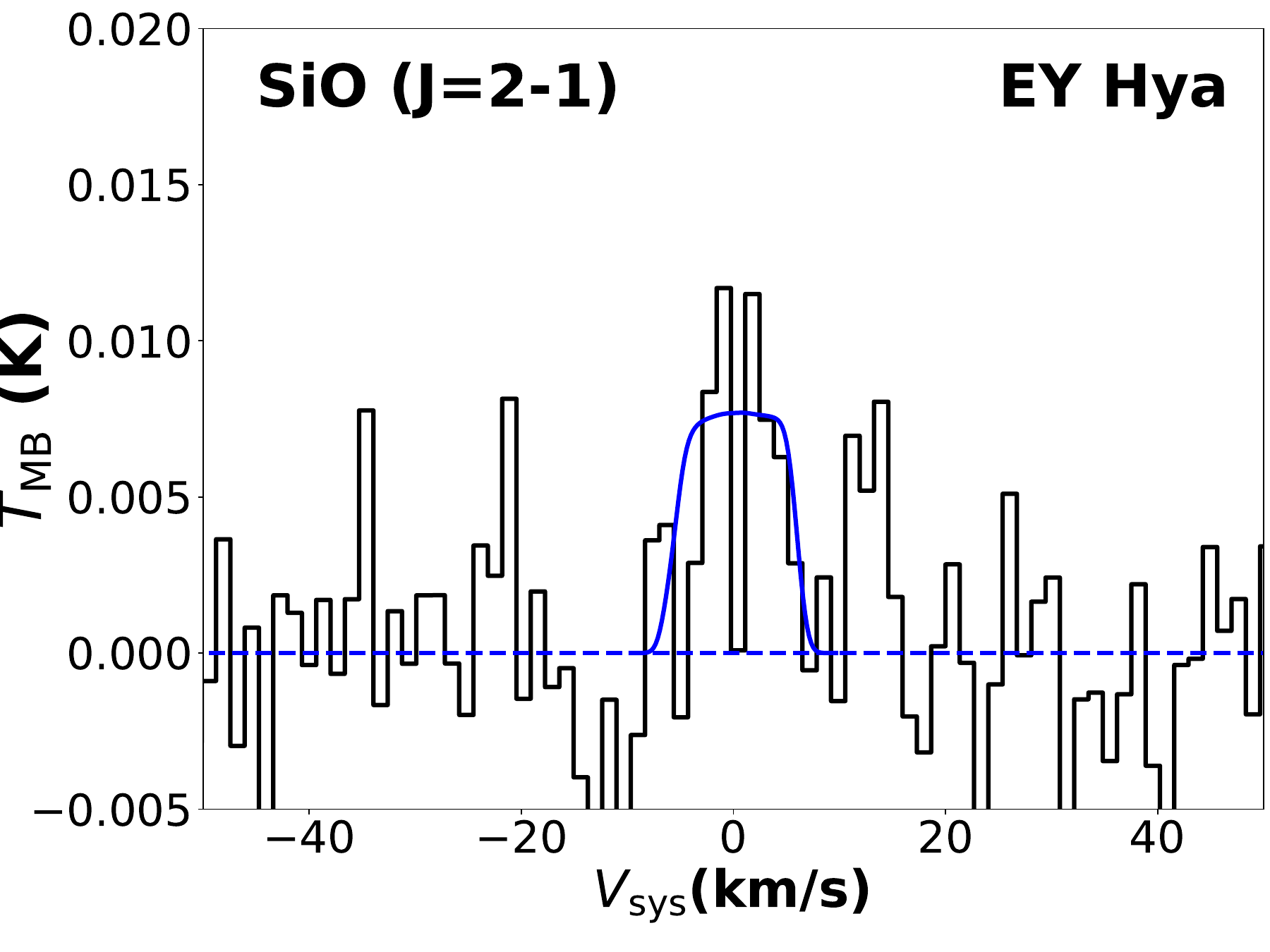}
     \end{subfigure}
     \begin{subfigure}[b]{0.33\linewidth}
         \centering
         \includegraphics[width=\linewidth]{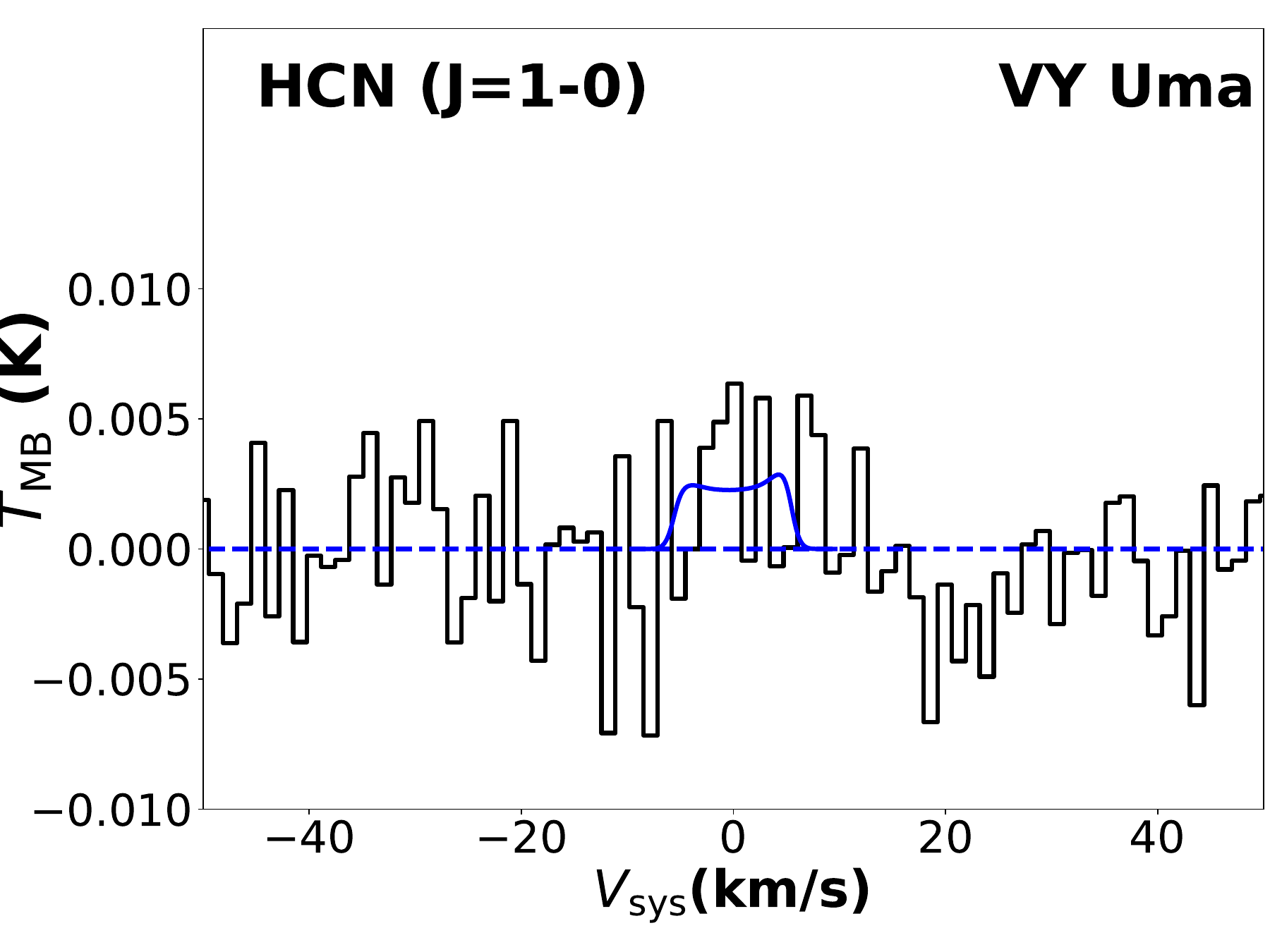}
     \end{subfigure}
     \begin{subfigure}[b]{0.33\linewidth}
         \centering
         \includegraphics[width=\linewidth]{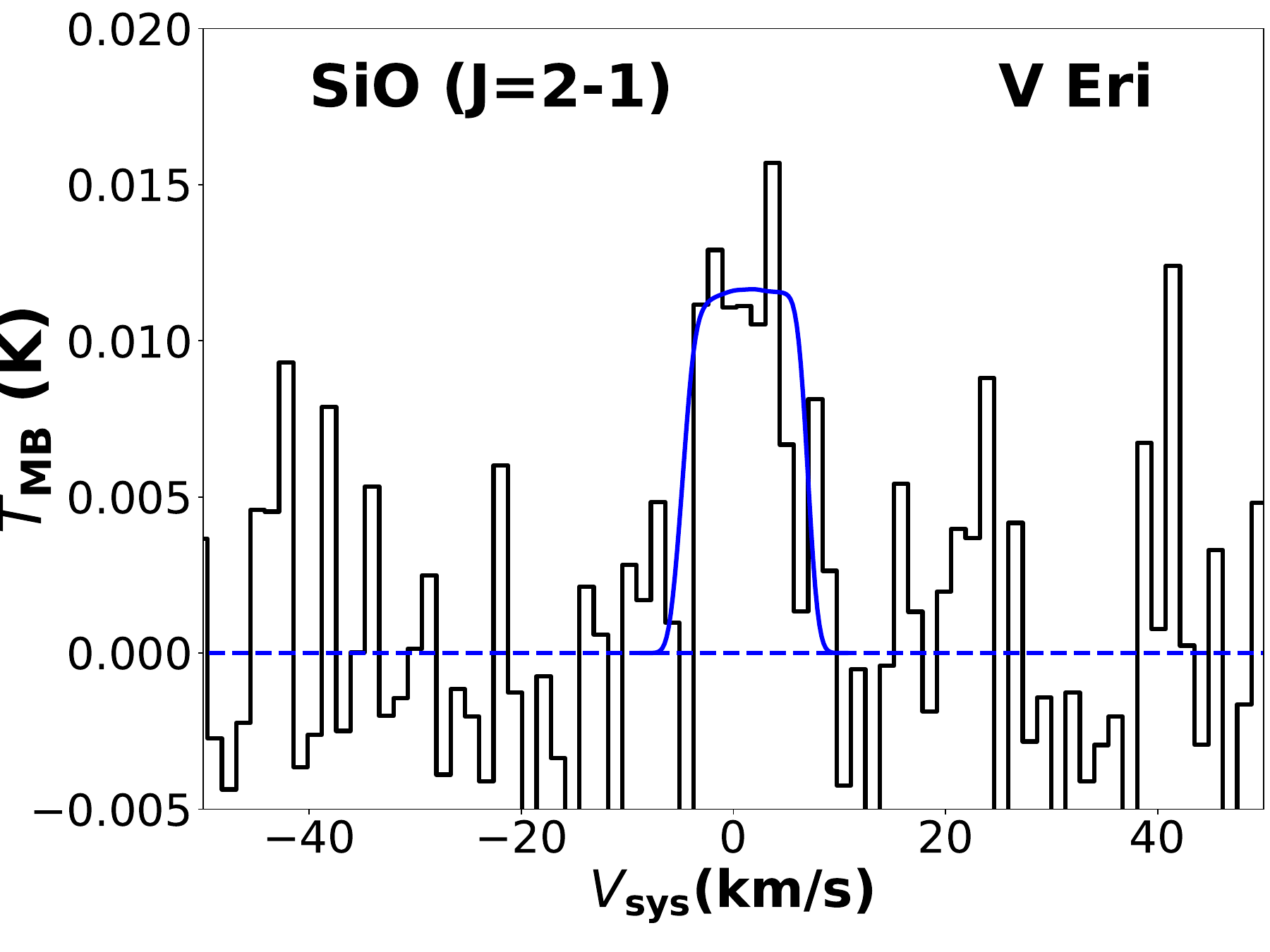}
     \end{subfigure}

     \begin{subfigure}[b]{0.33\linewidth}
         \centering
         \includegraphics[width=\linewidth]{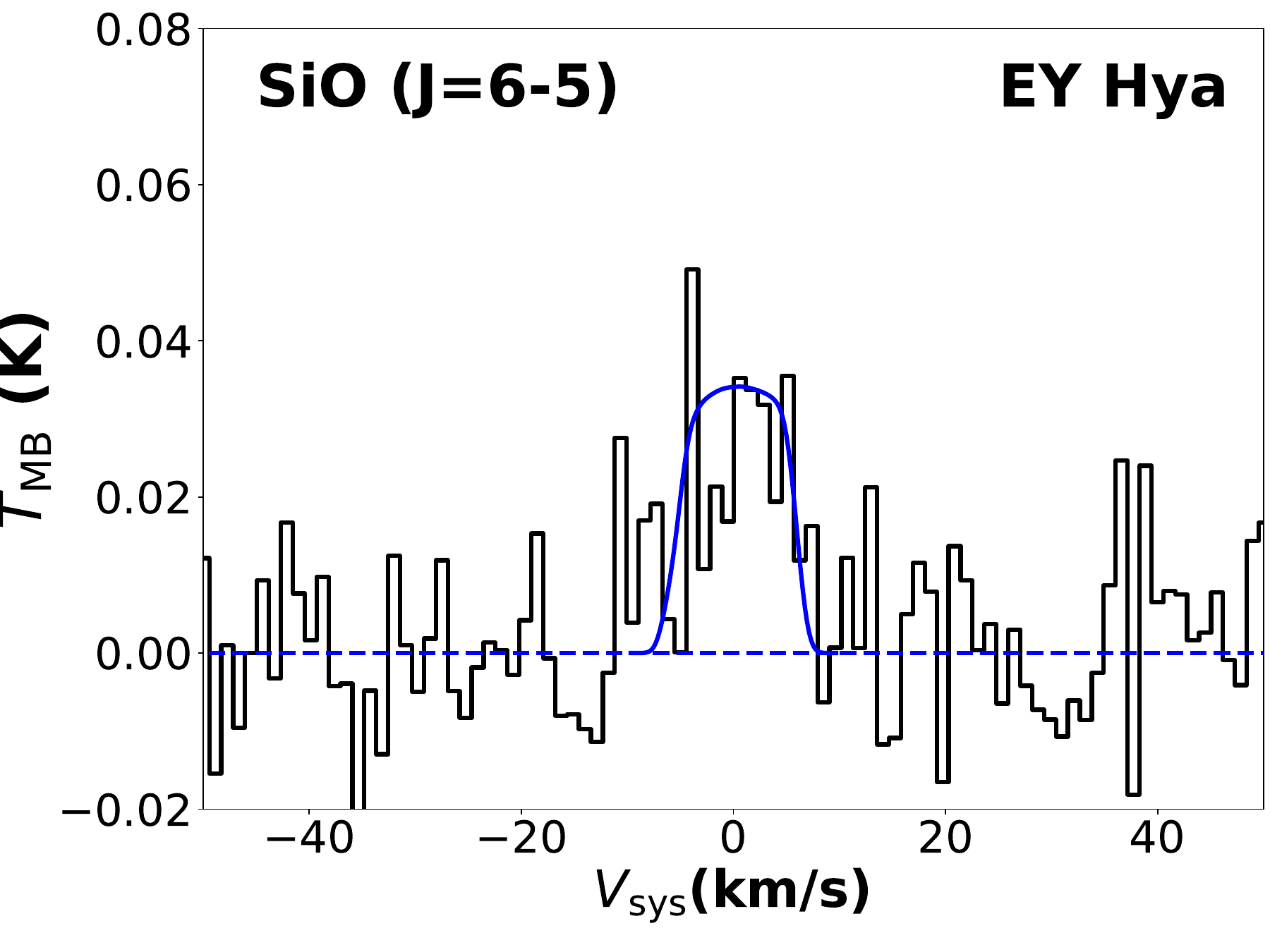}
     \end{subfigure}
     \begin{subfigure}[b]{0.33\linewidth}
         \centering
         \includegraphics[width=\linewidth]{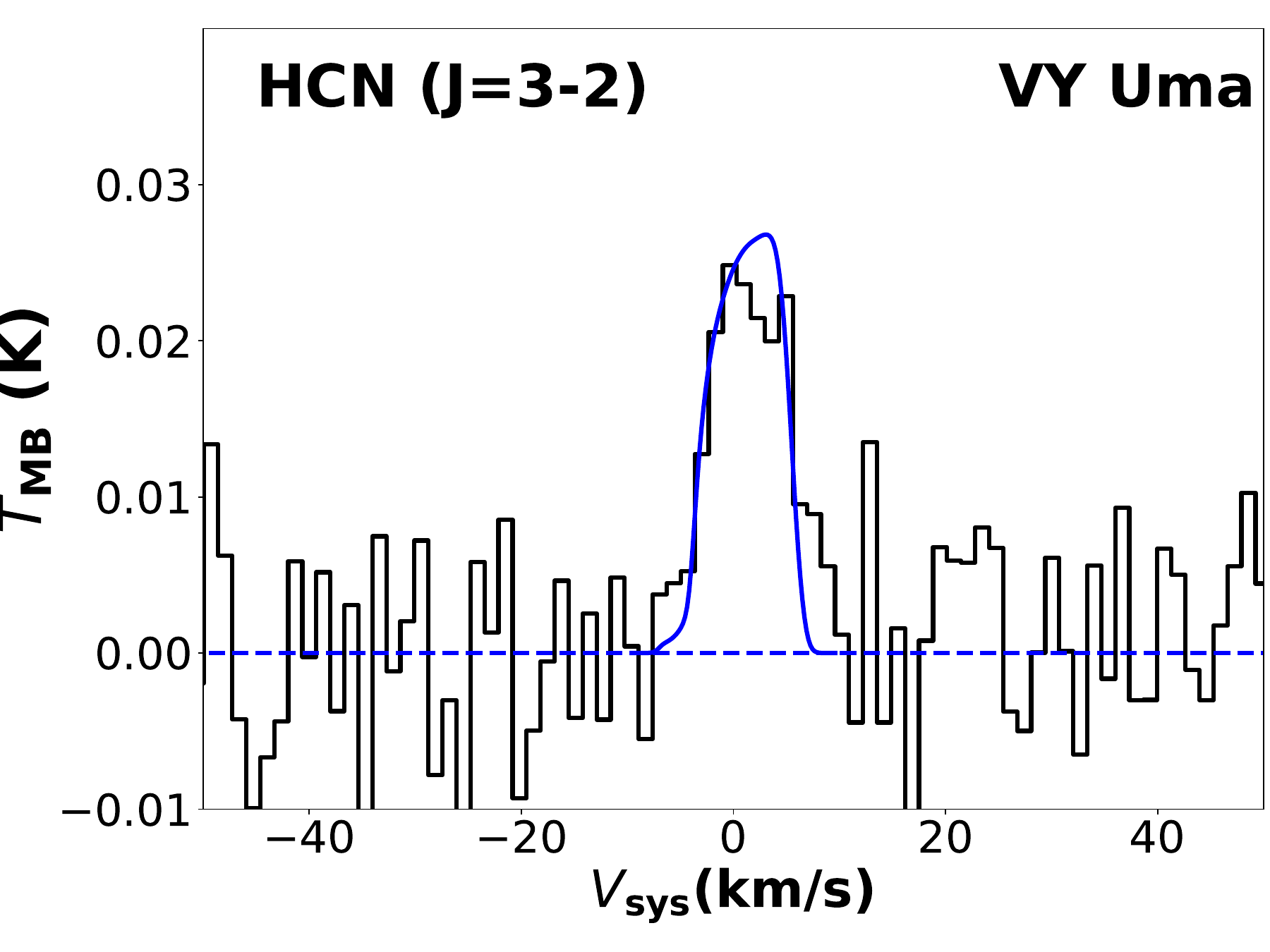}
     \end{subfigure}
     \begin{subfigure}[b]{0.33\linewidth}
         \centering
         \includegraphics[width=\linewidth]{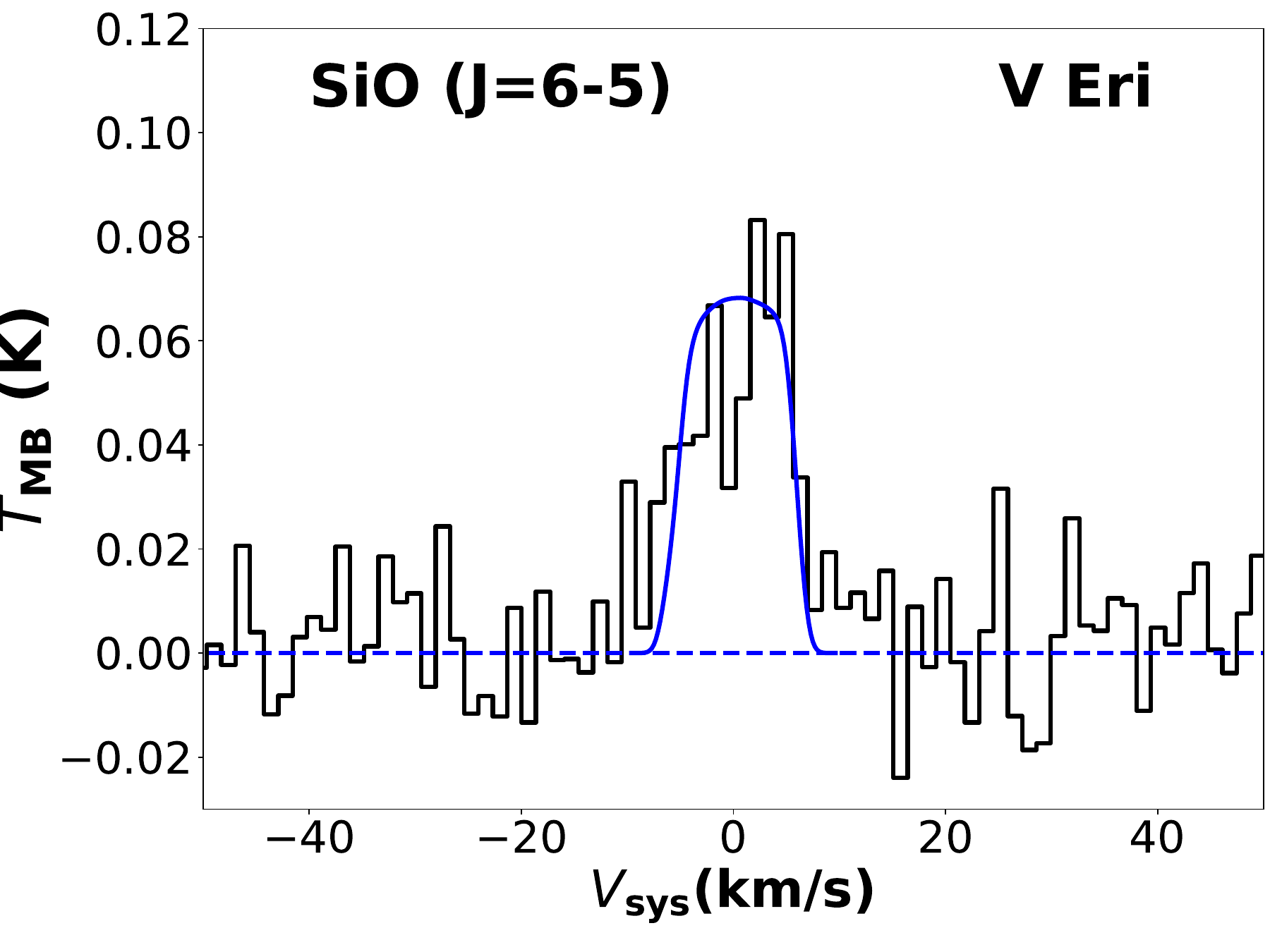}
     \end{subfigure}
        \caption{Spectra of \sio\, or \hcn\, for the rest of the sources: EY Hya (O-rich X-AGB), VY Uma (C-rich without X-ray detection), and V Eri (O-rich without X-ray detection), from left to right. Observed spectra are shown in black (velocity resolution is $\delta v$=1.4\,\kms). Synthetic spectra are shown in blue.}
        \label{fig:others_spectra}
\end{figure*}

We detected \hcouno\, and \hcodos\,  in the C-rich X-AGB T\,Dra (Fig.~\ref{fig:HCO+_T-DRA_spectra}); this molecular ion remained undetected in the other targets. In T\,Dra, lines of additional molecular species were also detected, denoting its comparatively richer chemistry, namely: \sio, \hcn, \hnc, \hctresn, \sicdos, \cch\, and \sis\, (Fig.~\ref{fig:T-DRA_spectra} and Table~\ref{tab:lines}). Moreover, three vibrationally excited \hcn\, lines were observed, two of which exhibited characteristic maser line features (Fig.~\ref{fig:T-DRA_spectra_vib}). Except for the \hcnuno\, $v_{2}$=$2$ maser line, previously reported in T\,Dra by \cite{Lucas_1988}, these are all first time detections in this target. We also detected \sio\, in the two O-rich AGB stars EY\,Hya and V\,Eri, and \hcn\, in the C-rich AGB star VY\,Uma (Fig.~\ref{fig:others_spectra}), all of them first detections.

An inspection of the spectra of T\,Dra on the two observed epochs revealed significant variability in the intensities of the \hcoplus\,lines, especially the \hcodos\, line, which varies by a factor $\sim$2-3. These variations exceed the estimated absolute flux calibration uncertainties ($\simeq$15\% at 3mm and $\simeq$25\% at 1mm). The \hcn \, maser lines also show clear variations (both in intensity and profile shape) between the two observing epochs. The lower signal-to-noise ratio of the \hcoplus\,lines prevent us from discerning any significant changes in their line profiles (if there are any; see Sect.~\ref{dis} for a discussion). For the other lines detected in T\,Dra in both epochs, no significant variability was observed. Therefore, the time-averaged spectra from the two epochs are presented in Fig.~\ref{fig:T-DRA_spectra}. The observations of the lines detected in T\,Dra, which show no variation between epochs, are presented separately for each epoch in Appendix~\ref{Two_epochs}.

The rotational lines identified, together with the main parameters deduced from their spectra are summarised in Table~\ref{tab:lines}. The detected thermal lines show the standard line profiles found in AGB CSEs \citep[see e.g.][]{Olofsson_1993}. On the other hand, the maser lines display the typical narrow spikes in their profiles \citep[see e.g.][]{Jeste_2022}.

We determined the velocity-integrated fluxes to each detected line, their full width at half maximum (FWHM) by fitting a Gaussian profile as well as their average expansion velocities by fitting the standard so-called Shell-type profile available in CLASS \citep[see e.g.][for details]{Alonso-Hernandez_2024}. The gas expansion velocities are slightly lower than those derived from the CO lines \citep[11.8 \kms\ for T\,Dra, 10.4 \kms\ for EY\,Hya, 6.6 \kms\ for VY\,Uma, and 10.2 \kms\ for V\,Eri, see][]{Alonso-Hernandez_2024}. These lower expansion velocities may reflect larger contributions to the line emissions from the inner regions of the envelope, where terminal velocities are not reached. 

We note that maser lines are, as expected, considerably narrower than CO and other thermal lines, a common well-known property in AGB CSEs. This occurs because HCN maser lines are formed in the innermost regions of the envelope, where the wind starts to accelerate \citep[see e.g.][]{Hofner_2018}. Furthermore, for the HCN thermal lines and the \sicdostres\, line, the FWHM and \vexp\ values presented in Table\,\ref{tab:lines} overestimate the true expansion velocity because of hyperfine splitting and severe line blending with \cch\ lines (Fig.~\ref{fig:T-DRA_spectra}), respectively.

\section{Analysis: Line radiative transfer models}\label{anal}

To constrain the abundances of the detected molecular species, we used a radiative transfer model to reproduce the observed line profiles. First, we modelled the CO lines to constrain the main physical parameters of the CSEs of our targets. These parameters were then used as inputs for customised models of the emission from \hcoplus\ and other identified molecular species. This approach allowed us to determine their average abundances and gain insights into their radial distributions.

The radiative transfer analysis was performed with the Madrid Excitation Code \citep[MADEX,][]{MADEX}. The model, which is described in detail by \cite{agundez_2012} and has been widely used \citep[see e.g.][]{agundez_2017, Velilla-Prieto_2017, Massalkhi_2024}, is based on an envelope structure formed by concentric spherical shells (multi-shell) and estimates the rotational level populations for each molecule by solving the statistical equilibrium equation under the Large Velocity Gradient (LVG) approximation \citep{Sobolev_1960}. This approximation assumes that the expansion velocity and its gradient are larger than the local velocity dispersion, which holds throughout the bulk of AGB CSE. A detailed discussion of the LVG approximation and its validity can be found in, e.g. \cite{Bujarrabal_2013a}.

We consider up to $J$=50 rotational levels for the radiative transfer calculations. In addition, the excitation produced by infrared pumping from the star is included for CO, \sio\, and \sis. We consider the first vibrational state in addition to the ground state (i.e $v$=0 and $v$=1). We use the rate coefficients for pure rotational transitions induced by inelastic collisions between CO and $\rm H_{2}$ as well as CO and He \citep[collisional rate coefficients from][respectively]{Yang_2010,Cecchi-Pestellini_2002}. We assumed the statistical ortho-to-para ratio of 3 for $\rm H_{2}$ and a Helium solar abundance of 0.17 relative to $\rm H_{2}$. For the rest of molecules, we adopted the following collisional rate coefficients: \sio\, \citep{Dayou_2006}, \hcn\,\citep{Abdallah_2012}, \hcoplus\, \citep{Flower_1999}, \hnc\, \citep{Dumouchel_2011}, \hctresn\,\citep{Wernli_2007, Wernli_2007b}, \sis\, \citep{Lique_2008}, \sicdos\,\citep{Chandra_2000}, \cch\, \citep{Spielfiedel_2012}, and $\rm N_{2}H^{+}$\, \citep{Daniel_2005}.

\subsection{CO line modelling: Methods and assumptions}\label{MADEX_CO}

\begin{figure*}[h!]
     \centering
     
     \begin{subfigure}[b]{0.33\linewidth}
         \centering
         \includegraphics[width=\linewidth]{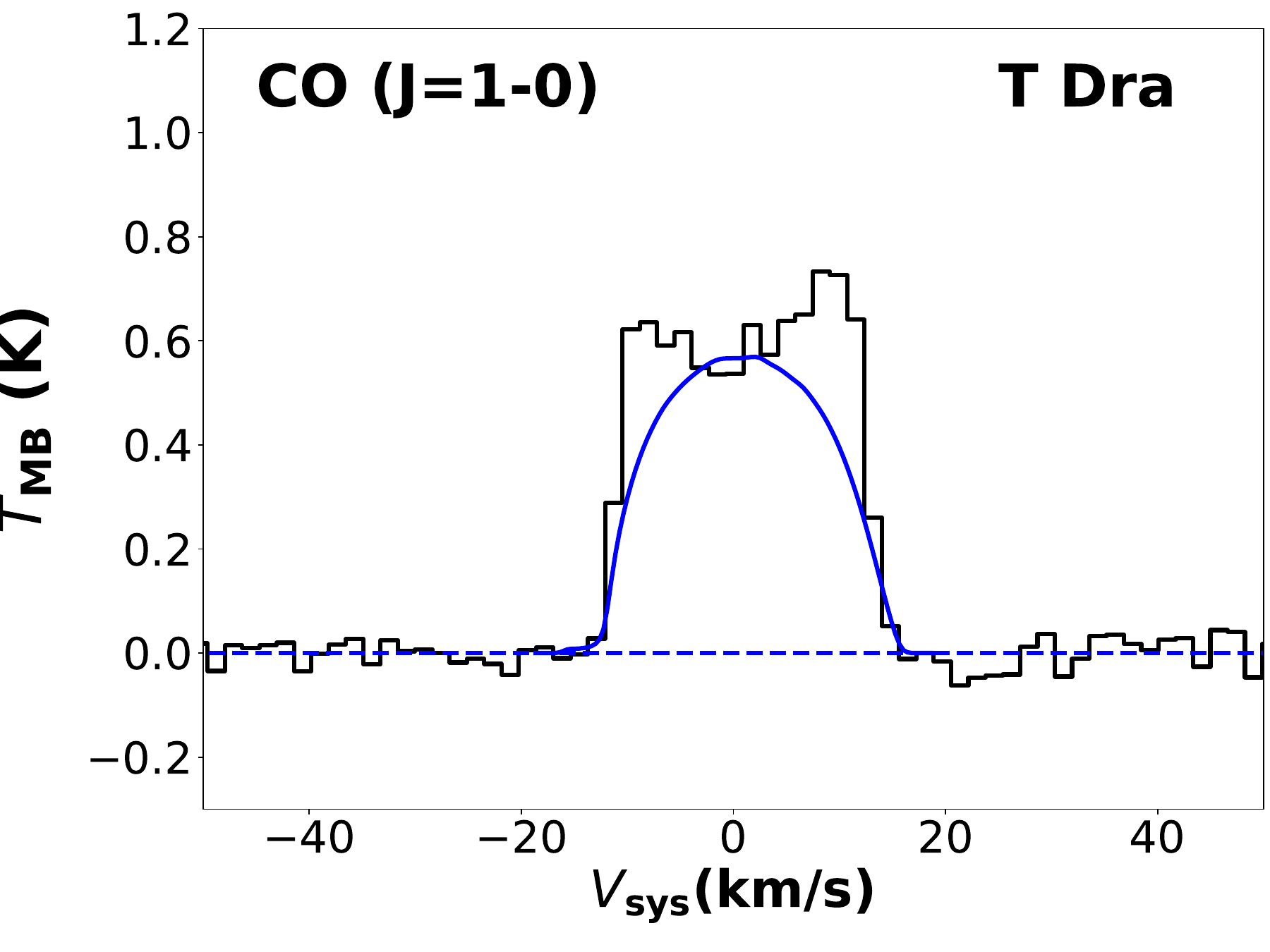}
     \end{subfigure}
     \begin{subfigure}[b]{0.33\linewidth}
         \centering
         \includegraphics[width=\linewidth]{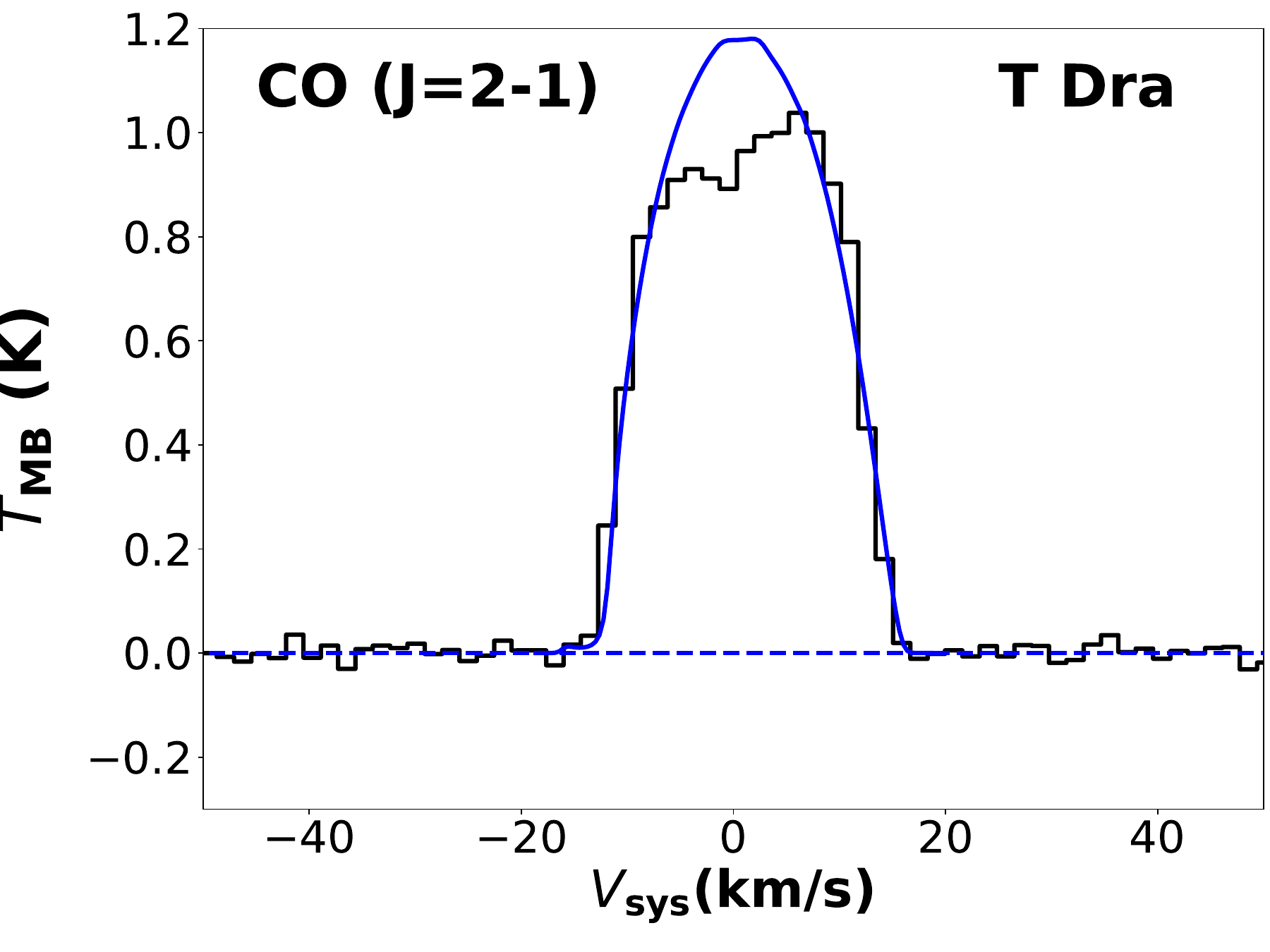}
     \end{subfigure}
     \begin{subfigure}[b]{0.33\linewidth}
         \centering
         \includegraphics[width=\linewidth]{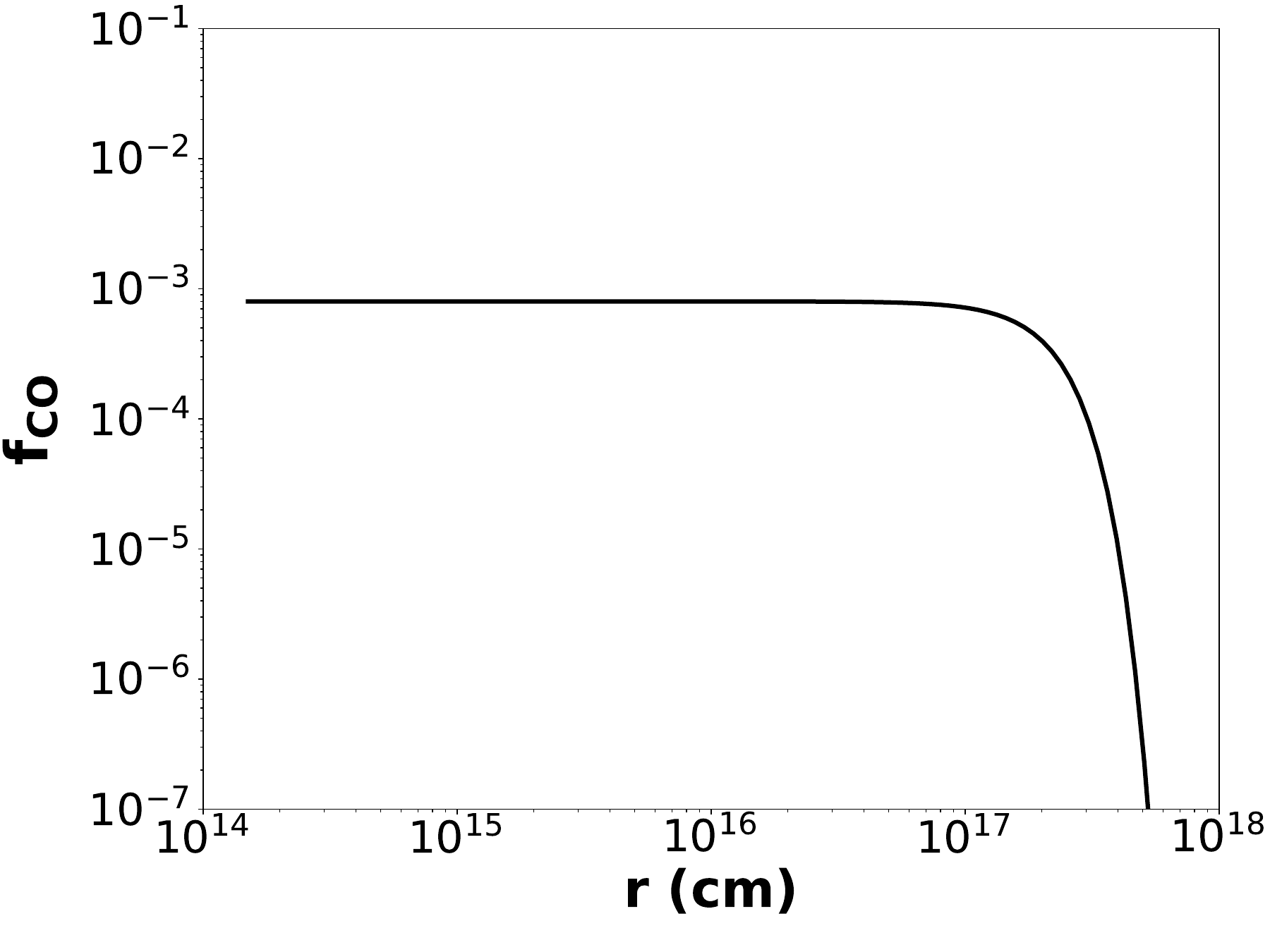}
     \end{subfigure}

     \begin{subfigure}[b]{0.33\linewidth}
         \centering
         \includegraphics[width=\linewidth]{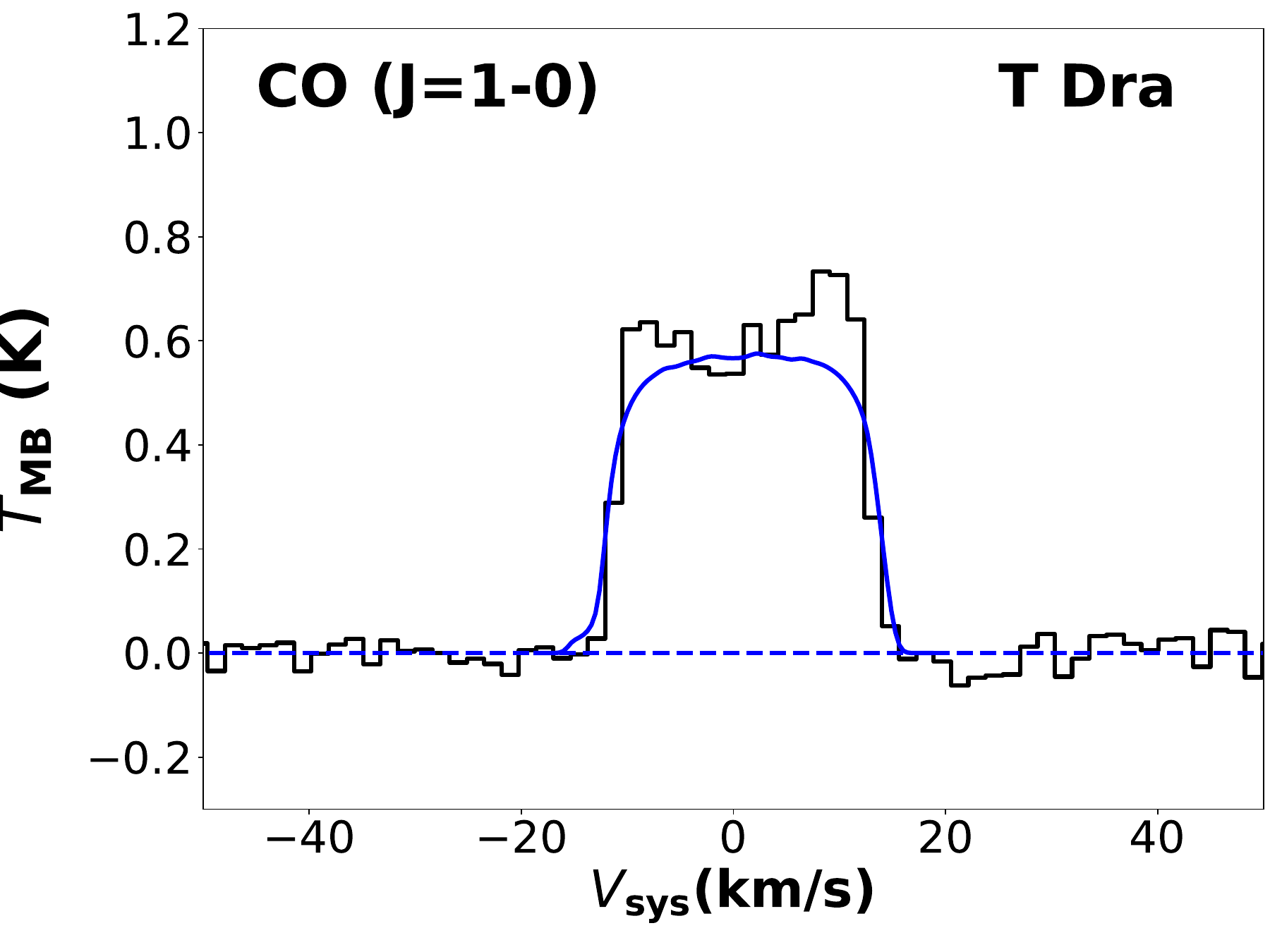}
     \end{subfigure}
     \begin{subfigure}[b]{0.33\linewidth}
         \centering
         \includegraphics[width=\linewidth]{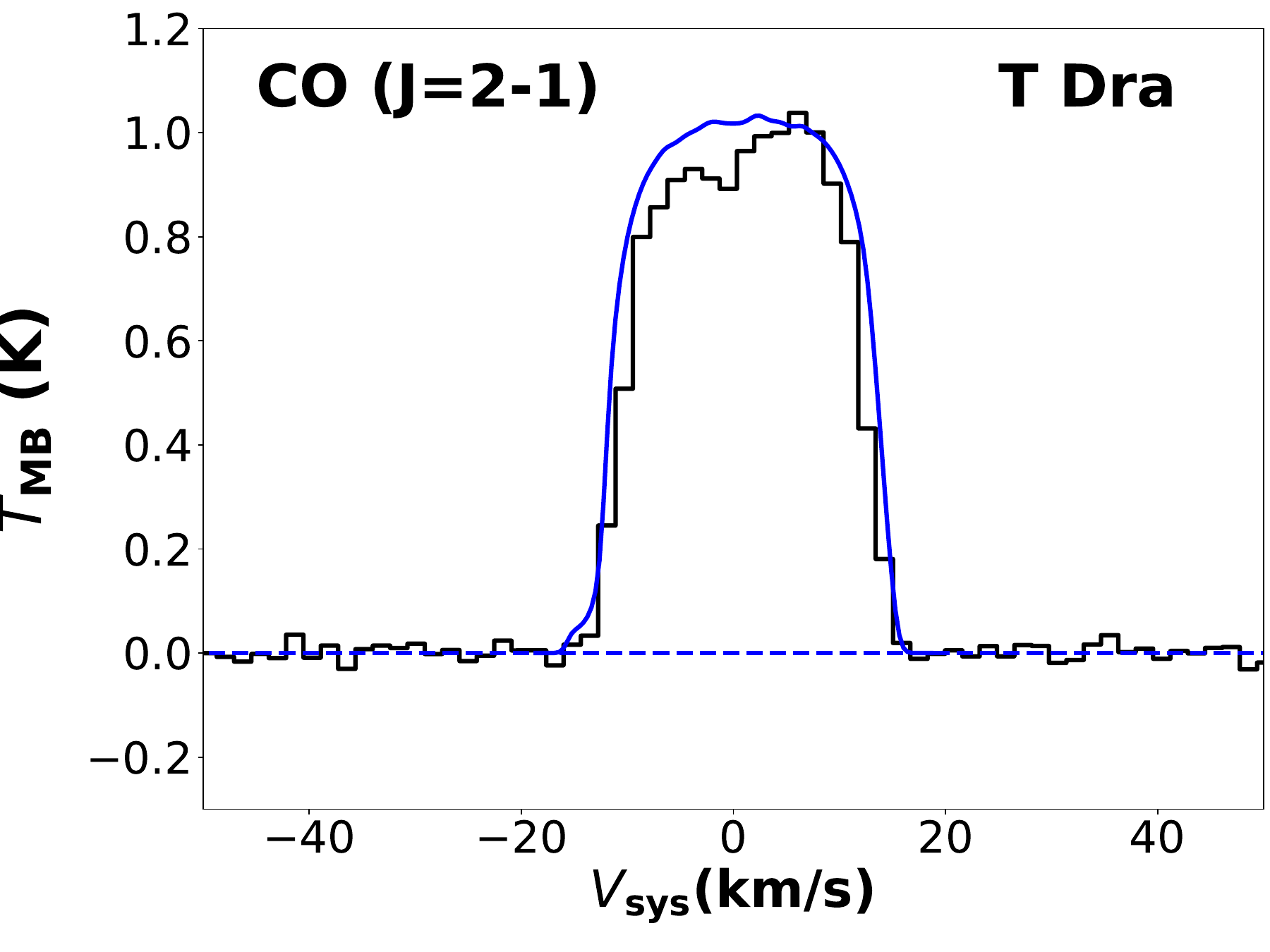}
     \end{subfigure}
     \begin{subfigure}[b]{0.33\linewidth}
         \centering
         \includegraphics[width=\linewidth]{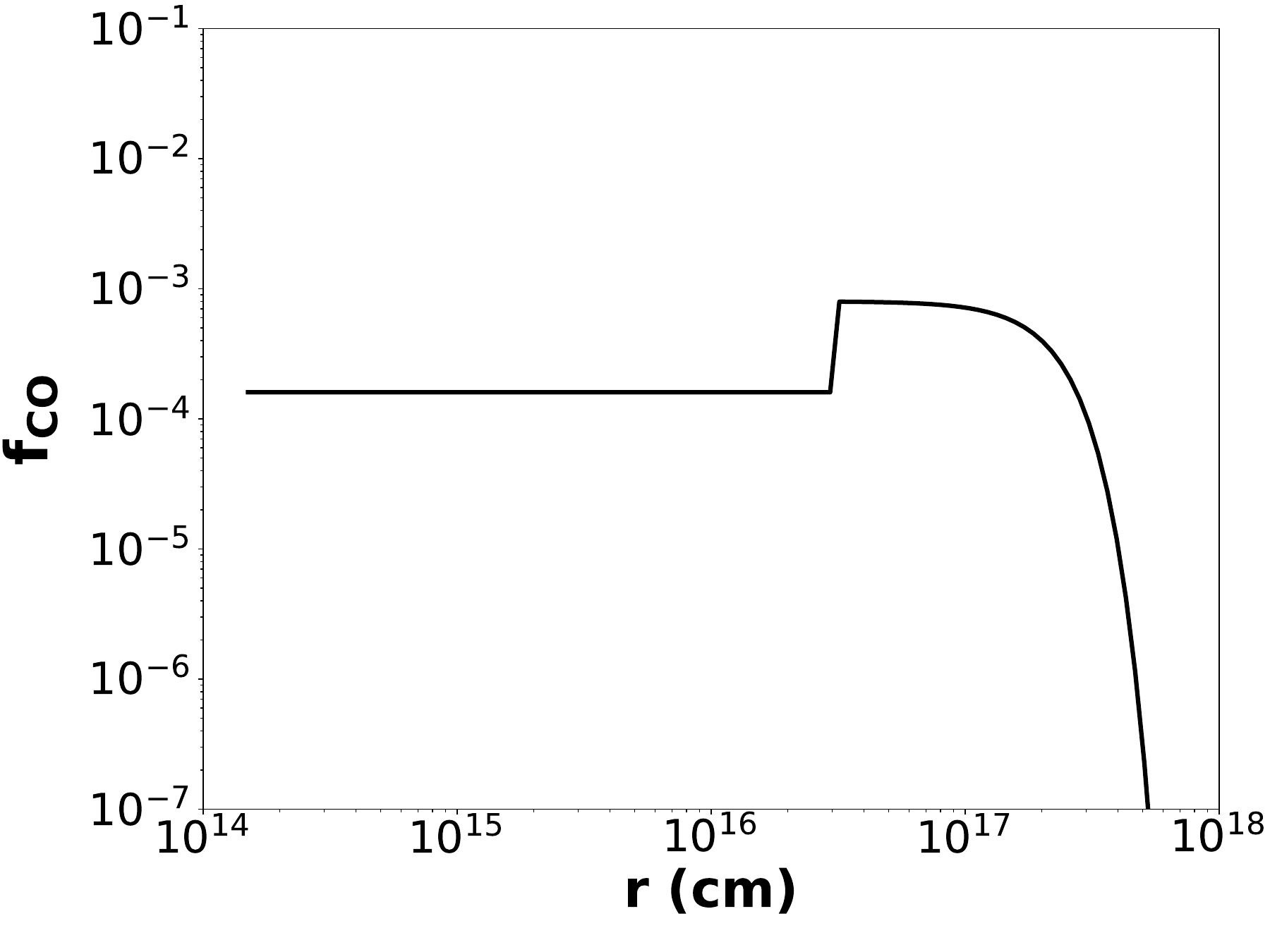}
     \end{subfigure}
        \caption{Model (a) assuming the standard CO radial abundance (top) and model (b) assuming a centrally depleted CO radial abundance likely caused by an enhanced CO dissociation driven by X-ray radiation (bottom). \doceuno\, (left) and \docedos\, (middle) observed and synthetic spectra in the C-rich X-AGB T\,Dra. Observed spectra are shown in black (velocity resolution is $\delta v$=1.6\,\kms). Synthetic spectra are shown in blue. Right panels: the different adopted CO abundance profiles for each model. As discussed in Sect.~\ref{anal}, model (b) fits the line shapes better than model (a).}
        \label{fig:CO_spectra_T-DRA}
\end{figure*}

\begin{figure*}[h!]
     \centering
     \begin{subfigure}[b]{0.33\linewidth}
         \centering
         \includegraphics[width=\linewidth]{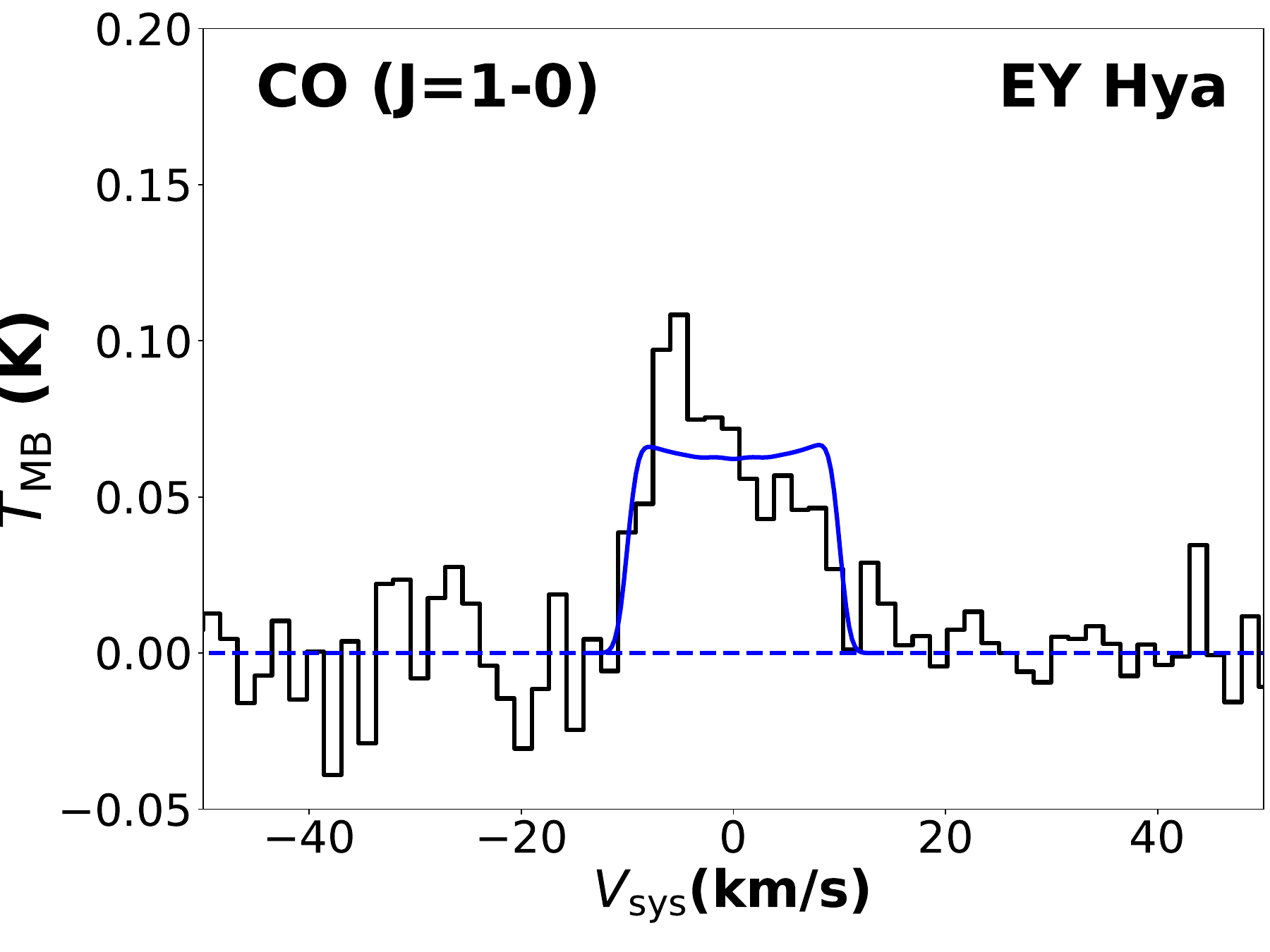}
     \end{subfigure}
     \begin{subfigure}[b]{0.33\linewidth}
         \centering
         \includegraphics[width=\linewidth]{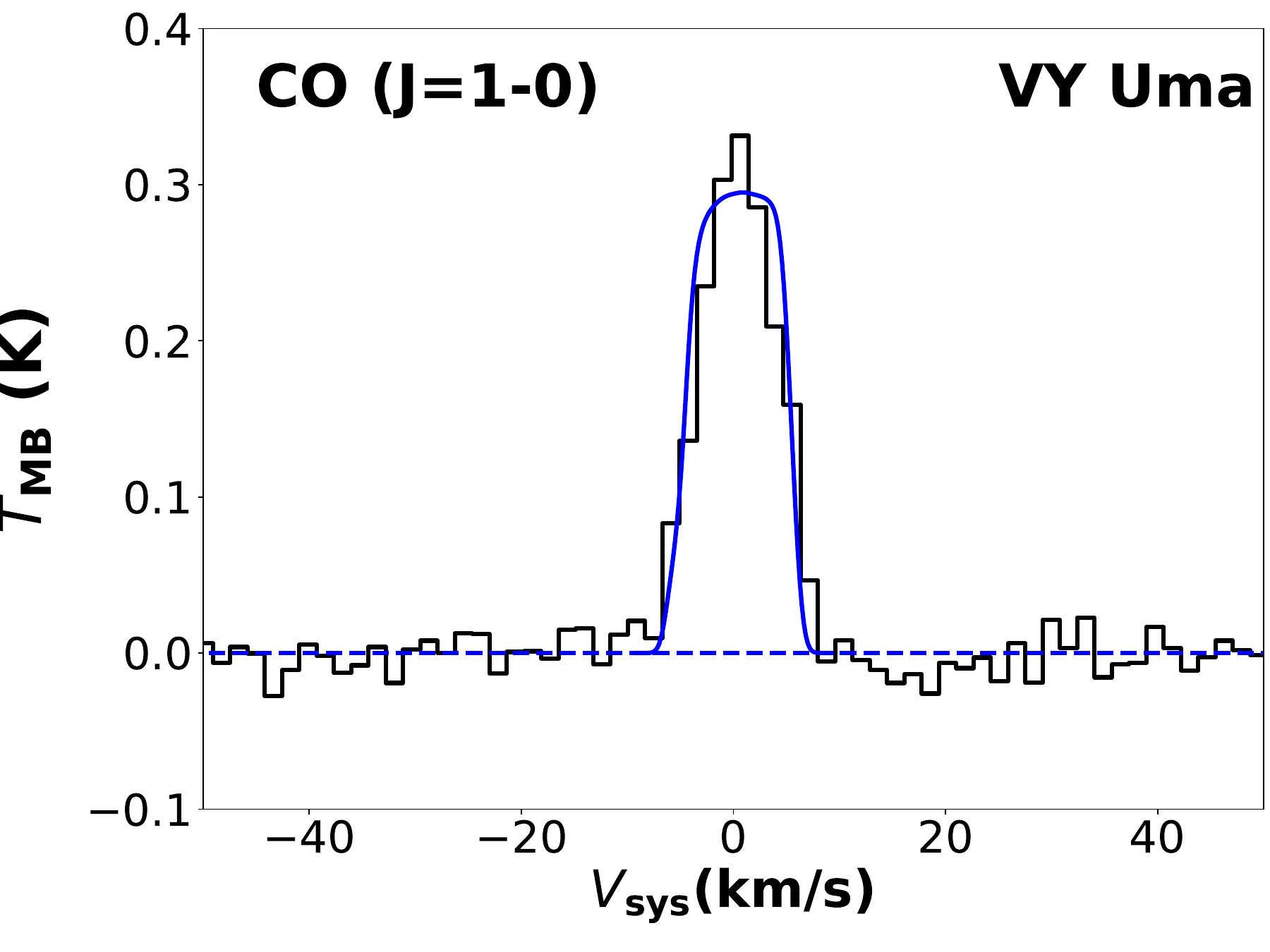}
     \end{subfigure}
     \begin{subfigure}[b]{0.33\linewidth}
         \centering
         \includegraphics[width=\linewidth]{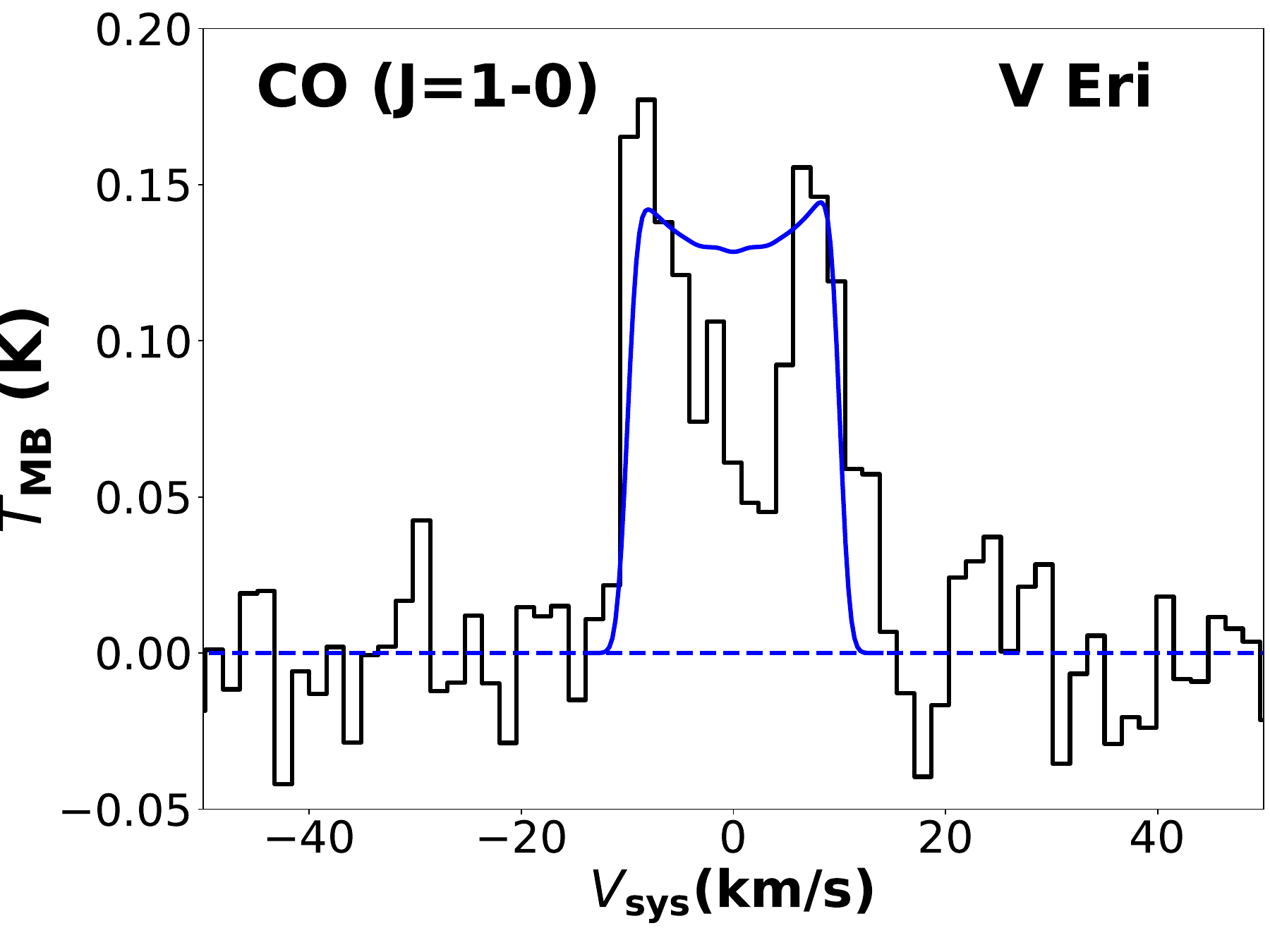}
     \end{subfigure}

     \begin{subfigure}[b]{0.33\linewidth}
         \centering
         \includegraphics[width=\linewidth]{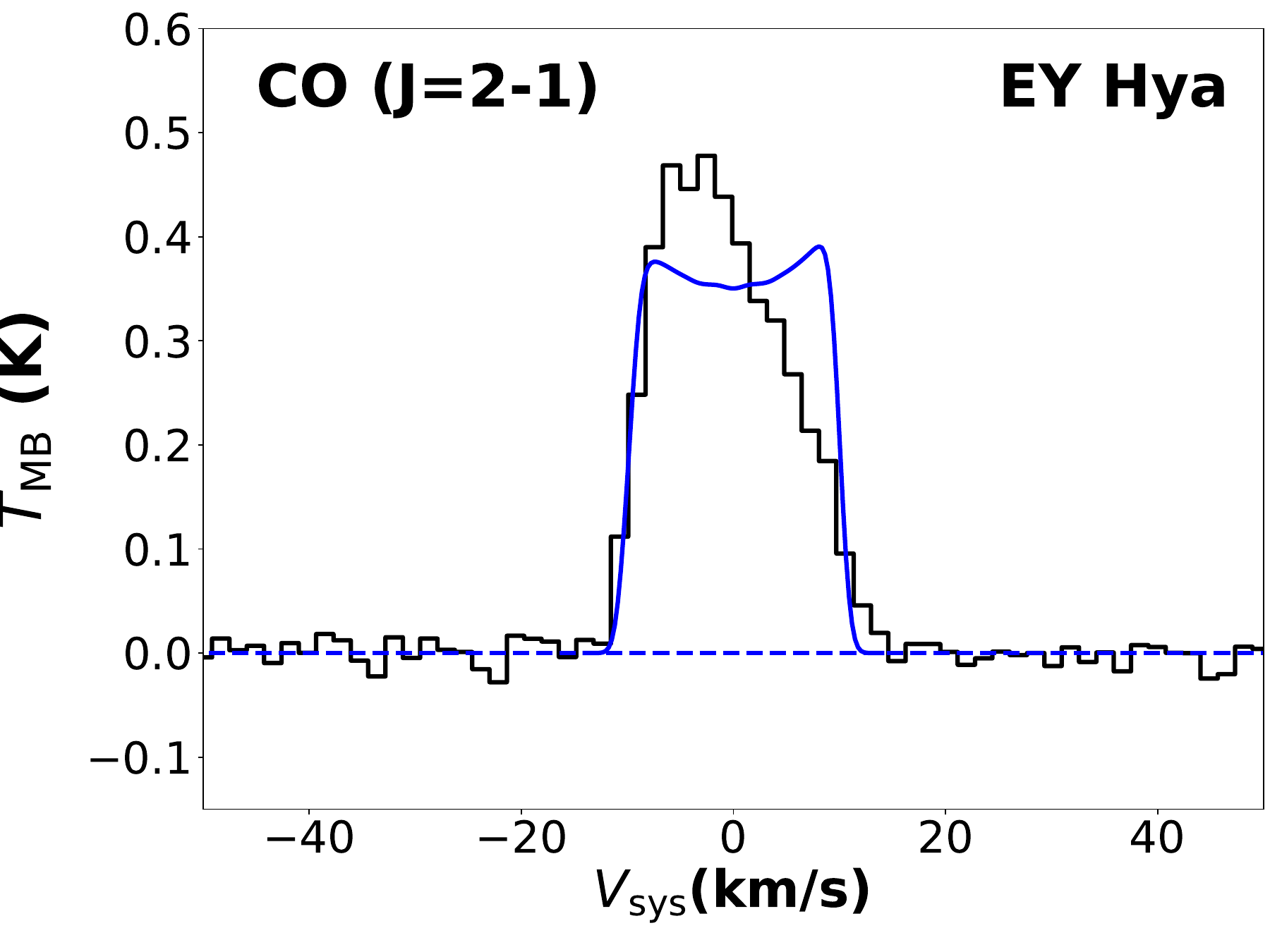}
     \end{subfigure}
     \begin{subfigure}[b]{0.33\linewidth}
         \centering
         \includegraphics[width=\linewidth]{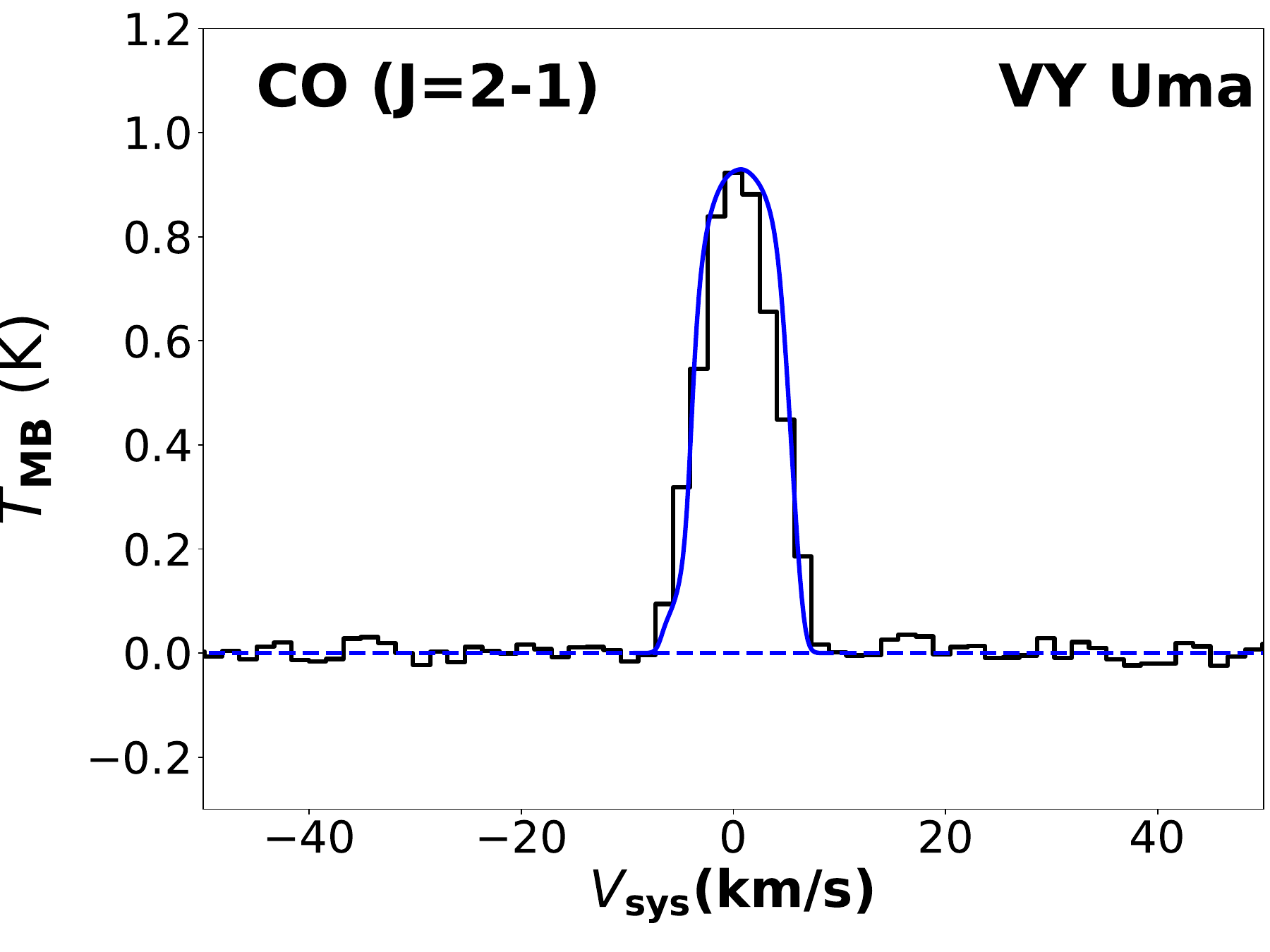}
     \end{subfigure}
     \begin{subfigure}[b]{0.33\linewidth}
         \centering
         \includegraphics[width=\linewidth]{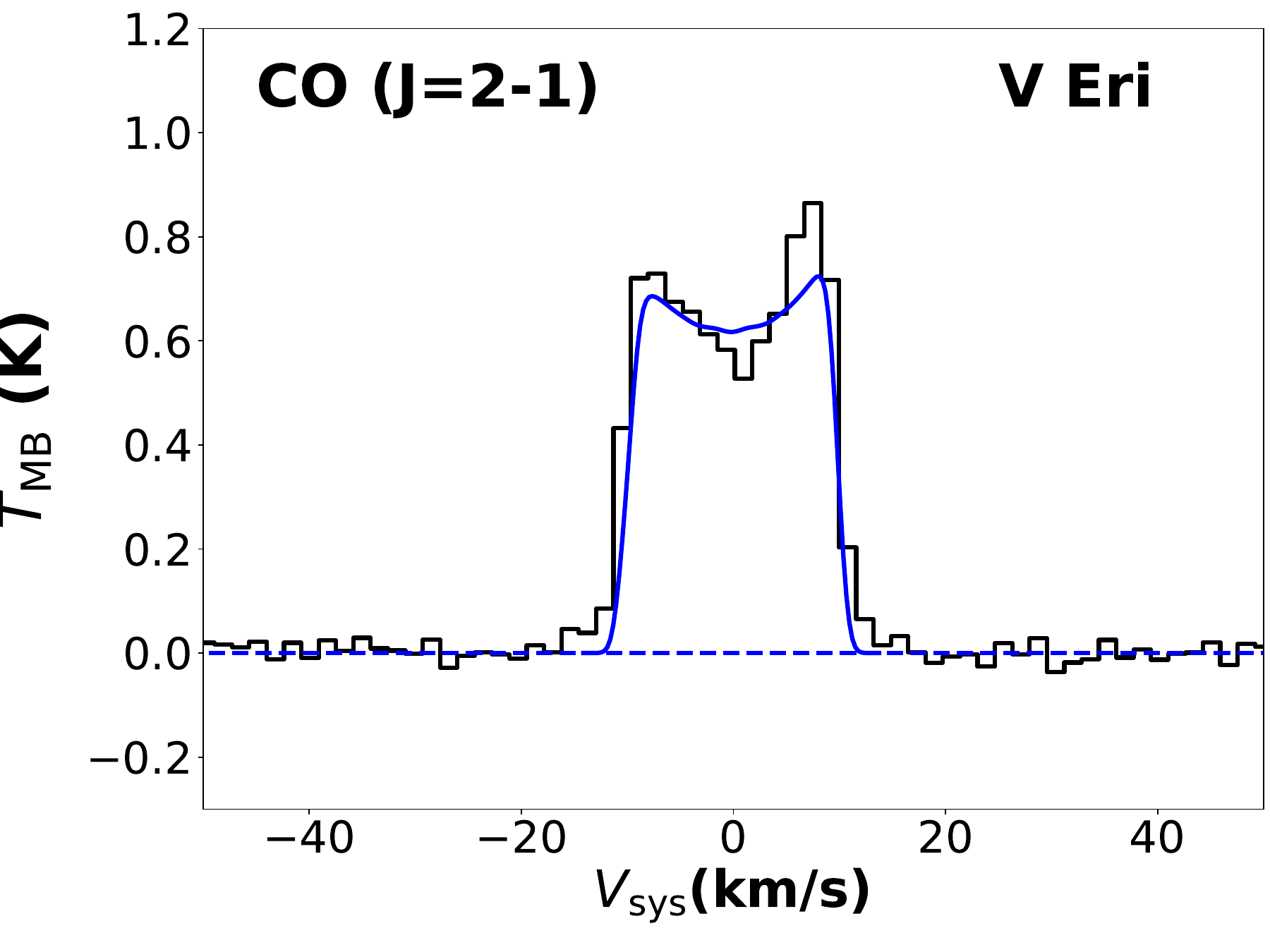}
     \end{subfigure}
        \caption{Spectra of CO for the rest of the sources: EY Hya (O-rich X-AGB), VY Uma (C-rich without X-ray detection), and V Eri (O-rich without X-ray detection), from left to right. Observed spectra are shown in black (velocity resolution is $\delta v$=1.4\,\kms). Synthetic spectra are shown in blue. As discussed in Sect. \ref{anal}, the observed line profiles can be relatively well fitted adopting a standard CO radial abundance.}
        \label{fig:CO_spectra}
\end{figure*}

We constrain the main physical envelope parameters by performing line radiative transfer modelling of the \doceuno\, and \docedos\, lines, which were first reported by \cite{Alonso-Hernandez_2024}. The observed CO line profiles and the best-model fit for the C-rich X-AGB T\,Dra is shown in Fig.~\ref{fig:CO_spectra_T-DRA} and for the rest of the targets in Fig.~\ref{fig:CO_spectra}.

We modelled the CSEs assuming spherical geometry and a constant mass-loss rate. Although spatial asymmetries produced by interactions with potential binary companions \citep[e.g. disks, spirals or bipolar outflows, see][]{Decin_2020} and/or a time variable mass-loss rate cannot be discarded \citep{Teyssier_2006, Maercker_2024}, the lack of spatially resolved observations prevented us from confirming or characterising such deviations (if present). Since our low-$J$ CO emission lines trace the bulk of the envelope, where the expansion velocity reaches the terminal expansion velocity, we assumed constant expansion velocities, and a typical value of the microturbulent velocity width of 1\,\kms~\citep[see e.g.][]{De_Beck_2012}. 

Under these simplifying assumptions the density radial profile is described as a power-law:

\begin{equation}
   \rho(r)=\left(\frac{\mloss}{v_{\rm exp}}\right)\frac{1}{4 \, \pi \,r^{2}}.
\end{equation}

As is typical in AGB CSEs studies, the radial profile of the kinetic temperature was assumed to follow a power-law function:

\begin{equation}
    T(r)=\tinner \left(\frac{\rinner}{r}\right)^{\beta} + T_{\rm CMB},
\end{equation}

\noindent where \tinner\, is the temperature in the inner edge of the envelope (\rinner), $\beta$ is the exponent of the temperature power-law, and $T_{\rm CMB}$ is the cosmic microwave background temperature ($2.7$ K). For the sources in this study, \tinner\, and \rinner\, were obtained by modelling the dust component of the CSEs to fit their Spectral Energy Distributions (SED, Alonso-Hernández in prep.). The parameter $\beta$ is commonly constrained to lie in the range $0.4\leq \beta \leq 1.0$ from previous studies \citep[see][and references therein]{De_Beck_2012,Ramos-medina_2018, Saberi_2020}. 

We adopted the classical CO radial abundance distribution described as

\begin{equation}
    f_{\rm CO}(r)=f^{\rm CO}_{0} \, \rm{exp}\left( - \rm {ln}(2) \left(\frac{r}{r_{1/2}} \right)^{\alpha} \right),
    \label{eqCO}
\end{equation}

\noindent where $f^{\rm CO}_{0}$ is the CO/\hh\ fractional abundance in the inner regions of the envelope. This (pseudo) Gaussian profile is known to be a good representation of the CO abundance distribution derived from detailed CO photodissociation models \citep[see][]{Mamon_1988, Groenewegen_2017, Saberi_2019}. The parameter $r_{1/2}$ is the CO photodissociation radius, i.e., the radial distance at which the CO abundance is reduced to $f^{\rm CO}_{0}/2$ due to its photodissociation by the UV ISRF. The parameter $\alpha$ determines how the abundance decreases in the outer regions of the CSE. 

The parameter $f^{\rm CO}_{0}$ is poorly constrained. The only available empirical measurement of $f^{\rm CO}_{0}$ in an AGB star is (6.7$\pm$1.4)$\times$$10^{-4}$ for the C-rich AGB star IRC+10216 \citep{Fonfria_2022}. Therefore, we assumed standard values of $f^{\rm CO}_{0}$ commonly used in the literature:  2$\times$$10^{-4}$ for O-rich AGB stars \citep[see e.g.][]{Bujarrabal_1989, Ramos-medina_2018} and 8$\times$$10^{-4}$ for C-rich AGB stars \citep[see e.g.][]{Knapp_1985, da_Silva_Santos_2019}.

The photodissociation radius and the $\alpha$ parameter were estimated by \cite{Groenewegen_2017} as a function of $f^{\rm CO}_{0}$, \mloss, \vexp, the excitation temperature (\tex) and the strength of the ISRF for a large sample of AGB CSEs. This parameter can vary significantly between different models  depending of the adopted parameters \citep{Groenewegen_2021} and sometimes differs from empirically derived angular sizes of AGB CSEs based on high-resolution interferometric observation \citep[see][]{Ramstedt_2020}. Therefore, to improve the line fitting, we treated $r_{1/2}$ as a free parameter rather than adopting a model-based estimate. Nevertheless, the best-fit values we derived are in relatively good agreement with those reported in \cite{Groenewegen_2017, Groenewegen_1994, Saberi_2020} for similar CO abundances, mass-loss rates, and expansion velocities.

In order to optimise the calculations and focus on the regions of interest, we truncated the envelope at a maximum radius (\rmax), which we considered to be approximately 3$\times$$r_{1/2}$, in order to make the modelling computations faster. We verified that using larger values of \rmax\, in the same model does not result in a noticeable change in the synthetic CO lines.

On the other hand, the $\alpha$ parameter is typically slightly above two in AGB CSEs \citep{Groenewegen_2017}. We adopted $\alpha$ values interpolated from the tabulated results in \cite{Groenewegen_2017}, using values of $f^{\rm CO}_{0}$, \vexp\, \mloss \, and \tex \, optimal for our targets \citep{Alonso-Hernandez_2024}, while assuming a standard ISRF ($\chi$=$1$). We confirmed that small variations in the $\alpha$ parameter do not produce significant effects on the synthetic lines profiles.

\begin{table*}[h!]

\renewcommand{\arraystretch}{1.2} 
\small
\centering

\caption{Envelope parameters.} 
\label{tab:MADEX_models}

\begin{adjustbox}{max width=\textwidth}
\begin{threeparttable}[b]

\begin{tabular}{l >{\centering\arraybackslash}p{1.20cm} >{\centering\arraybackslash}p{1.20cm} >{\centering\arraybackslash}p{1.00cm} >{\centering\arraybackslash}p{1.20cm} >{\centering\arraybackslash}p{1.00cm} >{\centering\arraybackslash}p{1.20cm} >{\centering\arraybackslash}p{1.20cm} >{\centering\arraybackslash}p{1.00cm} >{\centering\arraybackslash}p{1.20cm} >{\centering\arraybackslash}p{1.00cm} >{\centering\arraybackslash}p{1.20cm} >{\centering\arraybackslash}p{1.00cm}}
\hline\hline 
source  &     \rinner  &  \tinner & $\tau_{UV} \, $ & $L_{UV}$ & $X_{\rm ISRF}$ & $L_{X}$ & $L_{*}$ & $\alpha$  &  \mloss & $\beta$ &  \vexp & $r_{1/2}$   \\  
       &  (cm) & (K)  & & ($\mathrm{erg \, s^{-1}}$) &  & ($\mathrm{erg \, s^{-1}}$) & ($L_{\odot}$) & & ($ \my$) & & ($\mathrm{km \, s^{-1}}$) & (cm)  \\  
\hline

T Dra$^{(a)}$ &    1.5$\times 10^{14}$ & 1200 & 7.2 & 1.8$\times 10^{34}$ & 2.1$\times 10^{8}$ & 4.1e$\times 10^{31}$  & 9000 & 2.65 & 4.1$\times 10^{-6}$ & 1.00 &  14.0 &  2$\times 10^{17}$  \\ 
T Dra$^{(b)}$ &   1.5$\times 10^{14}$ & 1200  & 7.2 & 1.8$\times 10^{34}$ & 2.1$\times 10^{8}$ & 4.1e$\times 10^{31}$ & 9000 & 2.65 & 2.3$\times 10^{-6}$  & 0.75 &  14.0 &  2$\times 10^{17}$ \\
EY Hya        &    7.0$\times 10^{13}$ & 1350   & 1.0 & 3.3$\times 10^{31}$ & 3.9$\times 10^{5}$ & 1.1$\times 10^{31}$ & 6400 & 2.35 & 2.8$\times 10^{-7}$   & 0.35 &  10.0 & 3$\times 10^{16}$ \\
VY Uma        &    6.0$\times 10^{13}$ & 1450   & 0.5 & 1.6$\times 10^{31}$ & 1.9$\times 10^{5}$ & -- & 5300 & 2.40 & 1.0$\times 10^{-7}$   & 0.80 &   5.5 & 3$\times 10^{16}$ \\
V Eri         &    8.0$\times 10^{13}$ & 1250   & 1.0 & 1.4$\times 10^{31}$ & 1.7$\times 10^{5}$ & -- & 5800 & 2.35 & 2.8$\times 10^{-7}$   & 0.30 &  10.0 & 3$\times 10^{16}$ \\
\hline
\end{tabular} 
\begin{tablenotes}
\item \normalsize \textbf{Notes.} Column (1): Source, Col. (2-4): inner radius, temperature at inner radius and opacity estimated at UV wavelengths (200\,nm) from SED modelling (Alonso-Hernández in prep.), Col (5): UV luminosity from the time-average GALEX NUV and FUV fluxes corrected from extinction with $\tau_{UV}$, Col (6): ISRF scaling factor at $r = 5\times 10^{13}$cm, as an indicative value for AGB stellar radii, according to the UV luminosity (see Sect.~\ref{chem_model}), Col (7): X-ray luminosity from the X-ray fluxes (only for X-AGBs), Col (8): stellar luminosity from \cite{Alonso-Hernandez_2024}, Col (9): $\alpha$ parameter as defined by \cite{Groenewegen_2017}, Col (10-13): Mass-loss rate, temperature power-law index, expansion velocity and photodissociation radius from best-fit CO radiative transfer model. As discussed in Sect. \ref{anal}, there are two CO radiative transfer models for T\,Dra: (a) assuming a standard CO radial abundance and (b) assuming a centrally depleted CO abundance in the innermost regions of T\,Dra, likely caused by CO photodissociation driven by X-ray radiation (see Sect. \ref{anal}).
\end{tablenotes}

\end{threeparttable}
\end{adjustbox}

\renewcommand{\arraystretch}{1.0}

\end{table*}

In our search for the best-fit CO radiative transfer models, we varied \mloss, $\beta$, and $r_{1/2}$, while keeping the other parameters fixed. \vexp\ is readily constrained from the CO line width.  We aimed to reproduce the integrated areas of both the CO $J$=2--1 and $J$=1--0 lines, as well as the overall shape of their  profiles to a reasonable extent. Initially, we explored the parameter space ad hoc, starting with \mloss\, values close to those preliminarily estimated by \cite{Alonso-Hernandez_2024}. After identifying a reasonable fit, we refined our results by building a grid of models around this solution to determine the best-fit, defined as the model that minimised the sum of the squares of the difference between the integrated area of synthetic lines and the observed ones. We performed the minimisation procedure on a grid of \mloss\, and $r_{1/2}$ in ranges of one order of magnitude, around the preliminary solution, with logarithmic steps of 0.1, and $\beta$ in a range of 0.5, with linear steps of 0.05. 

\subsection{CO line modelling: Results}\label{co_model_results}

The derived CSE parameters from the CO radiative transfer modelling for our four targets are summarised in Table~\ref{tab:MADEX_models}.  Synthetic CO profiles are shown in Figs.~\ref{fig:CO_spectra_T-DRA} and \ref{fig:CO_spectra}.

In the case of the C-rich X-AGB T\,Dra, the best-fit model assuming an idealised CSE reproduces the most important CO line features: integrated intensity, line peak intensity, and line width (Fig.~\ref{fig:CO_spectra_T-DRA} top panels, model (a) in Table\,\ref{tab:MADEX_models}). It also yields an envelope size comparable to the rough estimate ($\sim$few arcsec) derived from coarse interferometric mapping \citep{Neri_1998}. However, the synthetic profiles of model (a) are distinctly parabolic and fail to reproduce the flatter shape of the observed lines. To generate synthetic CO profiles closer to the observed ones, we initially considered two alternative models:

($i$) A CSE with substantially lower mass-loss rates, which would produce optically thinner emission and, consequently, flatter profiles. However, for such low \mloss\ we could not identify a kinetic temperature profile that simultaneously fits the integrated fluxes of both CO lines and other molecular emission lines.

($ii$) A CSE with a central region depleted of molecular gas represented by larger inner radii (\rinner). This model can reproduce the observed profiles but requires \rinner $\gtrsim 10^{16}$cm, which is incompatible with the modelling of some of the observed lines, such as \sio\, and \hcn, which arise from relatively warm inner regions of the envelope.

We explored, and ultimately adopted a variation of scenario ($ii$) as our reference model, represented in the bottom panel of Fig.~\ref{fig:CO_spectra_T-DRA} (model (b) in Table\,\ref{tab:MADEX_models}). Instead of a completely detached envelope, the shell is filled with molecular gas but shows a CO abundance depression at the centre. Specifically, we propose a radial abundance profile with a lower constant CO abundance in the inner regions of the envelope, extending from \rinner\, up to a certain radius ($\simeq$3$\times10^{16}$\,cm). Beyond this point, the profile features an abrupt jump to a higher abundance followed by a (pseudo) Gaussian distribution as in Eq.\,\ref{eqCO}. The resulting model produced a flattening of the line profiles, which better matches the observed ones, yielding slightly lower $\beta$ and \mloss\, values compared to our original/base model. As we show in Sect.~\ref{chem}, this hypothesis is coherent with the CO radial abundance obtained from the chemical kinetics modelling, which predicts lower values in the most internal shells of the CSE. 

The simplified reference model we present for T\,Dra (model (b)) offers a reasonable compromise, providing a good data-model fit while maintaining a representative CO abundance profile that is qualitatively consistent with X-ray-driven chemical models. While further adjusting the CO cavity model parameters can slightly improve the match to the horned CO 1-0 profile, it did not lead to a satisfactory simultaneous fit for both CO transitions and produces unrealistic results for other molecular species (e.g. \sio\ and HCN, which are very sensitive to the density and temperature gradients). We opted for a simple model that balances these incompatibilities, as fine-tuning the parameters solely to match the CO profile shape would not necessarily enhance model reliability, given the broader uncertainties in our base assumptions and numerical treatment. Our goal is not to achieve a perfect match between observations, radiative transfer models, and chemistry models, but rather to propose a plausible scenario for T\,Dra.

We acknowledge that other options, such as more complex radial temperature and density prescriptions, distorted geometry, or assuming a lower distance to T\,Dra (which would imply optically thinner lines), could potentially improve the synthetic CO emission profiles. However, there is currently no empirical evidence to support these alternatives. In contrast, a centrally depleted CO abundance is a natural expectation for X-AGBs. 

For the remaining targets, we fitted the CO line profiles with a standard CO abundance distribution as described in Eq.~\ref{eqCO}. We achieved a good match to the integrated intensities and to the overall profile shapes in VY\,Uma and V\,Eri, in which the line profiles are nearly parabolic and double-horned respectively. In the case of the O-rich X-AGB EY\,Hya, the observed CO line profiles are markedly asymmetric, probably indicating deviations from a spherically symmetric CSE. Given the potential for more complex geometries in this case, we chose not to attempt a centrally CO-depressed abundance profile, as it would not improve the match to the observed profile.

The obtained \mloss\, values for the four targets, including the two models for T\,Dra, are close to those derived by \cite{Alonso-Hernandez_2024} using a simpler (rotational diagram) analysis. On the other hand, the obtained $\beta$ values range from 0.3 to 1, consistent with those typically found in the literature \citep[see][and references therein]{De_Beck_2012,Ramos-medina_2018, Saberi_2020, Massalkhi_2024}.

We performed a sensitivity analysis to determine the uncertainties associated with the fitted parameters. We estimated the uncertainties from the integration of the $\chi^{2}$ profile likelihood\footnote{We defined the $\chi^{2}$ likelihood function as, $\mathcal{L}(x)$$\propto$$e^{-\chi^{2}(x)/2}$, where $x$ are the free parameters of the model. Uncertainties were estimated as the 68\% confidence intervals from the normalised log-likelihood ($\mathrm{ln}(\mathcal{L}(x))$.} (so called "profile likelihood confidence intervals") in a linear space for $\beta$ and logarithmic space for the rest of parameters. A similar approach was previously described by \cite{Decin_2007}, although including the fit of the line profiles in the minimisation. We found that \mloss\, uncertainties are of a factor $\sim$2-3, while $\beta$ has uncertainties of $\sim$0.3, both of which are consistent with previous analyses \citep[see][]{Ramstedt_2008, Maes_2023}. On the other hand, the uncertainties in $r_{1/2}$ are also a factor $\sim$2-3, similar to the variations observed between the predictions of the different CO photodissociation models \citep[see][]{Groenewegen_2021}. Finally, the uncertainties associated with the fitted \vexp\, are estimated to be $\sim$0.5-1\,\kms.

\subsection{\hcoplus\, and other species}\label{MADEX_other}

Once the main physical properties of the CSEs are constrained from the CO radiative transfer modelling (see Sect.~\ref{MADEX_CO} and \ref{co_model_results}), we derive the abundance of \hcoplus\, and the other detected species by fitting their emission lines. As for the CO model, the expansion velocity (\vexp) is assumed to be constant across the emitting volume and it is directly constrained from the widths of the lines (Table~\ref{tab:abundances}). 

For the molecules in this study, as for CO, we assumed a radial abundance profile of the form:

\begin{equation}
    f_{\rm X}(r)=f^{\rm X}_{0} \, \rm {exp}\left( - \left(\frac{r}{r_{e}} \right)^{2} \right),
    \label{eq:gaussian}
\end{equation}

\noindent where $r_{e}$ is the effective radius, related to the $r_{1/2}$ parameter used for CO (when $\alpha$=2) by $r_{1/2}$=$r_{e}$$\sqrt{\ln2}$.  
This type of distribution has been shown to be a good representation of the abundance profile for parent species (formed in the inner envelope), such as \sio\, and \hcn\, \citep[see e.g.][]{Gonzalez-delgado_2003, shoier_2007, shoier_2013, Massalkhi_2024}. Although more complex profiles are expected for daughter species, particularly in the presence of internal UV/X-ray radiation (see Sect.\ref{chem}), we also adopt this simplified Gaussian function for them, as the study focusses on deriving the average abundance rather than attempting to obtain a detailed description of the radial abundance distribution (not possible with the current dataset).

We aimed to determine $f^{\rm X}_{0}$ and $r_{e}$ for each molecule and target. However, these two parameters are highly degenerate, making it possible to derive reasonably accurate values for both only in cases where multiple transitions are observed from energy levels that are well-separated. The minimisation procedure and uncertainties estimation for these parameters followed the same approach as for CO, described in the previous sections. We carried out the minimisation on a grid of $f^{\rm X}_{0}$ and $r_{e}$, exploring a range spanning one order of magnitude around the preliminary solution, with logarithmic steps of 0.1.

In our study, we determined $f^{\rm X}_{0}$ and $r_{e}$ simultaneously for molecules with observations of two or more transitions. For molecules with only one detected transition, or with all the detected transitions close in energy (i.e., \sis, \sicdos\, and \cch\, and in T\,Dra and \hcn\, in VY\,Uma), $r_{e}$ was adopted from the "model-std" chemical kinetics models (see Sect.~\ref{chem}).

The synthetic line profiles for all the molecules detected in the C-rich X-AGB T\,Dra are shown in Figs.~\ref{fig:HCO+_T-DRA_spectra} and \ref{fig:T-DRA_spectra}. The synthetic line profiles for the only two molecules detected in the rest of the targets (\sio\, and \hcn) are shown in Fig.~\ref{fig:others_spectra}.

The derived abundances and effective radii for each molecule are summarised in Table~\ref{tab:abundances}. The absolute uncertainties of $f^{\rm X}_{0}$ and $r_{e}$ are of a factor $\sim$2 and $\sim$1.5, respectively. For molecules where $r_e$ was assumed, abundance uncertainties increase to about one order of magnitude due to parameter degeneracy.

We note that correlations between \mloss\, and $\beta$ result in underestimation of these uncertainties. Moreover, additional uncertainties of up to half an order of magnitude are possible for molecules that are highly sensitive to density and temperature variations \citep[see][]{Maes_2023}. Finally, since $f^{\rm CO}_0$ is poorly constrained for AGB stars, especially for uv/X-AGBs, it introduces additional uncertainty, as the derived abundances depend on the H$_2$ density, which scales inversely with $f^{\rm CO}_0$.

From the \hcoplus\, non-detections (in EY\,Hya, VY\,UMa, and V\,Eri), we derived upper limits to the molecular abundance of this ion. For T\,Dra, we also estimated an upper limit for the abundance of $\rm N_{2}H^{+}$, another X-ray sensitive ion, based on the non-detection of the $\rm N_{2}H^{+}$($J$=1-0) line, which falls within the frequency range of our observations.

\subsection{Comparison with standard AGB stars}\label{comparison_obs}

We find that the average abundance of \hcoplus\, in T\,Dra is one order of magnitude larger than in IRC+10216, the best-studied C-rich AGB star and the only other one with a confirmed \hcoplus\, detection \citep[with an abundance estimate of 4.1$\times 10^{-9}$ with respect to \hh, see][]{Pulliam_2011}. This \hcoplus\, enhanced abundance is also comparable to or slightly larger than the abundances expected for O-rich AGB stars, which have naturally higher abundances of this molecular ion than C-rich AGBs, as shown by \cite{Pulliam_2011}. On the other hand, while the absolute uncertainty in $f^{\rm HCO^{+}}_{0}$ for T\,Dra may be a factor of $\sim$2, the significant change in line intensities between 2020 and 2024 suggests that the \hcoplus\, abundance in 2024 was approximately twice that in 2020.

The upper limits on \hcoplus\ abundance for the non-detections in the remaining targets are higher than, or comparable to, the expected/observed values for standard AGB CSEs, meaning that our non-detections are consistent with these expectations. For EY\,Hya, the O-rich AGB star with X-ray emission in our sample, the \hcoplus\ abundance upper limit ($< 10^{-7}$ relative to \hh) is too high to assess whether \hcoplus\ is enhanced compared to the abundances found in O-rich AGB CSEs \citep[around $10^{-8}-10^{-7}$, see][]{Pulliam_2011}.

As for \hcoplus, the average abundance of \hnc\, and \hctresn\, in T\,Dra are significantly larger than in IRC+10216, which are $\approx$10$^{-8}$ and $\approx$10$^{-7}$, respectively \citep{Daniel_2012, Siebert_2022}. However, both abundances are similar to average abundances estimated for three other C-rich AGB stars, including one (IRAS~07454$-$7112) with a mass-loss rate comparable to T\,Dra \citep[][]{Unnikrishnan_2024}. As shown by these authors, these two daughter species show ring-like distributions in AGB CSEs. Therefore our use of a Gaussian radial abundance profile may lead to an underestimation of the inferred abundances. 

The abundance of \hcn, a parent species, in the two C-rich AGB stars in our sample (with and without X-rays) are in good agreement with those previously estimated for C-rich AGB stars in their respective mass-loss rate ranges, considering the large scattering in the \hcn\, initial abundances and its relationship with the mass-loss rate \citep[see][]{shoier_2013}.

The \sio\, abundance in the O-rich AGB stars we study here is around one order of magnitude lower than in previous studies for similar mass-loss rates ($f^{\rm SiO}_{0}$=$10^{-6}$-$10^{-5}$),  as reported by \cite{Bujarrabal_1989, Gonzalez-delgado_2003, Massalkhi_2020, Massalkhi_2024}. However, these relatively low abundances are comparable to, and slightly larger than, those found in the O-rich AGB star $o$\,Cet, which is known to be in a multiple system and a strong X-ray emitter, with a \sio\, abundance of $f^{\rm SiO}_{0}$$\simeq$7$\times10^{-8}$ and a mass-loss rate of $\mloss$$\simeq$2$\times 10^{-7} \my$ \citep{Massalkhi_2024}.

In T\,Dra, the \sio\, abundance falls within the large range of values found in other C-rich AGB stars ($f^{\rm SiO}_{0}$$\approx$$10^{-7}$-$10^{-5}$). However, the effective radius of the SiO emitting region ($r_{e}$$\sim$1.8$\times 10^{15}$ cm) is significantly smaller than those estimated for similar mass-loss rates ($r_{e}$$\sim$$10^{16}$ cm) by \cite{shoier_2006} and \cite{Massalkhi_2019, Massalkhi_2024}. This lower effective radius is only expected for C-rich AGB stars with $\mloss$$\simeq$$10^{-8} \my$, two orders of magnitude lower than our estimates for T\,Dra, assuming the $r_{e}$-to-\mloss\,relationship shown in Fig. 5 of \cite{shoier_2006}. Therefore, T\,Dra is an outlier in this relationship or the $r_{e}$ was considerably underestimated.

The abundance of \sis\, in T\,Dra  is significantly lower than those found in C-rich AGB stars with similar mass-loss rates ($f^{\rm SiS}_{0}$$\sim$$10^{-6}$-$10^{-5}$) according to \cite{shoier_2007, Danilovich_2018, Massalkhi_2019, Massalkhi_2024}. Instead, this value is closer to those expected for C-rich AGB stars with lower mass-loss rates or for O-rich and S-type AGB stars ($f^{\rm SiS}_{0}$$\lesssim$$10^{-6}$). However, as mentioned before, this low abundance is highly uncertain and may result from an overestimated $r_{e}$ given that it is derived from a single detected line in this case. 

The \sicdos\, abundance in T\,Dra falls within the range reported for C-rich AGB stars ($f^{\rm SiC_{2}}_{0}$=$10^{-7}$-$10^{-5}$) by \cite{Massalkhi_2018}. 

The \cch\, abundance in T\,Dra is slightly lower than those estimated by \cite{Unnikrishnan_2024} in their study of a small sample of C-rich AGB CSEs ($f^{\rm C_{2}H}_{0}$=3$\times$$10^{-6}$-3$\times$$10^{-5}$). However, the abundance estimate for this species in T\,Dra, derived from a single detected line blended with a \sicdos\ line, is highly uncertain and may reflect an overestimation of $r_e$ in the radiative transfer modelling. 

The X-ray sensitive molecular ion \ndoshplus\ was not detected in any of our targets. Detections of this ion in circumstellar environments remain rare, with only a few cases reported in the post-AGB or Planetary Nebula phase. To our knowledge, OH\,231.8+4.2 -- a prominent fast bipolar outflow around an OH/IR star -- is the only known AGB CSE where \ndoshplus\ has been detected, with an abundance of $\sim$2\ex{-9} \citep{sanchez-contreras_2015}. Our upper limit for \ndoshplus\ in T\,Dra is one order of magnitude above this value. 

\begin{table}[h!]

\renewcommand{\arraystretch}{1.2}
\small
\centering

\caption{Initial fractional fractional abundances ($f^{\rm X}_{0}$), effective radii ($r_{e}$) and expansion velocities (\vexp) derived from radiative transfer modelling.}
\label{tab:abundances}

\begin{adjustbox}{max width=\textwidth}
\begin{threeparttable}[b]

\begin{tabular}{l >{\centering\arraybackslash}p{2.40cm} >{\centering\arraybackslash}p{2.40cm} r}
\hline\hline 
Molecule  & $f^{\rm X}_{0}$ & $r_{e}$ & \vexp \\  
       & index & (cm) & $\mathrm{km \, s^{-1}}$ \\  
\hline

\hline
\multicolumn{4}{c}{T Dra (C-rich with X-rays) } \\
\hline

\hcoplus (2024) & 3.0$\times10^{-8}$ & 3.0$\times10^{16}$ & 10.0\\
\hcoplus (2020) & 1.5$\times10^{-8}$ & 3.0$\times10^{16}$ & 10.0\\
HCN & 8.0$\times10^{-5}$ & 6.0$\times10^{15}$ & 12.0\\
H$\mathrm{C_{3}}$N & 1.2$\times10^{-5}$ & 1.7$\times10^{15}$ & 10.0\\
SiO & 2.5$\times10^{-6}$ & 1.8$\times10^{15}$ & 10.0\\
\sicdos $^{(*)}$& 1.3$\times10^{-6}$ & 3.0$\times10^{16}$ & 10.0\\
\cch $^{(*)}$& 5.0$\times10^{-7}$ & 1.0$\times10^{17}$ & 10.0\\
SiS $^{(*)}$& 2.0$\times10^{-7}$ & 5.0$\times10^{15}$ & 8.0\\
HNC & 1.8$\times10^{-7}$ & 5.0$\times10^{15}$ & 9.0\\
$\rm N_{2}H^{+}$ $^{(*)}$& $<$3.0$\times 10^{-8}$ & 1.0$\times10^{16}$ & 10.0\\
\hline
\multicolumn{4}{c}{EY Hya (O-rich with X-rays)} \\
\hline
\hcoplus $^{(*)}$& $<$1.0$\times10^{-7}$ & 1.0$\times10^{16}$ & 6.0\\
SiO  & 2.0$\times10^{-7}$ & 1.0$\times10^{16}$ & 6.0\\

\hline
\multicolumn{4}{c}{VY Uma (C-rich without X-rays)} \\
\hline
\hcoplus $^{(*)}$& $<$5.0$\times10^{-8}$ & 1.0$\times10^{16}$ & 5.5\\
HCN  & 1.5$\times10^{-6}$ & 8.0$\times10^{14}$ & 5.5\\

\hline
\multicolumn{4}{c}{V Eri (O-rich without X-rays)} \\
\hline
\hcoplus $^{(*)}$& $<$3.0$\times10^{-8}$ & 1.0$\times10^{16}$ & 6.0\\
SiO  & 2.0$\times10^{-7}$ & 7.0$\times10^{15}$ & 6.0\\
\hline
\end{tabular} 
\begin{tablenotes}
\item \normalsize \textbf{Notes.} Column (1): Molecule, Col. (2): initial abundance relative to $\mathrm{H_{2}}$, Col (3): effective radius, Col (4): expansion velocity. $^{(*)}$ indicates that $r_{e}$ and \vexp\, were assumed (see Sect.~\ref{MADEX_other}).
\end{tablenotes}

\end{threeparttable}
\end{adjustbox}

\renewcommand{\arraystretch}{1.0}

\end{table}

\section{Chemical modelling}\label{chem}

In this paper, we present new chemical kinetics models, accounting for the X-ray emission originating within the envelopes of the peculiar X-AGBs, presumably linked to the presence of an accreting companion. These models explore how internal X-rays influence chemical reactions and alter the abundances of key X-ray sensitive molecules such as \hcoplus, primarily through molecular photodissociation and the chemical processes it triggers. The models also include the effects of internal UV emission, which is a characteristic feature of all X-AGBs. By separately analysing the contributions of UV and X-ray emissions, we identify the chemical pathways driven by each and estimate their distinct impacts on molecular abundances.

The comparison between the empirically derived abundances of our uvAGBs, two with both UV and X-ray emissions and two with UV but no X-rays (used as a control sample), and those predicted by the chemical kinetics models allowed us to isolate the effects of X-rays from those of UV radiation and evaluate how these high-energy processes influence molecular species.

In the following section, we describe the model, our modelling approach, and the main results, including a brief overview of the dominant chemical reactions and a comparison with the fractional abundances and column densities empirically derived from line radiative transfer modelling.

\subsection{Chemical model with internal X-ray emission}\label{chem_model}

Our chemical kinetics model is based on that developed by \cite{agundez_2006}, which has been widely used to study the chemistry of AGB CSEs and other environments \citep[see e.g.][and references therein]{agundez_2012, agundez_2017, Velilla-Prieto_2015, sanchez-contreras_2015}. The chemical network in our code includes gas-phase reactions and processes induced by cosmic rays, UV photons, and X-rays, but it does not include reactions on dust grain surfaces, at the exception of the formation of H$_2$.

The model calculates the evolution of the chemical composition of the gas as it travels outwards across the circumstellar envelope. We assume that the gas expands isotropically with a uniform expansion velocity \vexp.  Another major input of our chemistry model is the set of initial abundances of the “parent” species. These are formed in deeper layers and are injected to the intermediate/outer envelope. The parent species, typical of O- and C-rich environments, and the initial abundances adopted in our model are given in Table~\ref{tab:code_abundances}.

\begin{table}[h!]

\renewcommand{\arraystretch}{1.2}
\small
\centering

\caption{Abundances of parent species adopted as input for our chemical kinetics model (see Sect.~\ref{chem_model}).} 
\label{tab:code_abundances}

\begin{adjustbox}{max width=\textwidth}
\begin{threeparttable}[b]

\begin{tabular}{>{\centering\arraybackslash}p{0.1\textwidth}>{\centering\arraybackslash}p{0.1\textwidth}>{\centering\arraybackslash}p{0.1\textwidth}>{\centering\arraybackslash}p{0.1\textwidth}}
\hline\hline 
\multicolumn{2}{l}{\hspace{1.2 cm}C-rich}     & \multicolumn{2}{l}{\hspace{1.2 cm}O-rich}   \\  
    Species   & Abundance & Species   & Abundances \\  
\hline

He$^{(a)}$ & $0.17$ & He$^{(a)}$ & $0.17$ \\
CO$^{(b)}$ & 8.0$\times 10^{-4}$ & CO$^{(g)}$ &  2.0$\times 10^{-4}$ \\
C$_{2}$H$_{2}$$^{(c)}$  & 8.0$\times 10^{-5}$ & H$_{2}$O$^{(h)}$  & 2.0$\times 10^{-4}$ \\
N$_{2}$$^{(d)}$ & 4.0$\times 10^{-5}$ & N$_{2}$$^{(d)}$ & 4.0$\times 10^{-5}$  \\
HCN$^{(c)}$& 2.0$\times 10^{-5}$ & SO$^{(i)}$ & 1.0$\times 10^{-6}$ \\
CH$_{4}$$^{(e)}$ & 3.5$\times 10^{-6}$ & NH$_{3}$$^{(j)}$ & 7.0$\times 10^{-7}$ \\
SiS$^{(c)}$& 1.0$\times 10^{-6}$ &  SO$_{2}$$^{(k)}$  & 5.0$\times 10^{-7}$ \\
SiO$^{(c)}$& 1.2$\times 10^{-7}$ & CO$_{2}$$^{(l)}$ & 3.0$\times 10^{-7}$  \\
H$_{2}$O$^{(f)}$ & 1.0$\times 10^{-7}$ & SiS$^{(m)}$ & 2.7$\times 10^{-7}$ \\
SiC$_{2}^{(c)}$& 5.0$\times 10^{-8}$ & SiO$^{(n)}$ & 1.7$\times 10^{-7}$ \\

\hline
\end{tabular} 
\begin{tablenotes}
\item \normalsize \textbf{Notes.} Column (1): parent species (C-rich), Col. (2): initial abundance relative to $\mathrm{H_{2}}$ (C-rich), Col (3): parent species (O-rich), Col (4): initial abundance relative to $\mathrm{H_{2}}$ (O-rich). 
\item \normalsize \textbf{References.} (a) solar abundance \citep[see][]{Asplund_2009}, (b) \cite{Knapp_1985}, (c) \cite{Agundez_2009}, (d) TE abundance \citep[see][]{Agundez_2010}, (e) \cite{Keady_1993K}, (f) \cite{Decin_2010}, (g) \cite{Bujarrabal_1989}, (h) \cite{Maercker_2008} scaled to CO abundance, (i) \cite{Bujarrabal_1994}, (j) \cite{Wong_2018}, (k) \cite{Massalkhi_2020}, (l) \cite{Tsuji_1997}, (m) \cite{shoier_2007}, (n) \cite{shoier_2006}.

\end{tablenotes}

\end{threeparttable}
\end{adjustbox}

\renewcommand{\arraystretch}{1.0}

\end{table}

The envelope is externally illuminated by the local interstellar radiation field \citep{Draine_1978} and the gas is shielded from the UV radiation by dust, with an adopted $N_{\rm H}$\,/\,$A_V$ ratio of 2.21$\times 10^{21}$\,cm$^{-2}$\,mag$^{-1}$ \citep{Guver_2009}. We adopt a cosmic ray ionisation rate of H$_2$ of 1.2$\times 10^{-17}$ s$^{-1}$. We use the chemical network previously used to model the growth of carbon chains in IRC +10216 in \cite{agundez_2017}.

In this study we have included the effects of an internal X-ray source following the approach described in \cite{Stauber_2005}. Therefore, we assumed that X-rays affect the chemistry by enhancing the H$_2$ ionisation rate over the cosmic rays value. We adopted the relationship between X-ray luminosity and H$_2$ ionisation rate from Figure\,3 of \cite{Stauber_2005}. Therefore, the H$_2$ ionisation rate, taking into account the geometrical dilution, is approximated as:

\begin{equation}
\zeta ({\rm H_2}) = 1.2 \times 10^{-17} + 10^{-15} \Big( \frac{L_X}{10^{30}} \Big) \times \Big( \frac{3 \times 10^{15}}{r} \Big)^2,
\end{equation}

\noindent where $\zeta$ (H$_2$) is in units of s$^{-1}$, $L_X$ is the stellar X-ray luminosity in units of erg s$^{-1}$, and $r$ is the radial distance from the star in units of cm. Regarding the ionisation and destruction rates for the rest of molecular species as well as ionisation rates for neutral atoms, they were assumed to have the same ratios with respect to the ionisation rate of H$_{2}$ as those produced by cosmic rays, these ionisation and destruction rates also scale proportional to the ionisation rate of H$_{2}$. Doubly-ionised species were not considered.

As discussed in previous works \citep[see e.g.][]{Wolfire_2022}, X-ray photons are energetic enough to photodissociate and photoionise any molecule. In addition, the hydrogen column densities of the CSEs considered in this study are relatively small ($N_{H}$$\ll$$ 10^{24}$cm$^{-2}$) and the attenuation expected within this wavelength range is negligible \citep[see][]{Stauber_2005}. However, we acknowledge that there is a more complex reality where wavelength-dependent attenuation is presents and depends on the density distribution of the CSE as well as the spectral shape of the internal X-ray emission. Given these complexities, our model should be considered as a first-order approximation.

Therefore, this radiation can penetrate the full extent of the envelope and efficiently lead to a variety of chemical reactions along the whole envelope. In some molecular species the overall effect is significant, increasing or decreasing the abundance by orders of magnitude, in comparison with the model without X-rays. We describe below the molecular species with the most noticeable changes in their radial abundance distributions and the chemical reactions that lead to these changes.

In the case of T\,Dra we estimated that the minimum X-ray luminosity needed to obtain noticeable effects in the chemistry is $\rm L_{X}$$\sim$$10^{27}$\,erg\,$s^{-1}$. For larger luminosities the X-ray emission effect increases in a proportional way until it reaches a value of $\rm L_{X}$$\sim$$10^{33}$\,erg\,$s^{-1}$. At these X-ray luminosities the photodissociation effect becomes so strong that the molecules become fully dissociated, and the gas transitions to an ionised atomic state.

Regarding the role of internal UV photons, we consider that they are geometrically diluted and attenuated by dust according to the $N_{\rm H}$\,/\,$A_V$ ratio previously given. We assumed  that the internal UV flux is described as the Draine field scaled to the internal UV flux, the internal UV fluxes were estimated as the average between the NUV and FUV fluxes shown in Table~\ref{tab:astronomy} corrected for circumstellar extinction according to the opacity estimated at UV wavelengths (200 nm) from the SED modelling (Alonso-Hernández in prep.). The UV luminosities and ISRF scaling factors estimated for each source are shown in Table~\ref{tab:MADEX_models}.

Most common molecules have dissociation and ionisation energies in the UV energy range (i.e. a few eV). Therefore, the UV radiation produces a selective effect of molecular photodissociation and photoionisation, being more efficient in molecules with low dissociation and ionisation energies, although there are additional indirect dissociation pathways \citep{Heays_2017}. Especially critical are the dissociation energies of \hh\, and CO \citep[both slightly larger than 11 eV, see][]{Heays_2017} as these are the molecules with larger abundances and whose dissociation products lead the chemical reactions of most interest in this paper.

Moreover, the optical opacity and the resulting attenuation produced by the CSEs plays a relevant role at UV wavelengths. In the cases with low density outflows the UV radiation can penetrate and affect the chemistry through the whole envelope suffering slightly attenuation. However, in the cases with high density outflows, the CSEs are optically thick and the UV radiation is highly attenuated, affecting only the chemistry in the innermost regions \citep[for further details, see][]{Van_de_Sande_2022, Van_de_Sande_2023}.

We designed a set of chemical kinetics models customised for each source with the UV/X-ray luminosities (from the information shown in Table~\ref{tab:astronomy}) and the CSE parameters estimated from the CO line modelling (see Table~\ref{tab:MADEX_models}). We designed four different kinds of models for each source: (i) Our ideal (most complete) model, which includes both X-rays and UV radiation from the central source ("model-x/uv"), (ii) a model of a standard AGB CSE with neither UV nor X-ray emission ("model-std"), (iii) a model with UV emission, but no X-rays ("model-uv"), and (iv) a model with only X-ray emission  ("model-x").

The synthetic abundance profiles and column densities obtained from our chemical kinetics models are shown in Figs.~\ref{fig:T-DRA_chemistry}-\ref{fig:bar_plot} and listed in appendix~\ref{column_densities}, respectively. Modelling details and results for each species are discussed individually in the following subsections.

\subsection{CO photodissociation}\label{CO-photodissociation}

\begin{figure}[h!]
     \centering
     \begin{subfigure}[b]{1.00\linewidth}
         \centering
         \includegraphics[width=\linewidth]{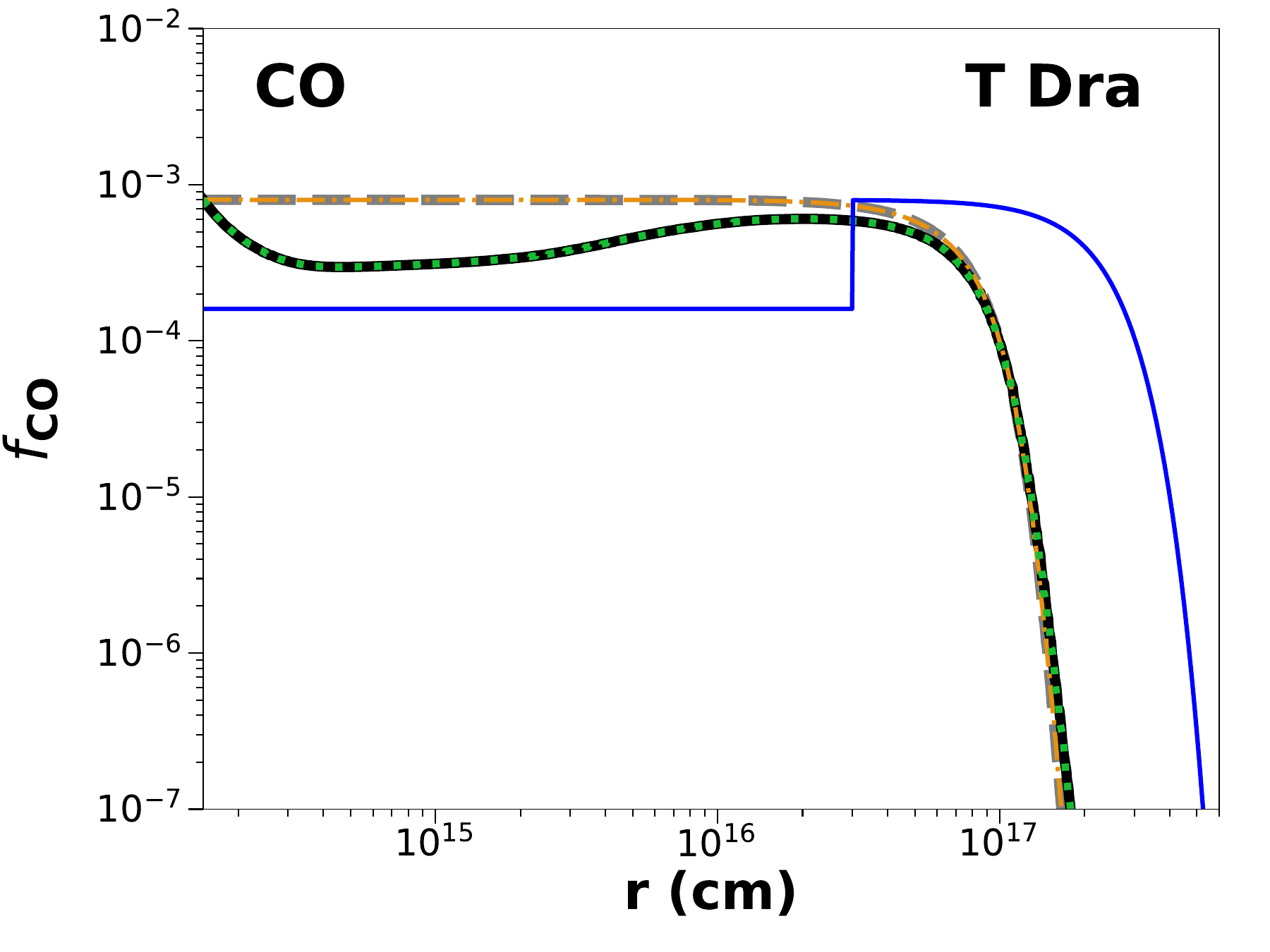}
     \end{subfigure}
     \begin{subfigure}[b]{1.00\linewidth}
         \centering
         \includegraphics[width=\linewidth]{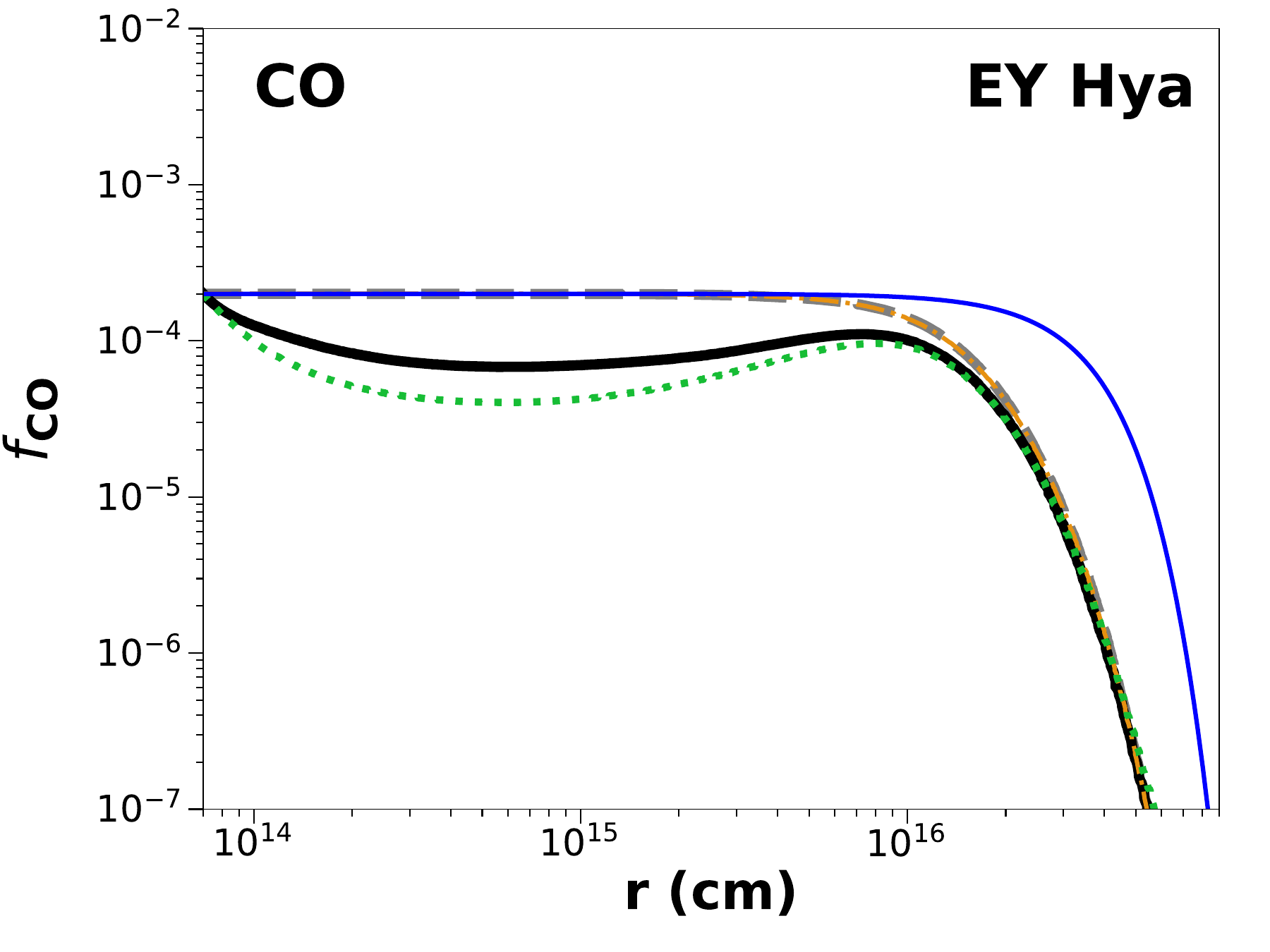}
     \end{subfigure}
        \caption{Comparison between chemical kinetics model predictions (Sect.~\ref{chem}) and the CO radial abundance distributions from our line radiative transfer analysis (Sect.~\ref{MADEX_CO}) for the two X-AGBs: T\,Dra (top) and EY\,Hya (bottom). The black solid line represents 'model-uv/x', the green dotted line represents 'model-x', the grey dashed line represents 'model-std', the orange dash-dotted line represents 'model-uv'. The blue solid line shows the derived CO radial abundance distribution from the radiative transfer modelling of the observed CO lines.}
        \label{fig:T-DRA_chemistry-CO}
\end{figure}

Fig.~\ref{fig:T-DRA_chemistry-CO} shows the radial abundance profiles of CO predicted by our chemical kinetics code for the two X-AGBs in this study, T\,Dra and EY\,Hya, together with the abundance distribution derived from the CO radiative transfer model for both targets (Sect.~\ref{MADEX_CO}).

CO is a parent species that forms in the inner regions of AGB CSEs through thermal equilibrium (TE) chemistry, primarily via the reaction between C and O atoms, and is subsequently transported outwards as part of the evolving chemical composition. 

The destruction of CO is mainly produced in the external regions of the envelope and is related, directly and indirectly, with the UV photons from the ISRF. The most effective destruction path of CO is photodissociation

\begin{equation}
  \begin{array}{l l}
    \rm CO \xrightarrow[]{\mathrm{h \nu}} C + O.
   \label{hhhplus}
  \end{array}
\end{equation}

Although in a lower degree, CO is also destroyed by photoionisation and chemical reactions induced by molecular ions (Sect.~\ref{hcoplus-formation}). The effect of UV radiation from the ISRF is well known as the primary factor responsible for the steep decline in CO abundance around $r_{\rm 1/2}$, thereby defining the outer boundary of most AGB CSEs through photodissociation.

The destruction of the CO in the inner layers of the envelope by the internal X-ray emission is clearly visible in Fig. \ref{fig:T-DRA_chemistry-CO}. High-energy photons photodissociate and photoionise CO molecules in these regions, resulting in a lower CO abundance. Due to their higher penetration capability, X-rays destroy CO much more efficiently and over a larger portion of the envelope compared to UV photons, which are rapidly attenuated by the high optical depths prevailing close to the star. As a result, the internal UV flux has a negligible effect in these two sources. In fact, "model-uv" and "model-std" are indistinguishable and, for the same reason (negligible effect of UV in the inner envelope), the predictions from "model-uv/x" and "model-x" fully overlap in Fig. \ref{fig:T-DRA_chemistry-CO}.

We conclude that an internal X-ray source can significantly alter the CO abundance in the inner regions of the envelope. The destruction of CO is directly proportional to the number of photons with sufficient energy to photodissociate it, and thus to the X-ray flux of the internal source. This prediction is in good qualitative agreement with the results from CO line radiative transfer modelling for T\,Dra, which suggest that the CO abundance could be centrally depleted -- by an uncertain factor of a few over an uncertain radius -- in this source. We have verified that increasing the X-ray luminosity in T\,Dra by a factor $\sim$2 \citep[well within the variability range observed in X-AGBs, see][]{Sahai_2015, Ortiz_2021} leads to a central CO abundance depletion in our chemical kinetics models that more closely matches the depletion inferred from CO radiative transfer modelling.

In terms of column density (Fig.~\ref{fig:bar_plot}), the model for T\,Dra including X-rays predicts a lower CO column density than the model without X-rays, though both remain slightly above the observed value—the X-ray model being in better agreement with observations. However, detecting the central depletion through column density measurements is not straightforward, as the column density integrates contributions from all envelope layers, potentially masking localised abundance variations. Regarding the O-rich X-AGB EY\,Hya, our chemical kinetics models also predict a central CO depletion with a $\sim$50\% reduction in the CO column density (Fig.~\ref{fig:bar_plot}), although the depletion is less pronounced in this case, as EY\,Hya's X-ray luminosity is a factor of five lower than that of T\,Dra.

\subsection{\hcoplus\ abundance enhancement}\label{hcoplus-formation}

\begin{figure*}[h!]
     \centering
     \begin{subfigure}[b]{0.33\linewidth}
         \centering
         \includegraphics[width=\linewidth]{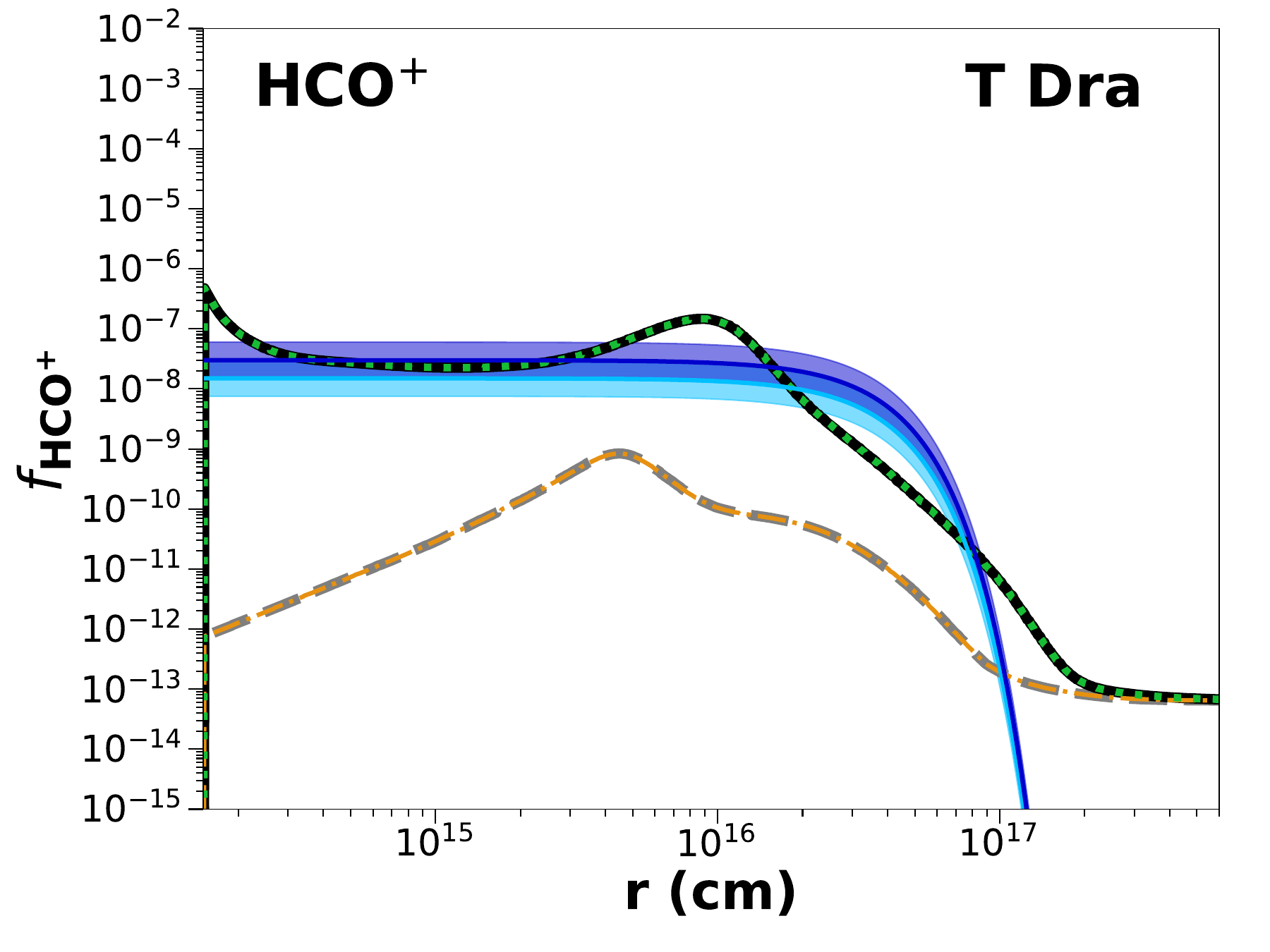}
     \end{subfigure}
     \begin{subfigure}[b]{0.33\linewidth}
         \centering
         \includegraphics[width=\linewidth]{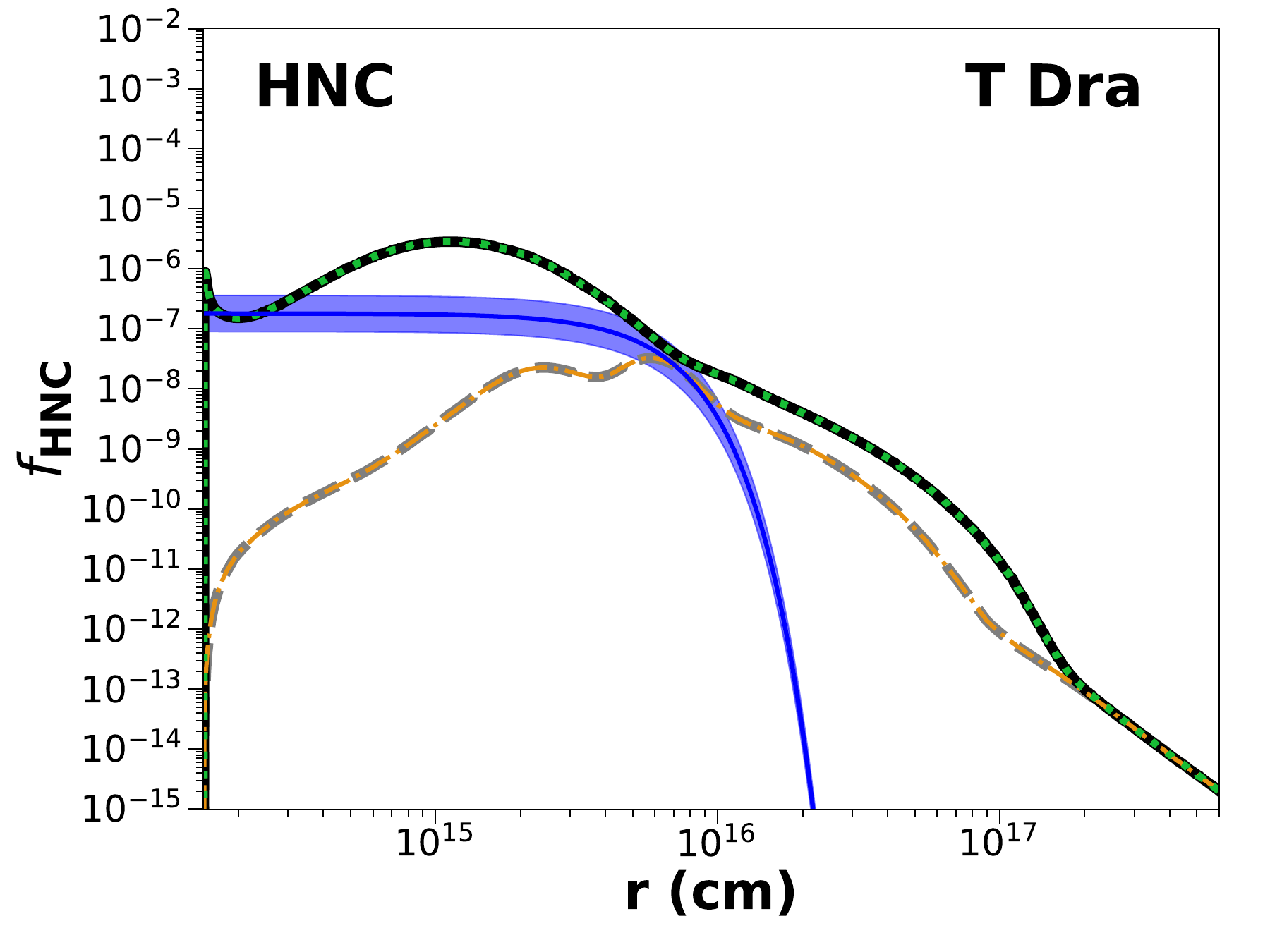}
     \end{subfigure}
     \begin{subfigure}[b]{0.33\linewidth}
         \centering
         \includegraphics[width=\linewidth]{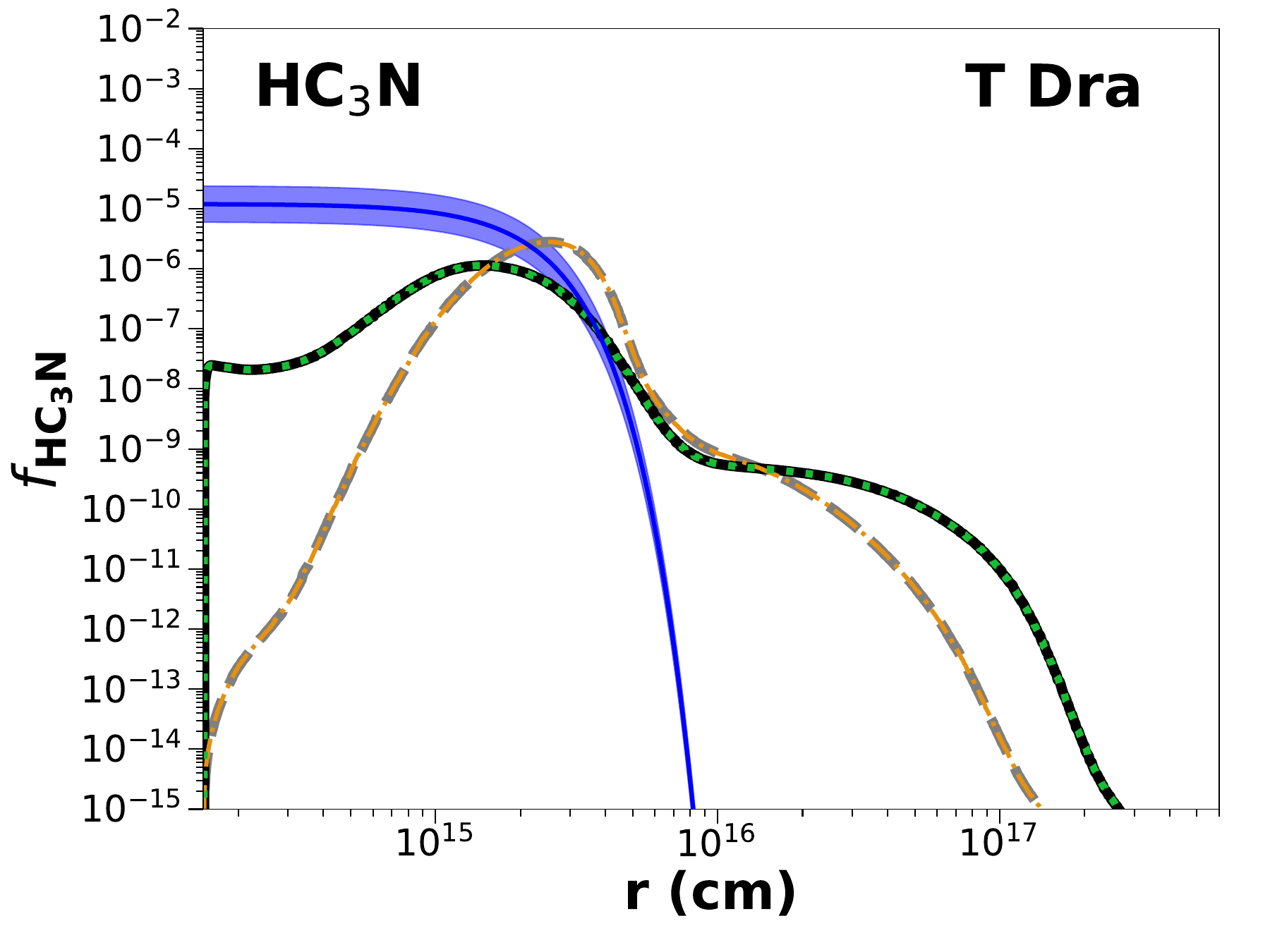}
     \end{subfigure}

     \begin{subfigure}[b]{0.33\linewidth}
         \centering
         \includegraphics[width=\linewidth]{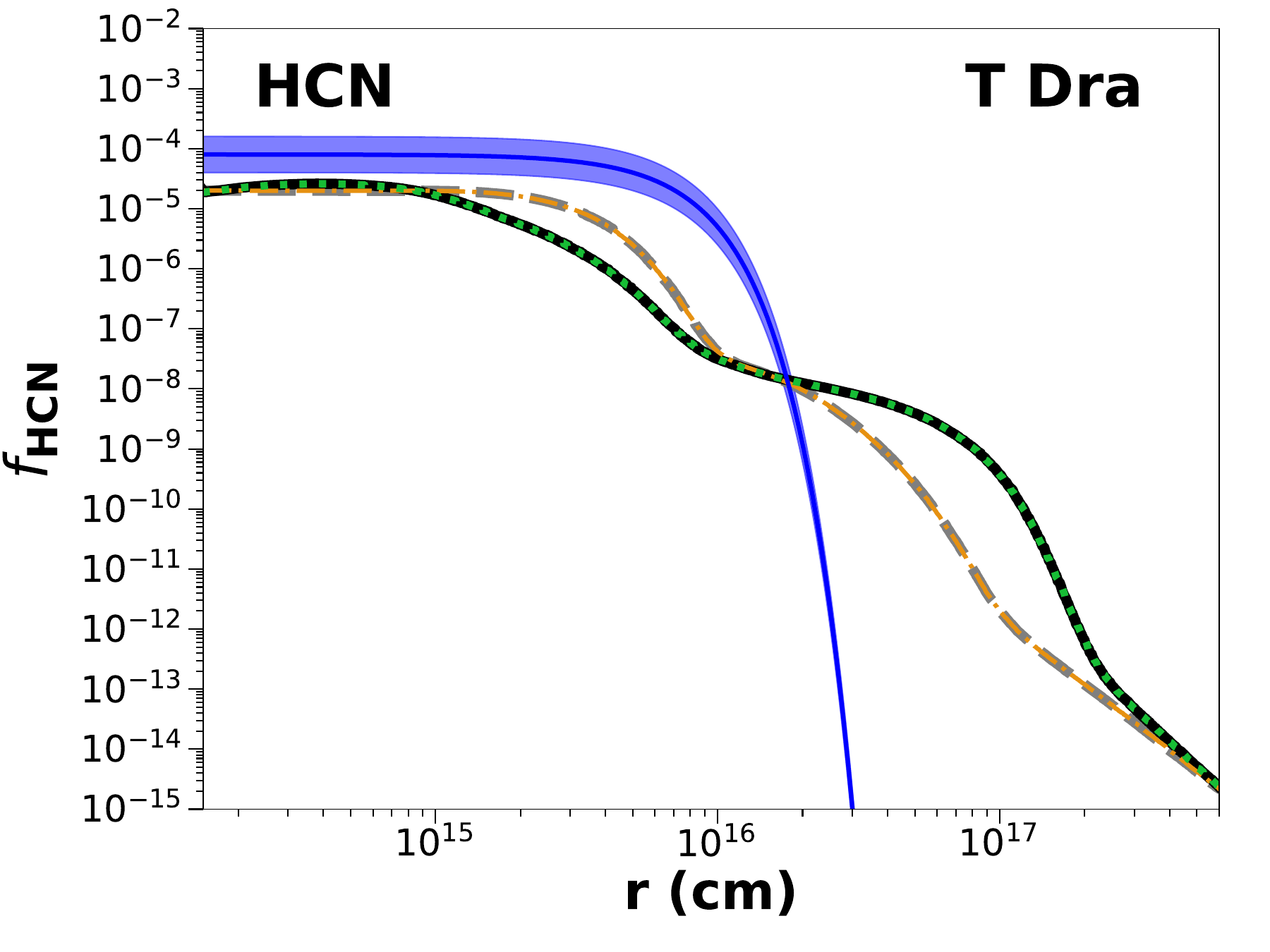}
     \end{subfigure}
     \begin{subfigure}[b]{0.33\linewidth}
         \centering
         \includegraphics[width=\linewidth]{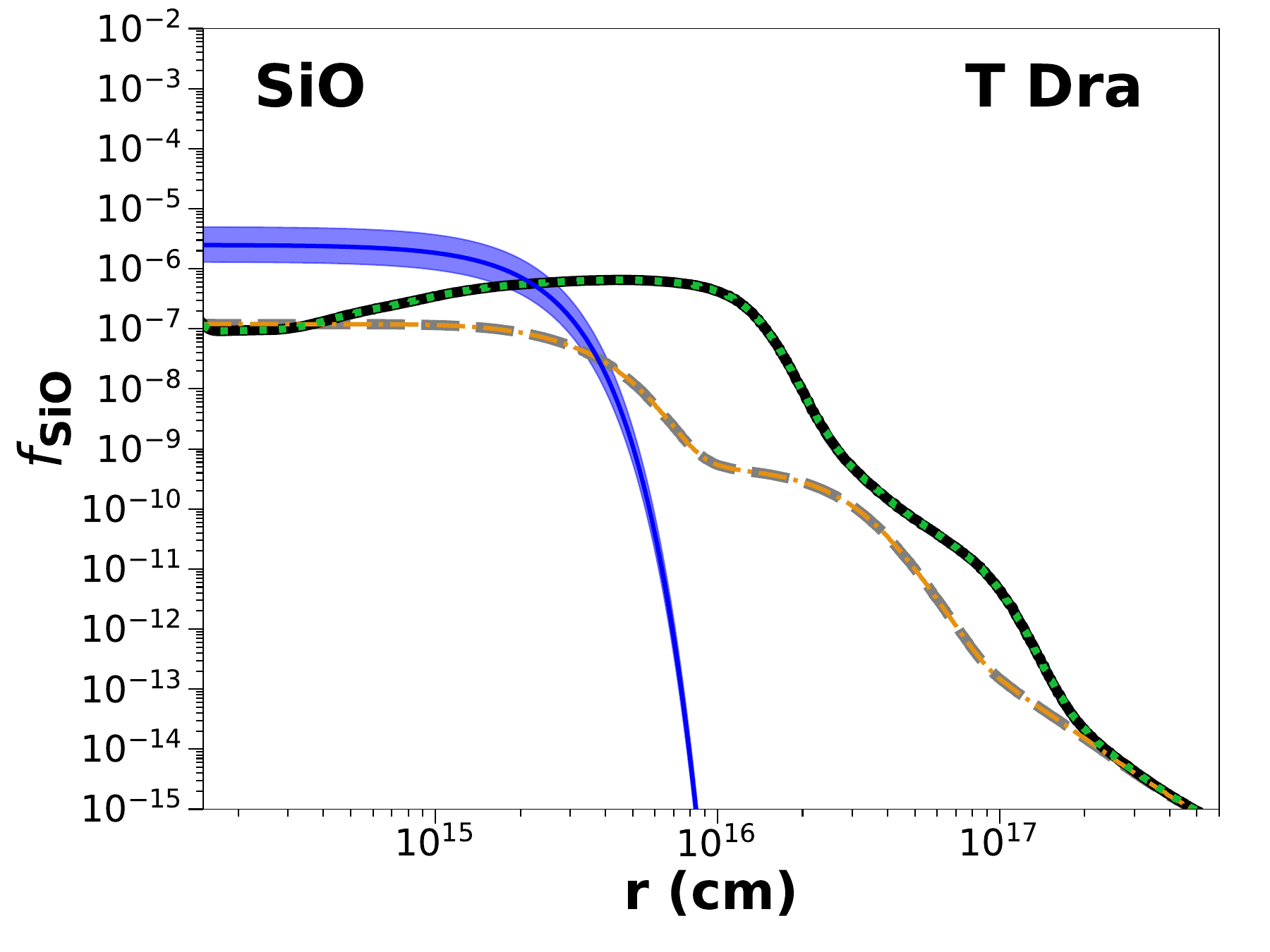}
     \end{subfigure}
     \begin{subfigure}[b]{0.33\linewidth}
         \centering
         \includegraphics[width=\linewidth]{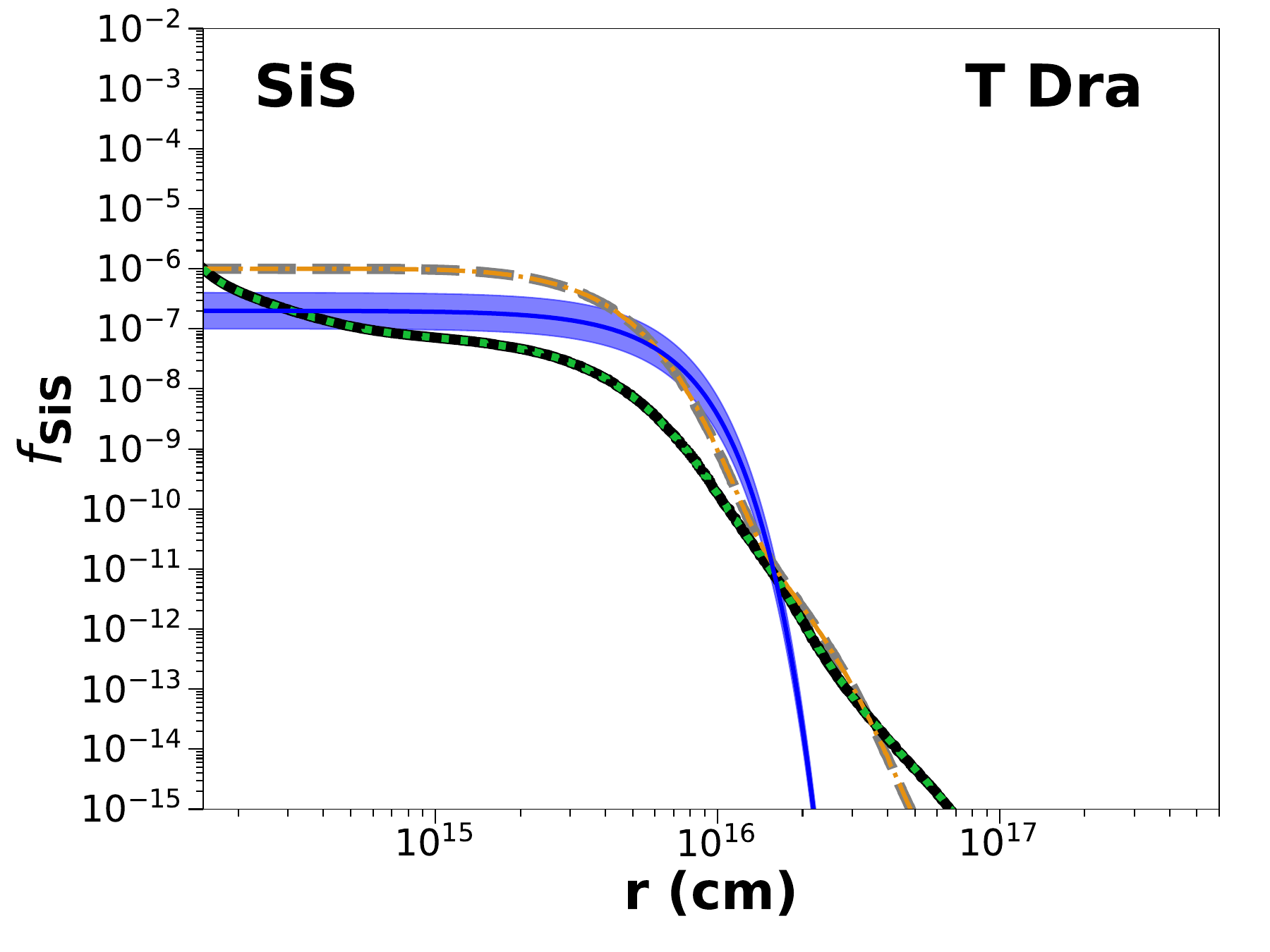}
     \end{subfigure}

     \begin{subfigure}[b]{0.33\linewidth}
         \centering
         \includegraphics[width=\linewidth]{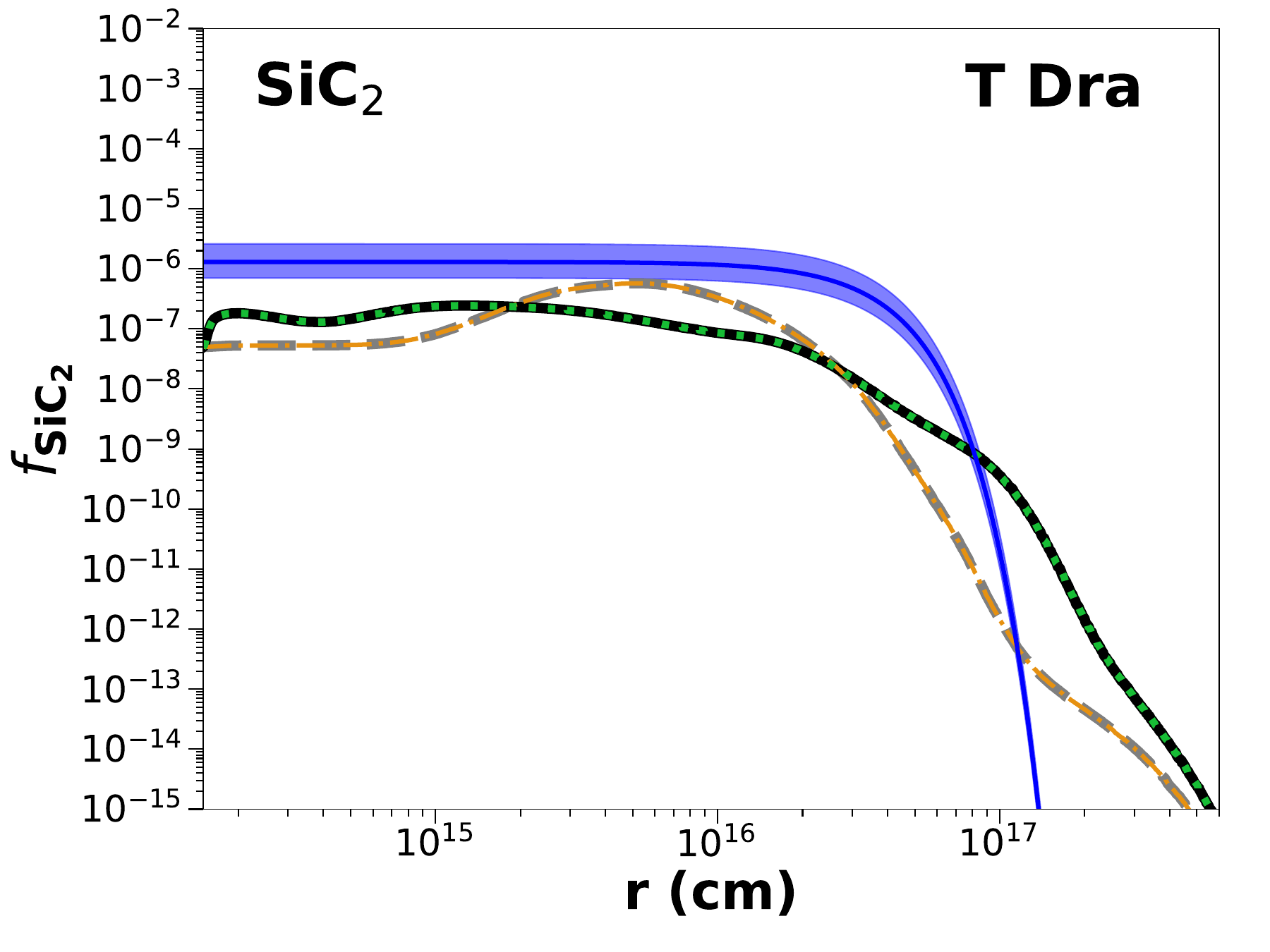}
     \end{subfigure}
     \begin{subfigure}[b]{0.33\linewidth}
         \centering
         \includegraphics[width=\linewidth]{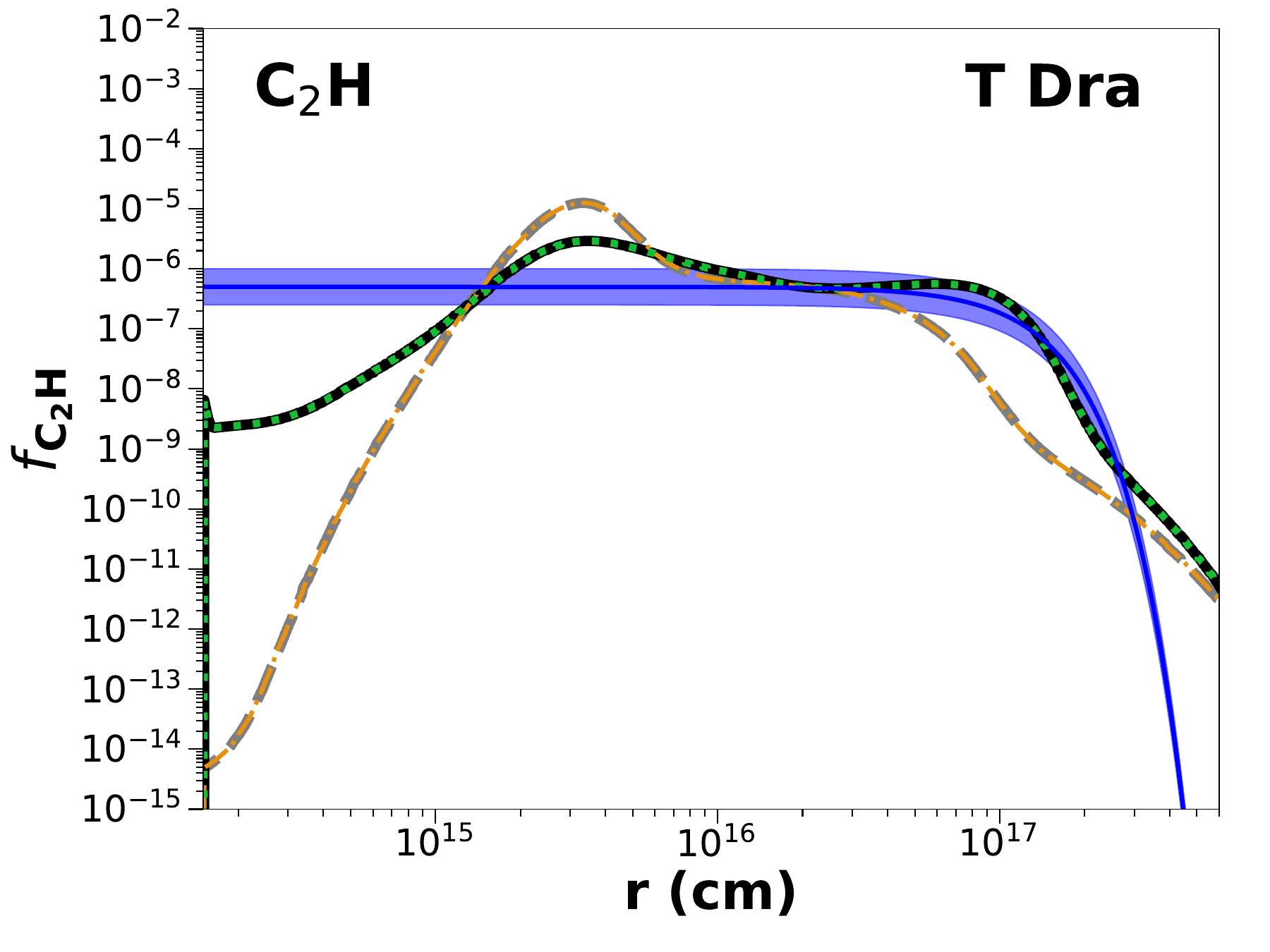}
     \end{subfigure}
     \begin{subfigure}[b]{0.33\linewidth}
         \centering
         \includegraphics[width=\linewidth]{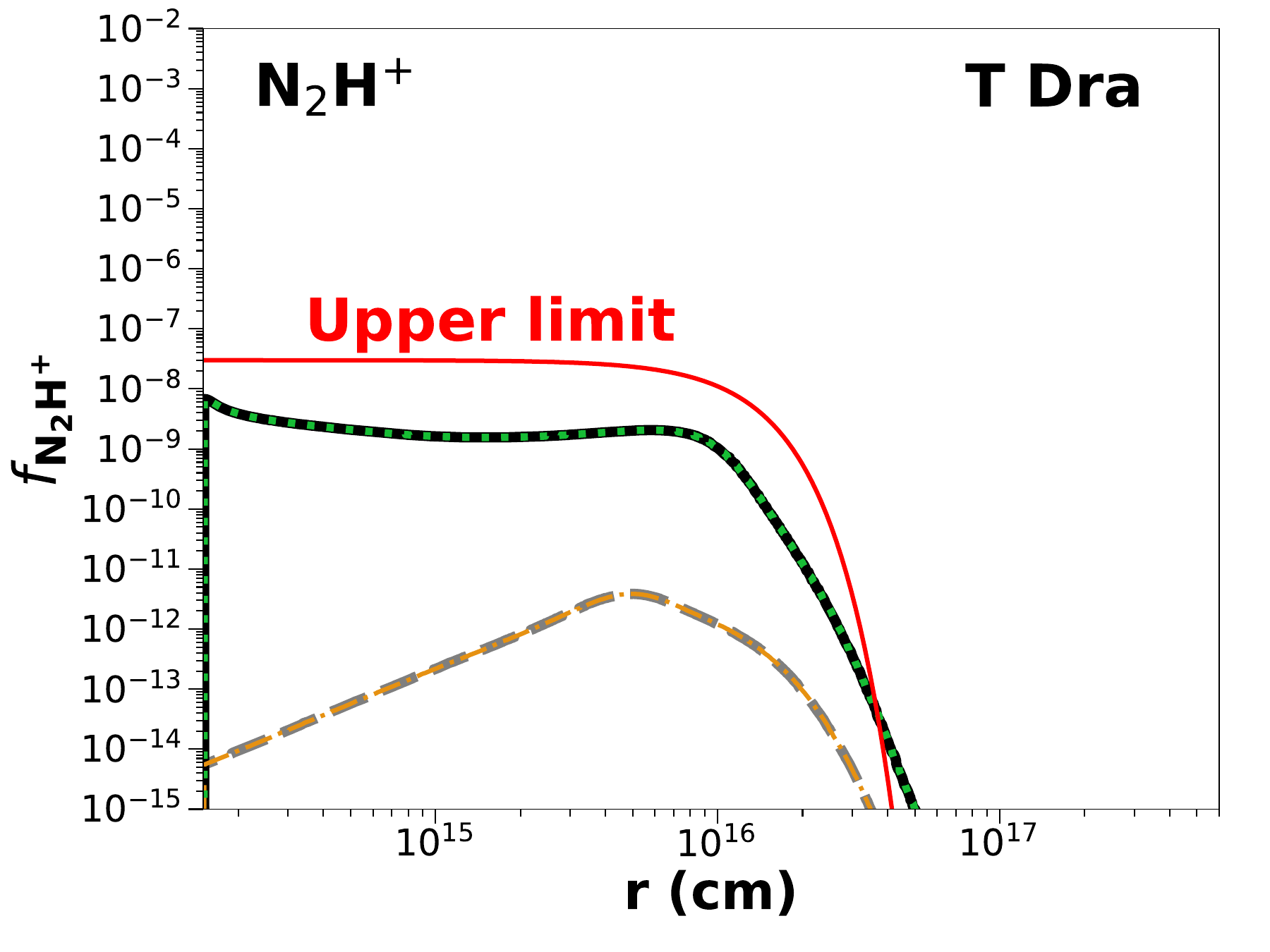}
     \end{subfigure}
        \caption{Comparison between chemical kinetics model predictions and empirically estimated abundances (see Table~\ref{tab:abundances}) for T\,Dra. The black solid line represents 'model-uv/x', the green dotted line represents 'model-x', the grey dashed line represents 'model-std', the orange dash-dotted line represents 'model-uv'. The blue solid lines shows the empirical estimations (\hcoplus\, has a dark blue line for 2020 observation and a light blue line for 2024 observations), and the shadowed areas indicate a factor of two for the uncertainties in the empirical abundances estimations. The red line represents the empirical abundance upper limit for $\rm N_{2}H^{+}$.}
        \label{fig:T-DRA_chemistry}
\end{figure*}

Fig.~\ref{fig:T-DRA_chemistry} shows the predictions from our chemical kinetics models in the case of T\,Dra for \hcoplus\ and the rest of the species detected by us in this target. As for CO, we consider the four different models with and without UV- and/or X-rays emission from the central source. However, as mentioned in the previous section, the effect of internal UV emission is negligible due to strong attenuation of this radiation in the inner envelope regions of T\,Dra.

Figure~\ref{fig:T-DRA_chemistry} shows a remarkable enhancement of the abundance of \hcoplus\, through the whole envelope, but especially from the inner regions, as an effect of X-ray emission: the predicted \hcoplus\ abundance from  model "model-x/uv" is at least $\sim$2 orders of magnitude larger than for "model-std" through the whole envelope. The average \hcoplus\ abundance in "model-x/uv" is also in excellent agreement with the average abundance in T\,Dra empirically determined in this work (Sect.~\ref{MADEX_other}, Table~\ref{tab:abundances}). The large \hcoplus\ abundance enhancement results in a three-order-of-magnitude increase in column density compared with models without X-rays, leading to a much better agreement with the observed column density for T\,Dra.

In standard AGB CSEs, \hcoplus\ predominantly forms in the outer envelope regions through photochemical processes driven by the ISRF, with CO as the main precursor. The dominant formation and destruction pathways for \hcoplus\ depend on the gas chemistry (i.e. C-rich or O-rich).

In C-rich envelopes, the expected \hcoplus\, abundance is about $\lesssim$4$\times 10^{-9}$ \citep[see][]{Pulliam_2011}, and its formation is led by protonation of CO by the chemical reaction

\begin{equation}
  \begin{array}{l l}
    \rm H_{3}^{+} + CO  \rightarrow  HCO^{+} + H_{2}, 
   \label{reacciones_hcoplus_1}
  \end{array}
\end{equation}

\noindent where \hhhplus\, is formed by cosmic ray ionisation of \hh\, and the interaction between \hh\, and \hhplus\, by

\begin{equation}
  \begin{array}{l l}
    \rm H_{2}  \xrightarrow[]{\mathrm{CR}}  H_{2}^{+} + e^{-},\\
    \rm H_{2}^{+} + H_{2} \rightarrow  H_{3}^{+} + H.
   \label{hhhplus}
  \end{array}
\end{equation}

In O-rich envelopes, the expected \hcoplus\, abundance is considerably larger, about $\sim$$10^{-8}$-$10^{-7}$ \citep[see][]{Pulliam_2011}, and it is related with the larger abundance of \water\ and \coplus\ \citep[by interaction between \cplus\ and oxygen-rich molecules, see][]{Li_2016}, thus intensifying the reactions

\begin{equation}
  \begin{array}{l l}
  \rm C^{+} + H_{2}O \rightarrow HCO^{+} + H,   \\
  \rm CO^{+} + H_{2} \rightarrow HCO^{+} + H, 
   \label{reacciones_hcoplus_2}
  \end{array}
\end{equation}

\noindent where $\rm C^{+}$ is produced by photoionisation of atomic carbon by the ISRF and $\rm CO^{+}$ is produced by interaction of $\rm C^{+}$ with OH

\begin{equation}
  \begin{array}{l l}
    \rm C \xrightarrow[]{\mathrm{h \nu}} C^{+} + e^{-},\\
    \rm C^{+} + OH \rightarrow  CO^{+} + H.\\ 
   \label{hhhplus}
  \end{array}
\end{equation}

In both chemical types, the \hcoplus isomer, \hocplus, which is produced by chemical reactions analogue to Eq.~\ref{reacciones_hcoplus_1} and ~\ref{reacciones_hcoplus_2}, is then used to produce \hcoplus\, mainly by interaction with \hh\ as follows 

\begin{equation}
  \begin{array}{l l}
    \rm HOC^{+} + H_{2} \rightarrow HCO^{+} + H_{2}. \\
   \label{reacciones_hcoplus}
  \end{array}
\end{equation}

Regardless of the chemistry or the presence of an X-ray source, the destruction of \hcoplus\, is driven by electron dissociative recombination.

\begin{equation}
  \begin{array}{l l}
   \rm HCO^{+} + e^{-} \rightarrow CO + H,\\
   \rm HCO^{+} + e^{-} \rightarrow C + OH.\\
  \end{array}
\end{equation}

In the presence of an internal X-ray source, which induces significant photoionisation of \hh\ and enhances \hhhplus, \hcoplus\,
is efficiently produced via the chemical reaction shown in Eq.~\ref{reacciones_hcoplus_1}. This X-ray-driven reaction dominates the production of 
\hcoplus\ throughout the entire envelope, except in the outermost regions, where the dominant chemical UV-driven pathways remain the same as in standard AGB CSEs.

\begin{figure*}[h!]
     \centering
     \begin{subfigure}[b]{0.33\linewidth}
         \centering
         \includegraphics[width=\linewidth]{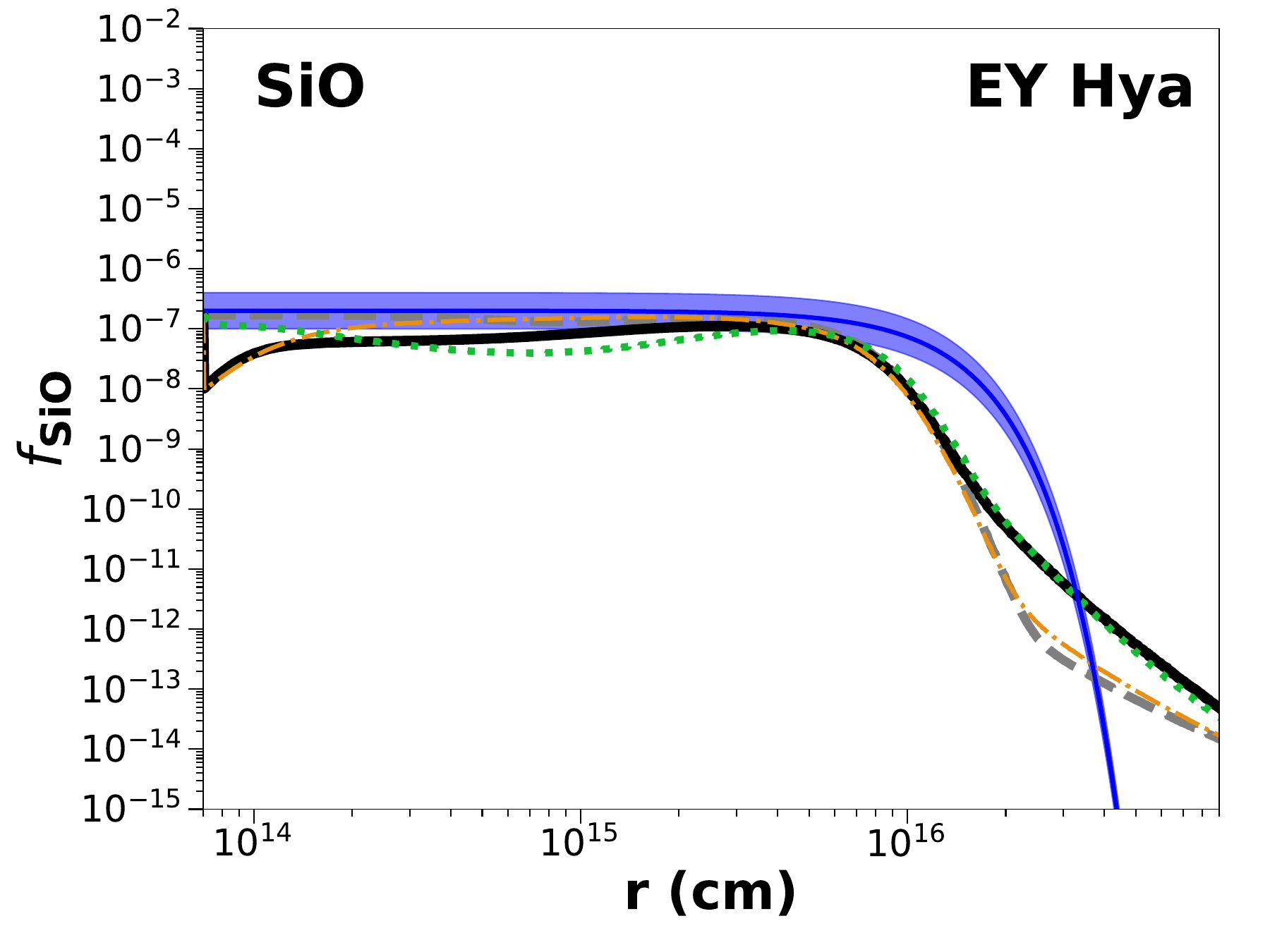}
     \end{subfigure}
     \begin{subfigure}[b]{0.33\linewidth}
         \centering
         \includegraphics[width=\linewidth]{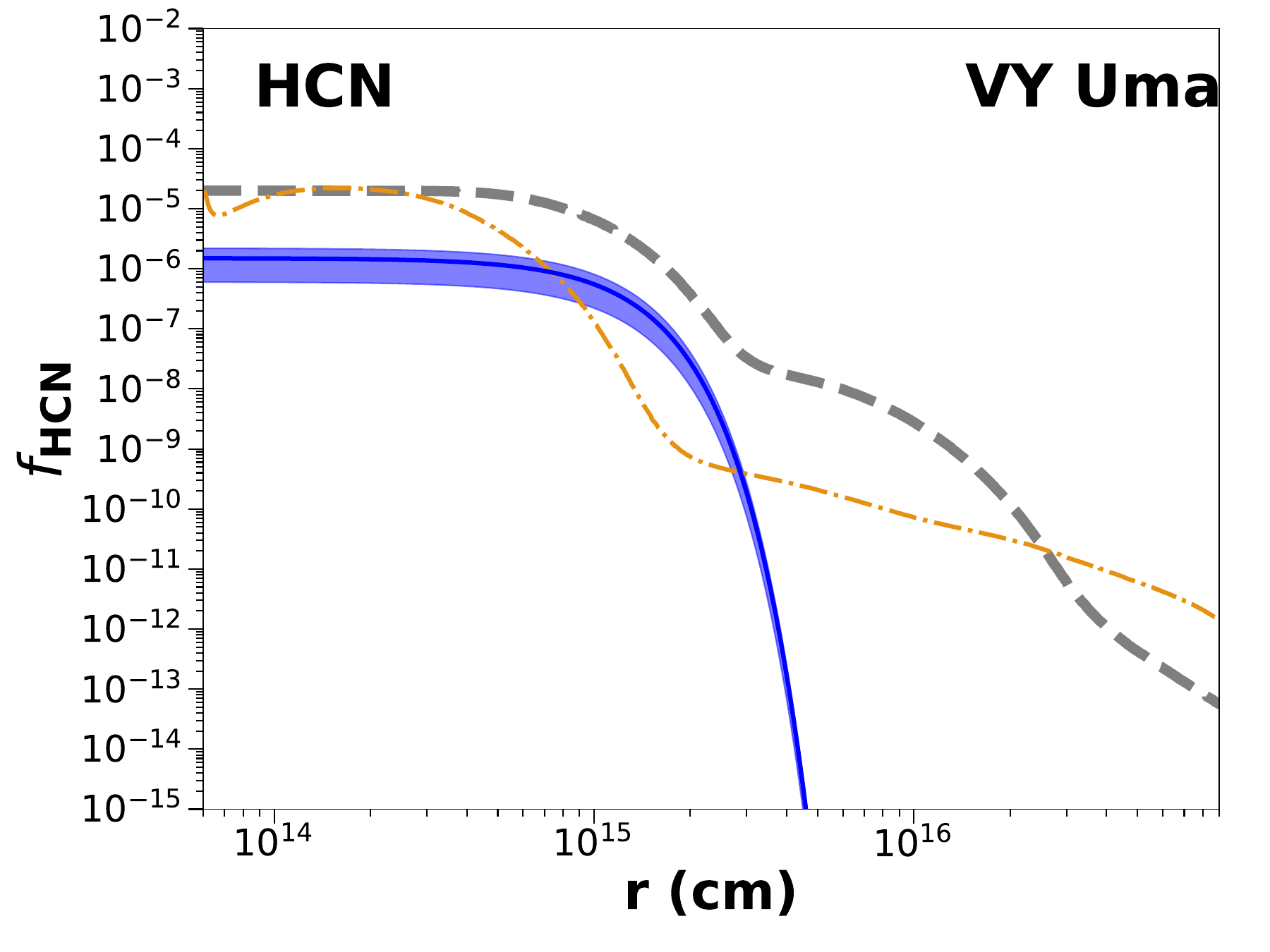}
     \end{subfigure}
     \begin{subfigure}[b]{0.33\linewidth}
         \centering
         \includegraphics[width=\linewidth]{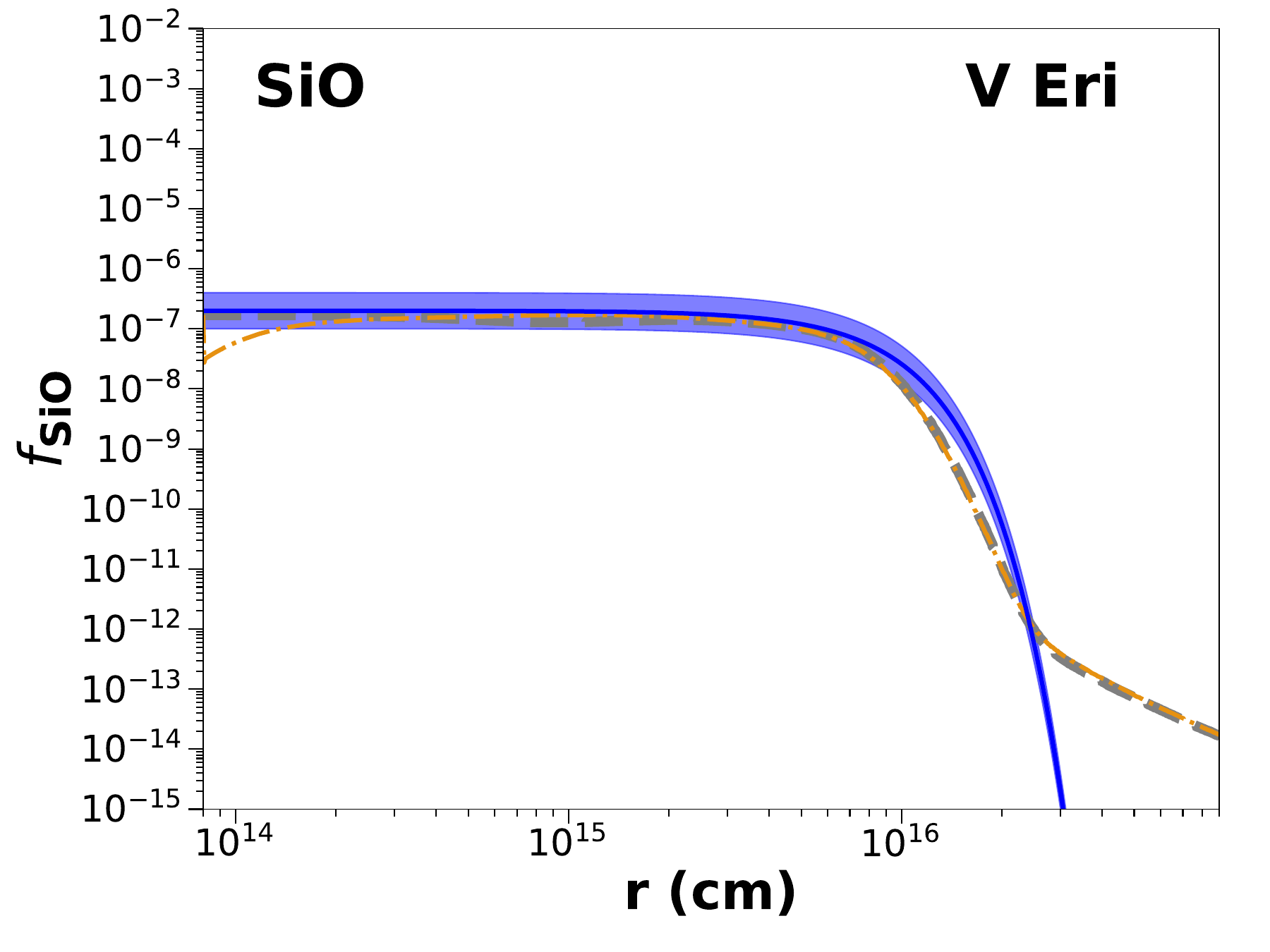}
     \end{subfigure}

     \begin{subfigure}[b]{0.33\linewidth}
         \centering
         \includegraphics[width=\linewidth]{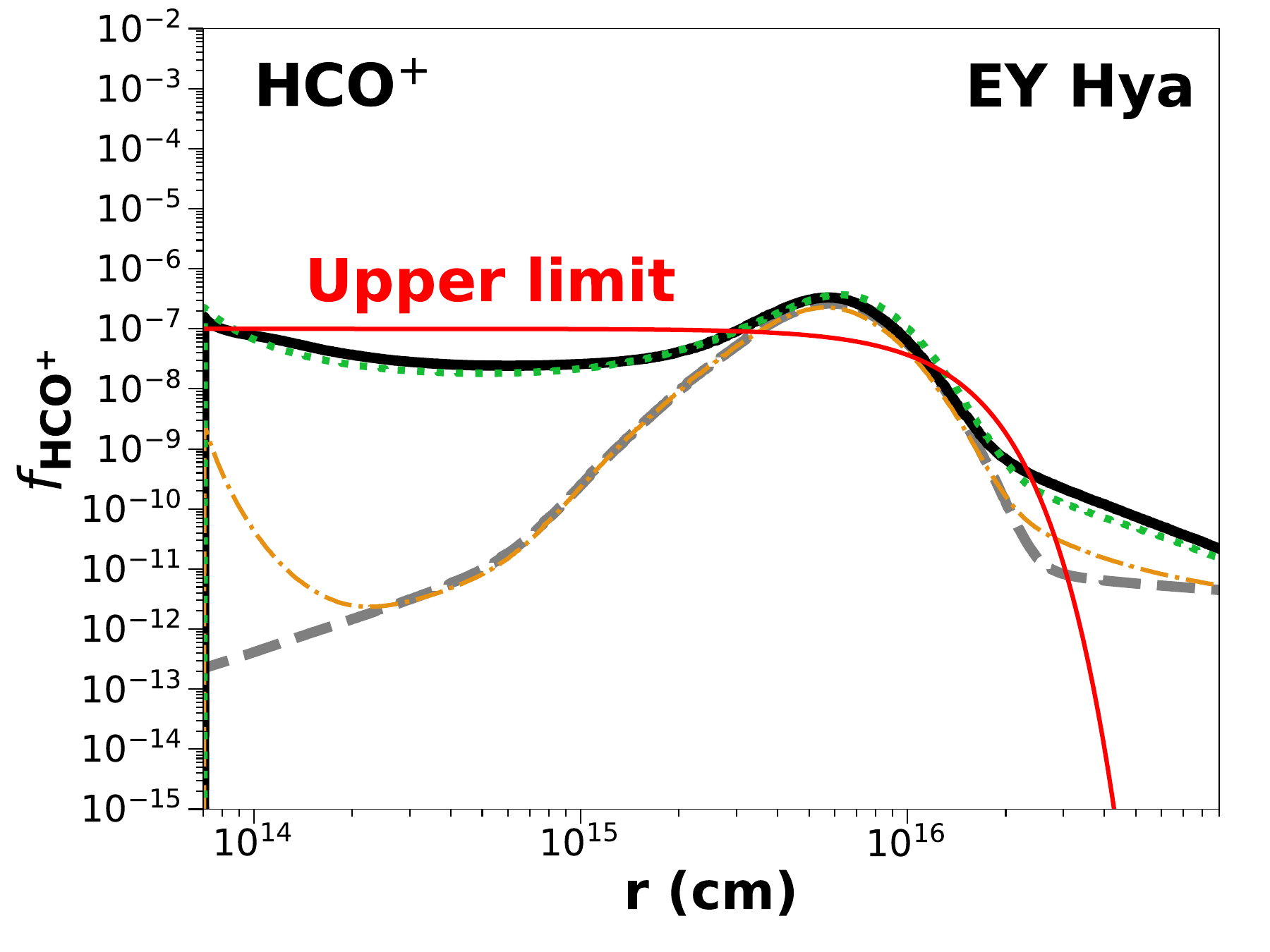}
     \end{subfigure}
     \begin{subfigure}[b]{0.33\linewidth}
         \centering
         \includegraphics[width=\linewidth]{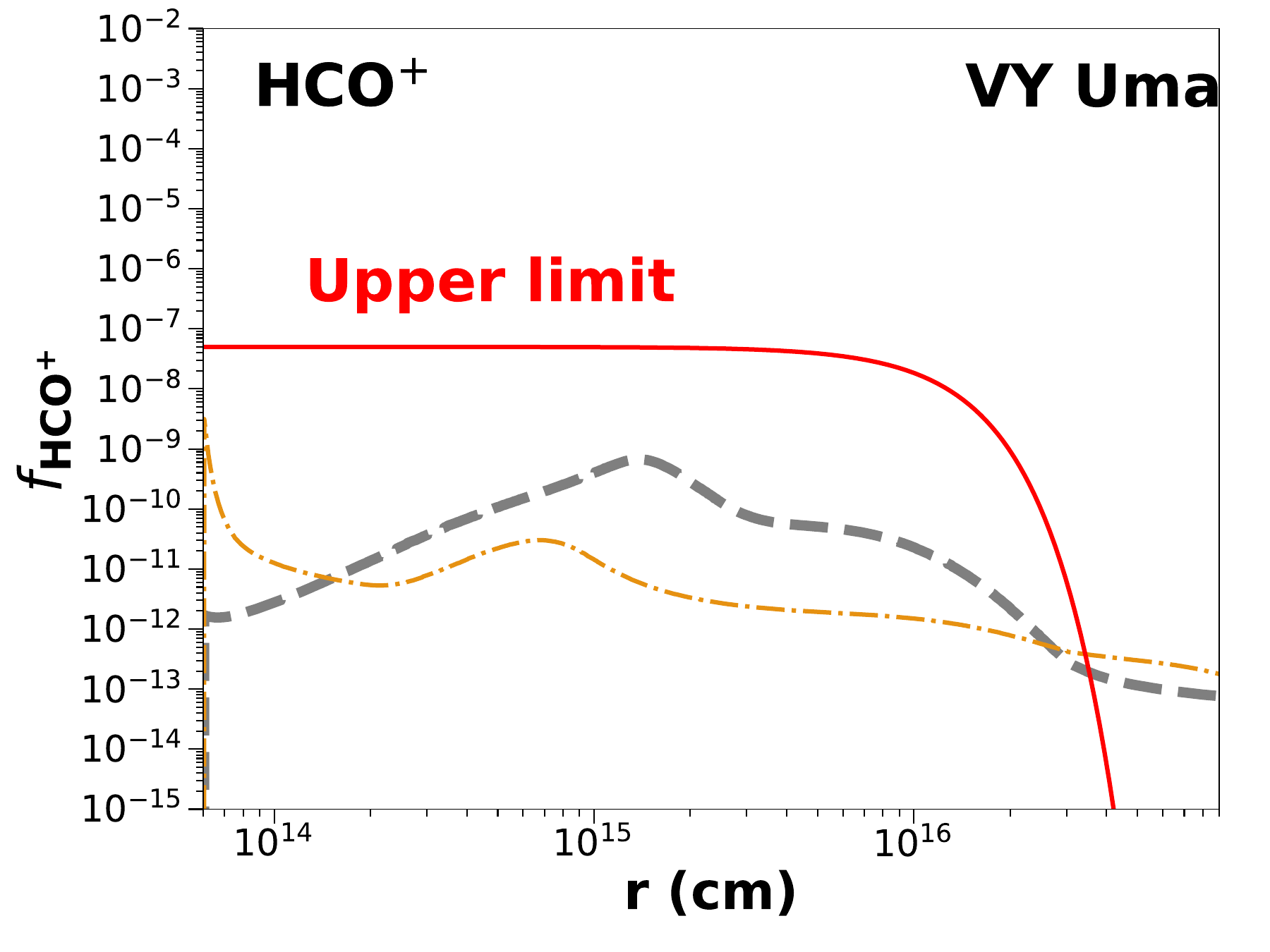}
     \end{subfigure}
     \begin{subfigure}[b]{0.33\linewidth}
         \centering
         \includegraphics[width=\linewidth]{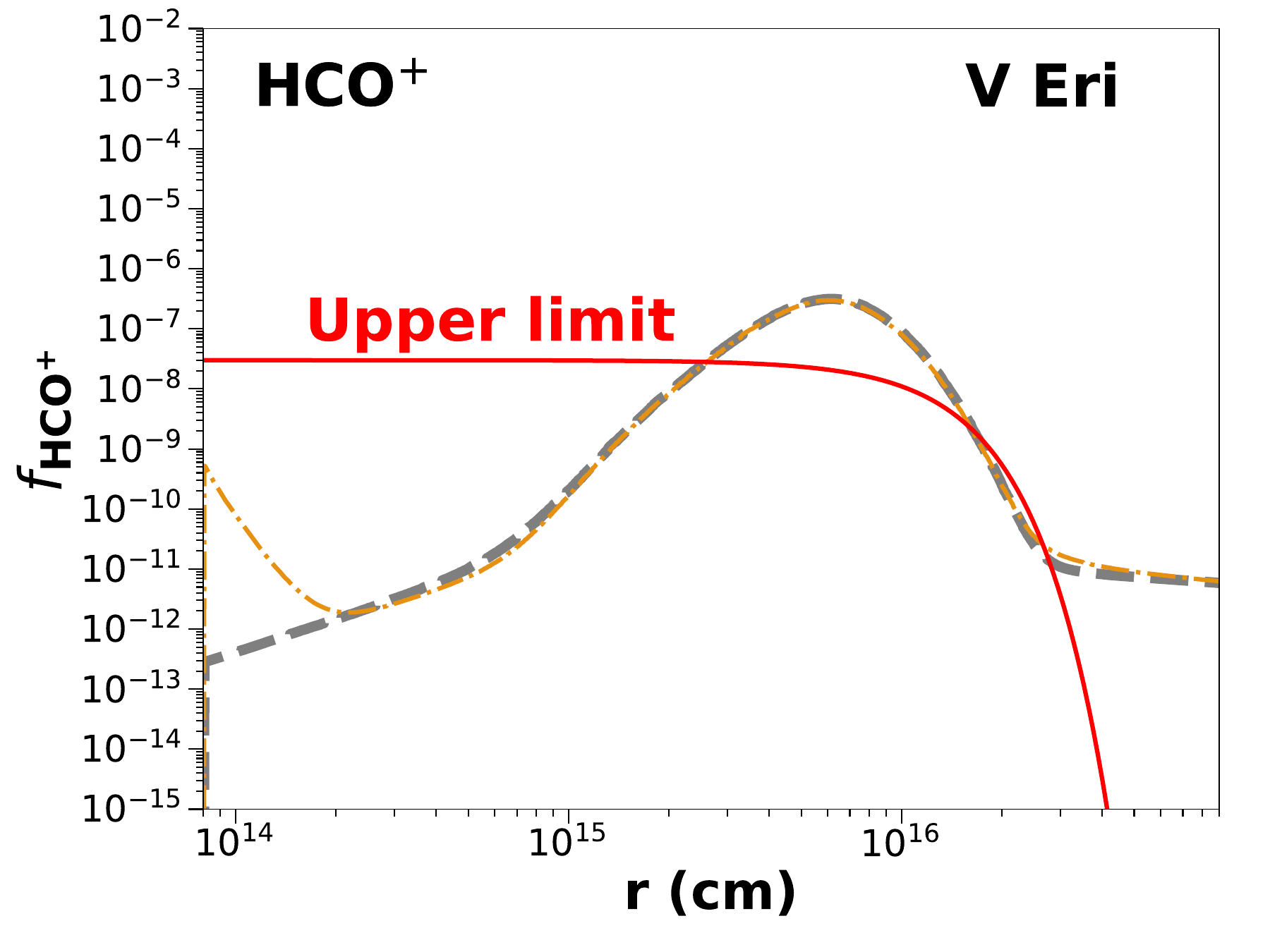}
     \end{subfigure}
        \caption{Comparison between chemical kinetics model predictions and empirically estimated abundances (see Table~\ref{tab:abundances}) for the rest of sources. The black solid line represents 'model-uv/x', the green dotted line represents 'model-x', the grey dashed line represents 'model-std', the orange dash-dotted line represents 'model-uv'. The blue solid lines shows the empirical estimations, and the shadowed areas indicate a factor of two for the uncertainties in the empirical abundances estimations. The red line represents the empirical abundance upper limit for \hcoplus.}
        \label{fig:OTHER_chemistry}
\end{figure*}

\subsection{\hnc\, enhanced abundance}

Similar to \hcoplus, the abundance of HNC is enhanced by up to four orders of magnitude in the inner regions of T\,Dra due to the effect of the internal X-ray source (Fig.~\ref{fig:T-DRA_chemistry}). Models without X-rays fail to reproduce the relatively high average \hnc\ abundance observed in T\,Dra, whereas including an internal X-ray source is leads to a much better agreement. In terms of column density,  the model with X-rays predicts an increase of at least two orders of magnitude compared to the model without X-rays, resulting in significantly improved agreement with the observed value (see Fig.~\ref{fig:bar_plot}). 

In a standard C-rich AGB star envelope, the chemical reaction that dominates the production of \hnc in the absence of X-ray emission is \hcnhplus protonation through

\begin{equation}
  \begin{array}{l l}
   \rm HCNH^{+} + e^{-} \rightarrow HNC + H,
   \label{reac_hcnhplus}
  \end{array}
\end{equation}

\noindent where the molecular ion \hcnhplus\ is formed in the outer regions of the envelope by cosmic ray ionisation via

\begin{equation}
  \begin{array}{l l}
    \rm HCN + H_{3}^{+} \rightarrow HCNH^{+} + H_{2}.
  \end{array}
\end{equation}

Under the presence of an internal X-ray source, the production of \hhhplus\ and, subsequently, \hcnhplus\ and \hnc\ abundances are enhanced throughout the whole envelope.

In any case, the destruction of \hnc\ is driven by the interaction with ISRF photons in the outer regions of the envelope or internal UV/X-ray photons in the inner layers via

\begin{equation}
  \begin{array}{l l l}
   \rm HNC \xrightarrow[]{\mathrm{h \nu}} CN + H.
  \end{array}
\end{equation}

\subsection{\hctresn\, enhanced abundance}\label{HCtresN_reactions}

Internal X-ray emission produces a notable enhancement of \hctresn\ abundance in the inner regions of the envelope of T\,Dra, close to (yet slightly lower than) the average abundance empirically derived  (Fig.~\ref{fig:T-DRA_chemistry}). Models excluding X-rays fail to reproduce the elevated \hctresn\ levels derived from our observations, with the largest discrepancies occurring at the inner regions. 

In terms of column densities, the total values predicted by chemical models with and without X-rays remain comparable (both below the observed value). This is because significant contributions from the middle and outer envelope partially counterbalance the pronounced abundance differences found in the inner regions (see Fig.~\ref{fig:bar_plot}). The fact that both models underestimate the observed column density could be partially attributed to the adopted input abundances, which may be unrealistically low.

In a standard C-rich AGB star, the production of \hctresn\, involves different chemical paths in the different regions of the envelope \citep[see][]{agundez_2017}, although the reaction that dominates on average is

\begin{equation}
  \begin{array}{l }
   \rm CN + C_{2}H_{2} \rightarrow HC_{3}N + H.\\
   \label{reac_hctresn}
  \end{array}
\end{equation}

However, in the presence of an internal high-energy radiation source, X-ray and the most energetic UV photons can efficiently dissociate \hcn, which has a relatively low dissociation energy \citep[slightly larger than 8 eV, see][]{Heays_2017}. Therefore, the \cn\, abundance is enhanced by \hcn\, photodissociation via

\begin{equation}
  \begin{array}{l }
   \rm HCN \, \xrightarrow[]{\mathrm{h \nu}} \, CN  \, + H,  
   \label{hcn_photo}
  \end{array}
\end{equation}

\noindent which subsequently increases the abundance of \hctresn\, significantly by the chemical reaction \ref{reac_hctresn}. In addition, the X-ray-driven enhancement of \hnc\, also produces more \hctresn\, through
 
\begin{equation}
  \begin{array}{l}
   \rm HNC + C_{2}H \rightarrow HC_{3}N + H.
   \label{HNC_to_HCtresN}
  \end{array}
\end{equation}

In the presence of internal X-ray radiation, these chemical reactions dominate, significantly enhancing the production of \hctresn\ throughout the entire envelope. In contrast, only the first reaction (\ref{reac_hctresn}) is dominant in standard AGB CSEs without X-ray emission. Although a fraction of the UV photons are also energetic enough to photodissociate \hcn, following the reaction shown in Eq.~\ref{hcn_photo}, the density in the innermost regions of T\,Dra is high and does not allow the UV photons to penetrate the envelope and produce an \hctresn\, enhancement unless the envelope has a high level of porosity (see Sect.~\ref{porosity}). 

In both scenarios (with or without X-ray emission), the destruction of \hctresn\ in AGB CSEs occurs primarily through photodissociation regardless of the region within the envelope

\begin{equation}
  \begin{array}{l l l}
   \rm HC_{3}N  \xrightarrow[]{\mathrm{h \nu}} C_{3}N + H, \\
   \rm HC_{3}N \xrightarrow[]{\mathrm{h \nu}} CN + C_{2}H.
  \end{array}
\end{equation}

We highlight that the chemical reactions that enhance the \hctresn\, abundance can be also extrapolated to $\rm{HC_{X}N}$, where X is an odd number, with equivalent molecules in the respective reactions.

\subsection{Other molecular species}\label{other_molecules}

In addition to \hcoplus, \hnc, and \hctresn, which are the most X-ray sensitive species in this study and have been discussed in detail, other molecules are also affected to varying degrees by the presence of X-rays.

\hcn\, is an abundant species with intense emission lines commonly detected in AGB CSEs. In this study, we detect \hcn\, in the C-rich stars T\,Dra and VY\,Uma (Figs.~\ref{fig:T-DRA_chemistry} and \ref{fig:OTHER_chemistry} respectively). \hcn\, abundance can be reduced due to UV and X-ray photodissociation, or it may be increased by the liberation of N atoms from $\rm N_{2}$ photodissociation when the outflow has low density and/or high porosity, similarly as shown by \cite{Van_de_Sande_2022}. In any case, 
the overall effects of X-rays on HCN are moderate. The predicted abundances and column densities for both models, with and without X-rays, match the observations reasonably well, with the slight underestimation of the abundance possibly stemming from an input value that is too small.

\sio\, is another parent species with intense emission lines typically found in O-rich AGB CSEs. Indeed, this molecule is the only species detected in the envelopes of our O-rich AGB stars, EY\,Hya (X-ray emitting) and V\,Eri. For O-rich sources, the average \sio\, suffers a slight decrease of its overall abundance due to UV and X-ray photodissociation  \citep[see][]{Van_de_Sande_2022}, especially in the case of EY\,Hya due to the internal X-ray source (leading to a reduction of a factor $\sim$3 of the \sio\, column density with both UV and X-ray emission). This reduction of the \sio\, abundance is in good agreement with the relatively low \sio\, abundance found in these sources in comparison with standard O-rich AGB stars (see Sect.~\ref{comparison_obs}).

Among our C-rich AGB stars, \sio\, is detected only in the X-AGB T\,Dra, but not in VY\,Uma, which lacks X-rays. In C-rich envelopes, there is a complex interplay of phenomena in which \sio\, is destroyed by internal energetic photons and produced from free O atoms released from CO photodissociation due to the X-ray radiation. The observed average \sio\, abundance in T\,Dra aligns better with predictions from X-ray chemical kinetics models, which show an enhanced \sio\, abundance in the outer and intermediate envelope layers. However, the value of $r_e$ derived from our radiative transfer analysis is significantly smaller than the expected spatial extent of \sio\, from previous studies of C-rich envelopes (see Sect.~\ref{comparison_obs}). This discrepancy could indicate either a real feature or an underestimation of $r_e$ due to the limited number of lines used in the modelling, suggesting that the average SiO abundance inferred for T\,Dra could be an upper limit. 
In terms of column density, the predictions with and without X-rays for T\,Dra, which are both one order of magnitude lower than the value observed could also reflect that the SiC$_{2}$ (or other Si-bearing species) input abundance in the chemical kinetics model is too low for T\,Dra, as it could also be the case of other parent species (HCN most notably). 

According to our chemical kinetics models (Fig.~\ref{fig:T-DRA_chemistry}), the other Si-bearing molecules detected in the C-rich X-AGB T\,Dra, \sis\ and \sicdos, show a moderate sensitivity to X-rays. \sis\ experiences an overall reduction in abundance due to X-ray photodissociation, which is in agreement with the relatively low \sis\ abundance (see Sect.~\ref{comparison_obs}). On the other hand, \sicdos\ experiences a slight X-ray-driven enhancement across most of the envelope, although the peak value slightly decreases. However, there are no significant differences between the average \sicdos\ abundance predictions from the models with and without X-rays. In terms of column densities, the model predictions (with and without X-rays) better match the observed values for SiS than for SiC$_{2}$, which also points out that the SiC$_{2}$ input abundance in the chemical kinetics model might be underestimated for T\,Dra.

Our models also predict some X-ray-driven enhancement in the abundance of \cch, both  in the inner and outer envelope regions mainly produced by $\rm C_{2}H_{2}$ photodissociation: 

\begin{equation}
  \begin{array}{l l}
   \rm C_{2}H_{2} \xrightarrow[]{\mathrm{h \nu}} C_{2}H + H,
  \end{array}
\end{equation}

\noindent similarly to the production of $\rm C_{2}H$ in the outer shells of C-rich CSEs, although in this case by ISRF photodissociation \cite[see][]{agundez_2017}. The empirically inferred \cch\, abundance is not especially accurate and cannot be used to discern between the predictions from the models with and without X-ray emission. In terms of column densities, the models for T\,Dra with and without X-rays produce similar values that are comparable, within uncertainties, to the observed values, with a slightly better match for the models including UV radiation.

We additionally explored the effects of internal X-ray emission in the abundance of \ndoshplus\, which is extremely sensitive to X-rays as shown in Figs.~\ref{fig:T-DRA_chemistry} and \ref{fig:bar_plot}. However, despite the relatively high abundance enhancement of this ion in the presence of X-rays, the emission of \ndoshplus\ is expected to be weak and below the detection limits of this study, consistent with its non-detection in T\,Dra. The \ndoshplus\ abundance enhancement is led by the reaction

\begin{equation}
  \begin{array}{l l}
   \rm N_{2} + H_{3}^{+} \rightarrow  N_{2}H^{+} + H_{2}.
  \end{array}
\end{equation}

\subsection{UV effects considering a porous envelope}\label{porosity}

\begin{figure*}[h!]
     \centering
     \begin{subfigure}[b]{0.33\linewidth}
         \centering
         \includegraphics[width=\linewidth]{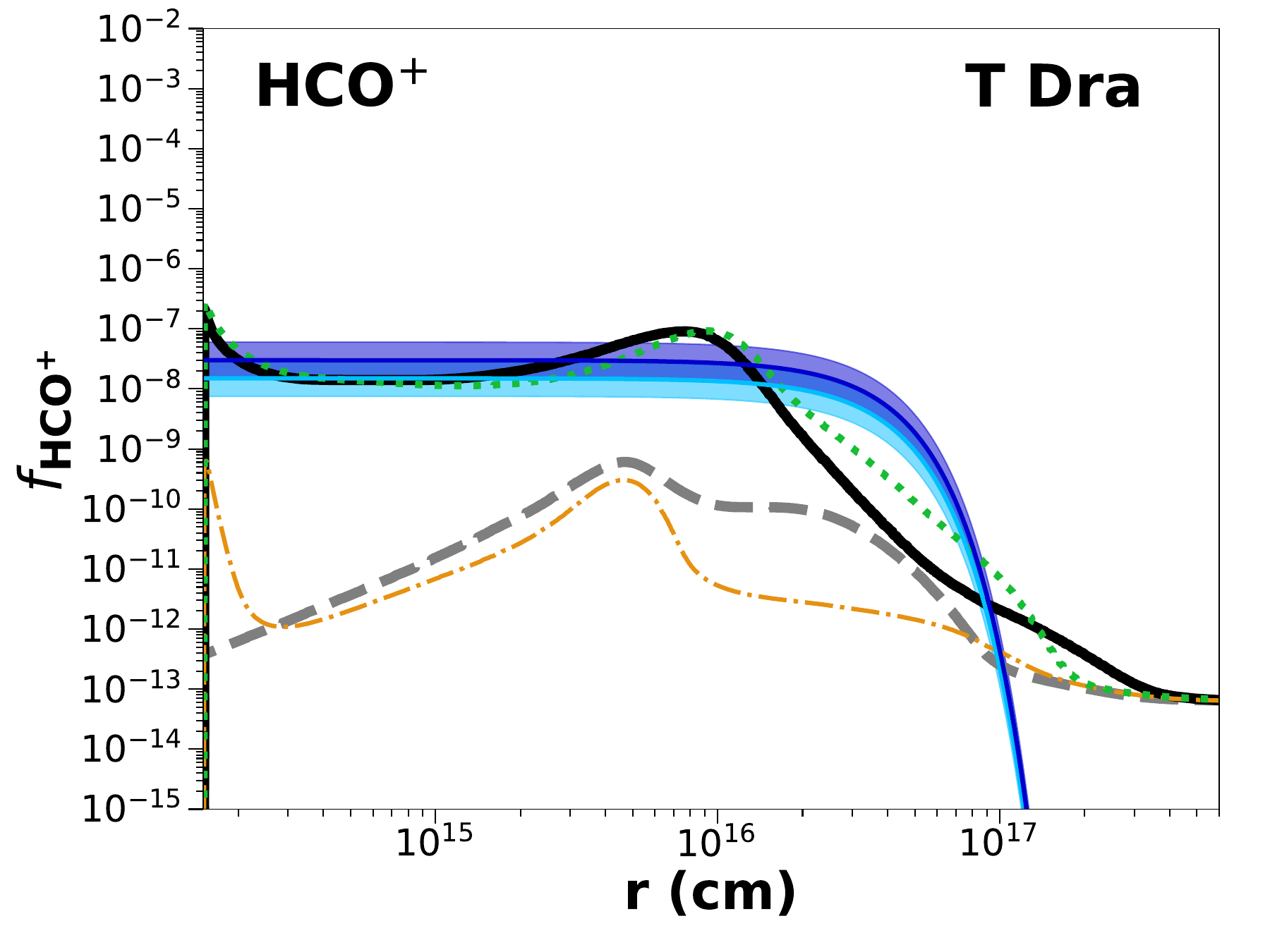}
     \end{subfigure}
     \begin{subfigure}[b]{0.33\linewidth}
         \centering
         \includegraphics[width=\linewidth]{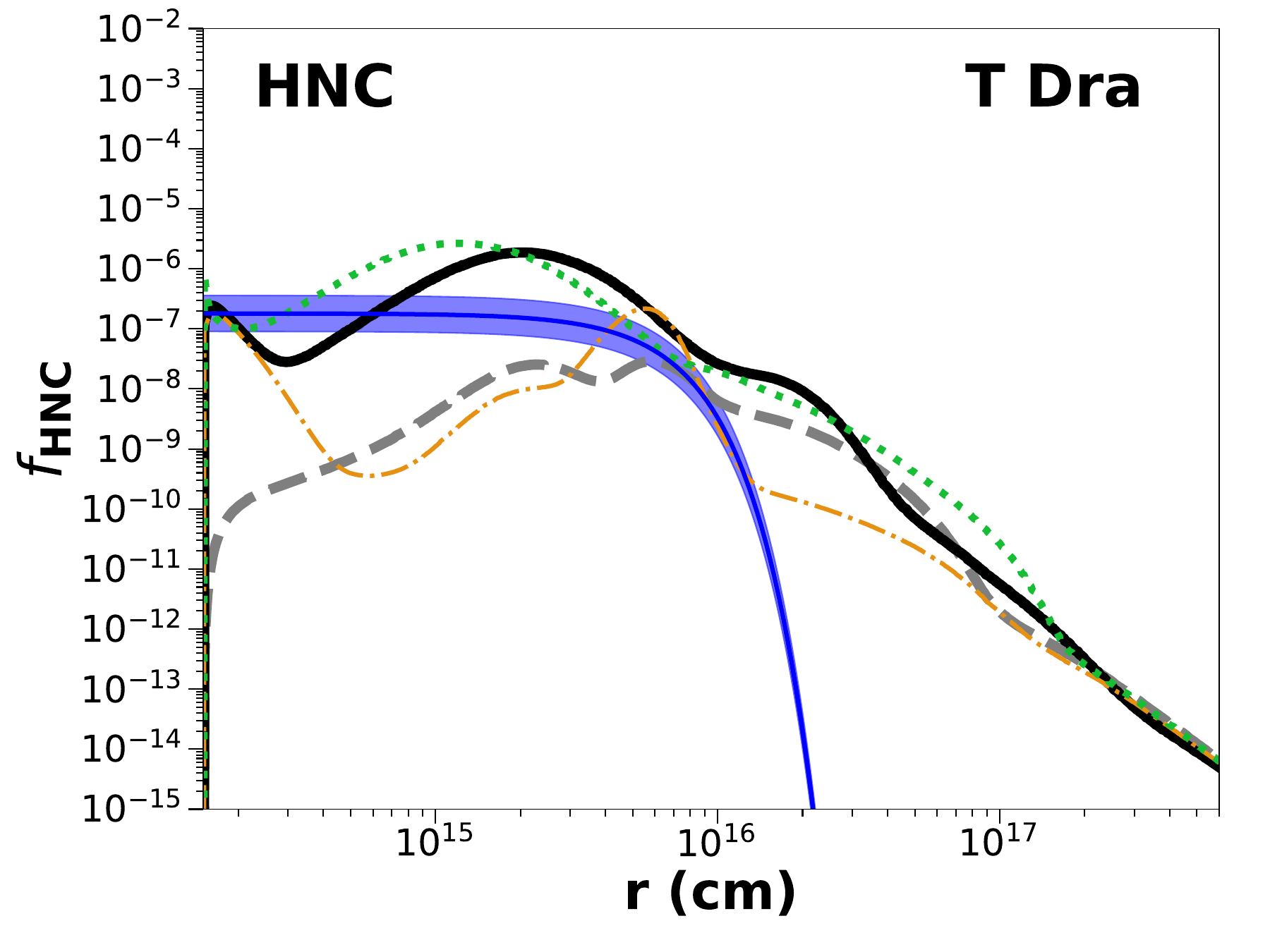}
     \end{subfigure}
     \begin{subfigure}[b]{0.33\linewidth}
         \centering
         \includegraphics[width=\linewidth]{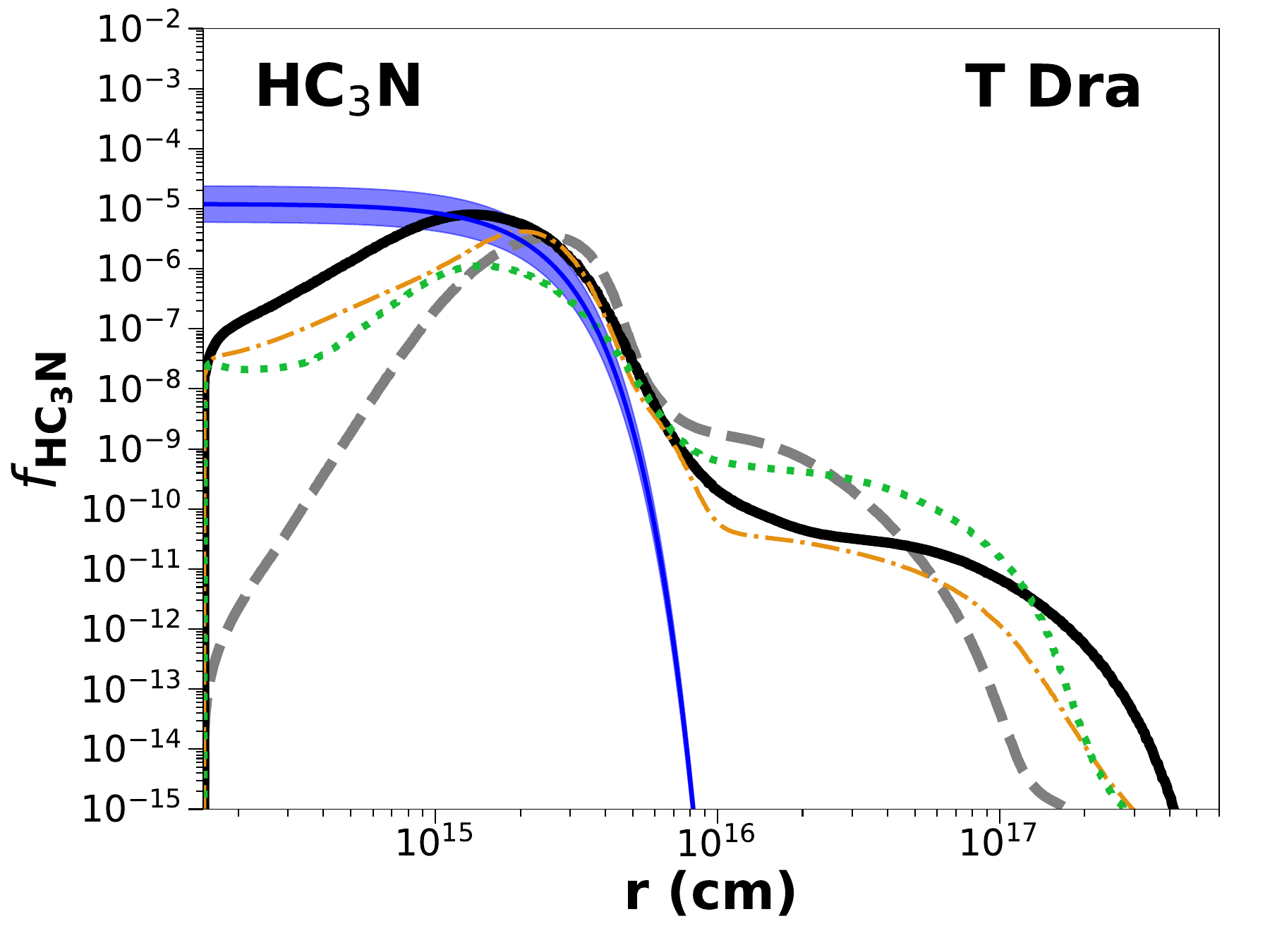}
     \end{subfigure}
        \caption{Comparison between the chemical kinetics model predictions and the empirically estimated abundances (see Table~\ref{tab:abundances}) for T\,Dra considering the case of a porous envelope. The black solid line represents 'model-uv/x', the green dotted line represents 'model-x', the grey dashed line represents 'model-std', the orange dash-dotted line represents 'model-uv'. The blue solid lines shows the empirical estimations (\hcoplus\, has a dark blue line for 2020 observation and a light blue line for 2024 observations), and the shadowed areas indicate a factor 2 for the uncertainties in the empirical abundances estimations.}
        \label{fig:T-DRA_chemistry2}
\end{figure*}

As shown earlier, in the smooth idealised CSE chemical kinetics model adopted here, the presence of an internal UV radiation source has minimal impact on the chemistry for objects with dense, optically thick envelopes such as T\,Dra (Fig.~\ref{fig:T-DRA_chemistry}). For optically thinner envelopes, the influence of internal UV radiation on the chemistry becomes more significant. For VY\,Uma, which has the lowest mass-loss rate in our sample, effects still remain marginal (Fig.~\ref{fig:OTHER_chemistry}). In contrast to UV, X-rays, with their greater penetration capability, produce more significant chemical effects over a broad range of mass-loss rates and envelope opacities.

To explore an extreme scenario in which the effects of UV photons on the CSE chemistry are maximised—helpful for distinguishing which effects can be confidently attributed to X-rays rather than UV photons—we additionally adopted the porosity model approach from \cite{Van_de_Sande_2018}, considering a one-component outflow where the interclump medium is void. In this approach, the porosity of the envelope is characterised by two parameters, the clump volume filling factor $f_{\rm vol}$ and the characteristic length scale near the stellar surface $l_*$. The related parameter $h_*$=$l_*/f_{\rm vol}$, described as the porosity length at the stellar surface, represents the mean free path of the UV photons.

While these parameters are not well constrained empirically for any object \citep[see][]{Van_de_Sande_2018}, we adopted $l_*$=$5\times10^{12}$ cm, $f_{\rm vol}$=0.5 and $h_*$=$1\times10^{13}$ cm, to achieve a significant penetration of the UV photons through the envelope maintaining similar density conditions through envelope. It is important to emphasise that there is no empirical evidence supporting a porous CSE in T\,Dra. Our goal is merely to explore an extreme scenario where UV radiation has maximal impact, allowing us to assess which molecules, under such conditions, could reach abundances comparable to those expected in X-ray-irradiated sources.

The chemical kinetics model presented here is customised for the C-rich X-AGB  T\,Dra, which is the best characterised object in our sample, exhibiting the richest chemistry and a \hcoplus\ detection. The envelope parameters are the same than those considered in the smooth envelope model (model (b) -- Table.~\ref{tab:MADEX_models}). However, in the porous envelope scenario, the effective optical depth has been reduced from $\tau_{UV}$=7.2 (smooth envelope) to $\tau_{UV}$=1.7. To ensure consistency with the detected UV flux, we adopt a UV luminosity of $L_{UV}$=7.0$\times 10^{31}$ $\mathrm{erg \, s^{-1}}$ and scale the ISRF (at $r$=5$\times 10^{13}$ cm) as $X_{\rm ISRF}$=8.2$\times 10^{5}$.

We focus on modelling \hcoplus, \hnc, and \hctresn, which are X-ray sensitive species detected in T\,Dra with abundances exceeding the typical values for standard AGB CSEs (see Fig.~\ref{fig:T-DRA_chemistry2}). Nevertheless, the expected column densities for the porous model with UV emission across all detected species are shown in Fig.~\ref{fig:bar_plot} shows and tabulated the for all the porous models in Appendix~\ref{column_densities}.
 
We find that \hcoplus\ and \hnc\ show enhancement in the innermost regions of the envelope when porosity is included, leading to factors $\sim$2 and $\sim$20 in their column densities (see Fig.~\ref{fig:bar_plot}). However, neither the predicted abundances nor the column densities reach the empirically inferred average/peak values or spatial extent.
In contrast, models with internal X-rays produce significantly larger enhancements—orders of magnitude in both central abundances and column densities. The close match between these X-ray models and the empirical abundances indicates that X-rays primarily drive the production of \hcoplus\ and HNC, despite model limitations. Since porosity alone fails to explain the observed enhancements, internal X-ray emission remains as the most plausible explanation.

On the other hand, the \hctresn\, abundance is expected to be significantly enhanced (around 1 order of magnitude in the \hctresn\, column density), in this case the effect of UV photons is similar to the X-rays, producing a central abundance enhancement \citep[similarly as proposed for IRC+10216 by][]{Siebert_2022}. In this case, both the central abundance and the column densities are more similar than those empirically inferred (although still 1 order of magnitude lower). However, this might be related with an underestimation of the assumed central abundances of some parent species in the chemical kinetics model or with the limitations of the radiative transfer analysis (e.g. the assumed abundance distribution) and the associated uncertainties. Therefore, its abundance enhancement cannot be discarded to be produce by both UV and X-ray internal emissions.

It is important to highlight that \ndoshplus\ is an ideal molecule for distinguishing the effects of X-rays and UV radiation, as they exert opposing influences on its abundance. While X-rays enhance the abundance of \ndoshplus\ by driving processes that increase its formation, UV radiation has a destructive effect, photodissociating N$_{2}$ and thus limiting the formation of \ndoshplus\ (see Sect.~\ref{other_molecules} and Fig.~\ref{fig:bar_plot}). However, the challenge stands in the fact that \ndoshplus\ lines are very weak and therefore difficult to detect. We note that this behaviour holds true only if the envelope is optically thin or porous, as it then allows UV radiation to penetrate and have a significant destructive effect on the molecule.

Finally, using this base envelope model, we explore even further the effects of UV emission across different ranges of UV luminosities and porosity. We found that, even in extreme cases with UV luminosities several orders of magnitude larger and extremely porous envelopes, it is not possible to recreate an abundance enhancement of \hcoplus\, and \hnc. Furthermore, in these extreme cases there would be produced a large destruction of UV sensitive molecules (e.g. \sio\, and \hcn) whose abundances would not be in agreement with those empirically constrained.

\begin{figure*}[htbp!]
     \centering
     \includegraphics[width=\linewidth]{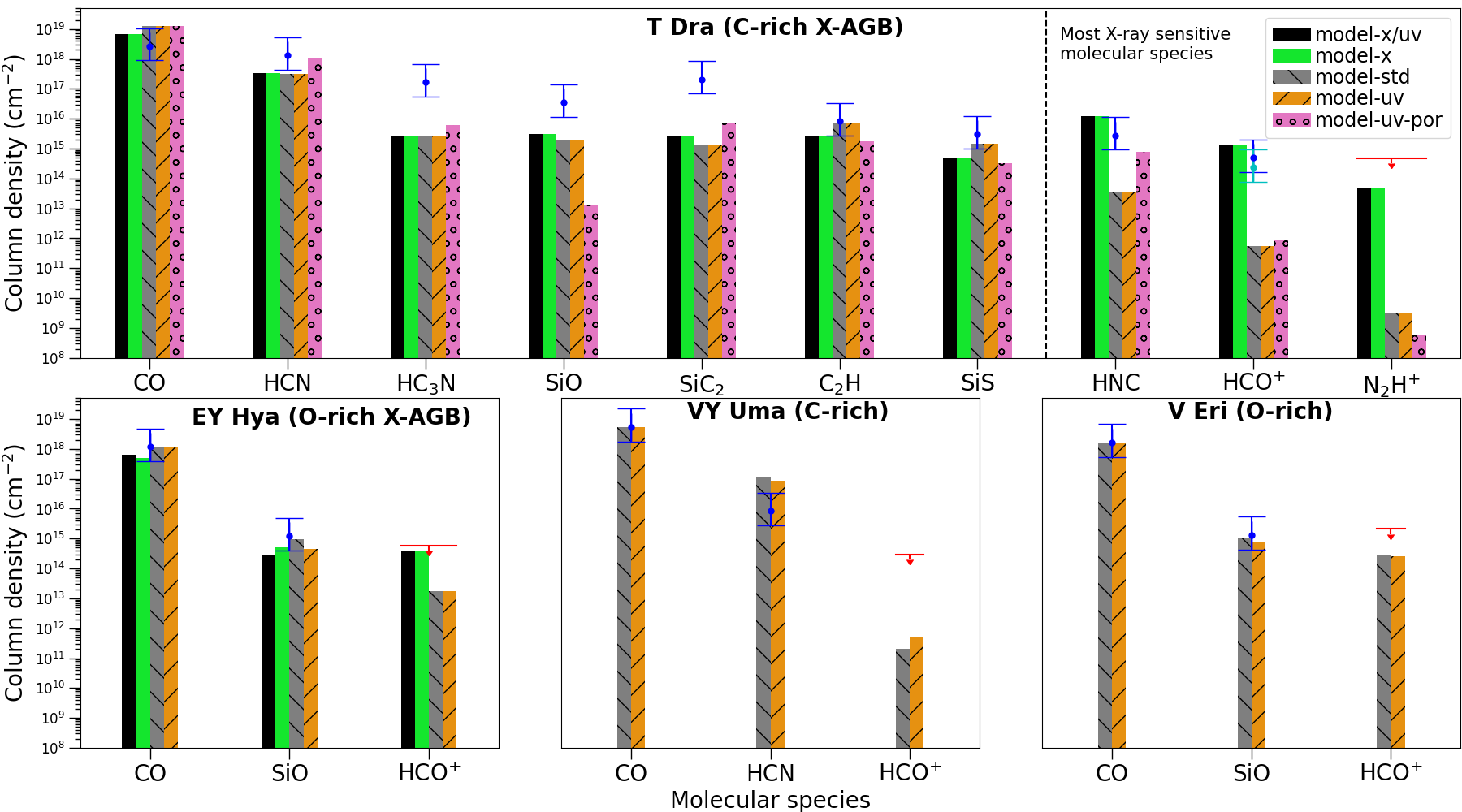}
        \caption{Column densities estimated from our radiative transfer models (values and upper limits represented by segments and arrows respectively) compared with those predicted by our chemical kinetics models (represented by the bars) for different species and targets explored in this work (see Sects.~\ref{anal} and \ref{chem}). The blue and cyan segments represent detections obtained in 2020 and 2024, respectively; red arrows indicate upper limits. The size of the segments correspond to column density uncertainties of a factor of three. The black solid bar represents 'model-x/uv', the green solid bar represents 'model-x', the grey striped bar represents 'model-std', the orange striped bar represents 'model-uv' and the pink ringed bar represents 'model-uv' when considering a porous envelope ('model-uv-por').}
        \label{fig:bar_plot}
\end{figure*}

\section{Discussion}\label{dis}

As demonstrated, the presence of an internal source of UV or X-ray emission can drastically reshape the radial abundance distribution of both parent and daughter species in AGB CSEs. X-rays, with their high penetration capability, ionise and dissociate molecules, initiating a wide range of chemical reactions throughout the envelope, affecting a broad range of layers. The most relevant is the photoionisation of molecular hydrogen, leading to the formation of the reactive \hhhplus\ radical, which drives key chemical processes that strongly influence the molecular composition of the CSE. In contrast, UV photons have a more localised effect, primarily affecting molecules such as SiO and \ndoshplus\ (overall destruction) and HCN and \hctresn\ (overall enhancement), among the species explored in this work, in optically thin (inner layers) or porous regions of the envelope \citep[see also][]{Van_de_Sande_2018, Van_de_Sande_2022}. 

The effects of internal X-ray might be particularly evident in some parent species, where photodissociation can significantly reduce their abundances in the innermost envelope regions. This leads to the appearance of centrally depleted regions, as seen in the CO distribution of T\,Dra. In contrast, daughter and intermediate molecules (e.g. \hnc, \hctresn, and \hcoplus), which typically show ring-like/shell-like radial abundance distributions \citep[see e.g.][]{Daniel_2012, Unnikrishnan_2024}, might display centrally enhanced abundances under the presence of an internal X-ray source, which would result in flatter abundance distributions (i.e. Gaussian-like). A similar behaviour was described by \cite{Van_de_Sande_2022} for some particular UV sensitive molecules.

The enhanced inner-envelope abundance of \hctresn\ in T\,Dra is consistent with similar findings for IRC$+$10216 by \cite{Siebert_2022}, who indicated that the presence of UV photons from a hot companion star could explain the estimated overabundance of \hctresn\, in the inner regions of IRC$+$10216. This highlights how internal high-energy sources can drive significant changes in the chemistry of CSEs.

In addition to molecules included in this study, X-rays can also promote the formation of complex molecules such as toluene, as suggested by \citet{Monfredini_2016}, making them of great interest for future astrochemical studies.

The abundance estimates for daughter species are particularly sensitive to the physical parameters of the envelope, especially the temperature distribution \citep[see e.g.][]{Maes_2023}. However, the radiative transfer analysis performed in this study yields results consistent with expectations for standard AGB stars with similar mass-loss rates, particularly for non-UV/X-ray sensitive molecules. The agreement between these estimates and predictions suggests that systematic effects related to the physical description of the envelope are not playing a major role and are not significantly biasing abundance estimates. 

In principle, an alternative explanation for the high \hcoplus\, abundance found in T\,Dra might be the presence of shock waves that induce the formation of molecular ions as happens, e.g. in OH\,231.8$+$4.2 \citep{sanchez-contreras_1997, sanchez-contreras_2015}. However, we excluded this possibility due to the lack of high-velocity wings in the CO and \hcoplus\, line profiles and the low expansion velocities measured in the line profiles (\vexp$\lesssim$15\,\kms).

Since the enhanced amount of \hcoplus\, is related with the X-ray emission produced inside the envelope, by the presumed stellar companion or accretion disc \citep{Sahai_2015}, the variability of \hcoplus\,is also likely related with a variability of this internal X-ray emission \citep[which can reach an order of magnitude over monthly timescales, see][]{Ortiz_2021} as happens in other astronomical sources such as protoplanetary discs \citep[see e.g.][]{Cleeves_2017, Waggoner_2023} in which the \hcoplus\, formation mechanism is also led by the chemical reaction (\ref{reacciones_hcoplus_1}).

It is worth to mention that a future monitoring of \hcoplus\, and other lines from X-ray sensitive molecules (e.g. $\rm H_{3}O^{+}$ or $\rm N_{2}H^{+}$) in this target, ideally accompanied by simultaneous observations with sensitive X-ray telescopes, is desirable to determine the relationship between X-ray emission and \hcoplus\, production in AGB circumstellar environments.

\section{Summary and conclusions}\label{summ}

We have presented the results of a detailed study exploring the circumstellar chemistry of four AGB stars that, unlike most of their class, possess internal sources of high-energy radiation (UV and/or X-ray emission). Our work focused on detecting the X-ray sensitive molecular ion \hcoplus. The stars in our sample included T\,Dra (C-rich) and EY\,Hya (O-rich), which are both emitters of X-ray and UV radiation, as well as VY\,UMa (C-rich) and V\,Eri (O-rich), which only exhibit UV emission. Observations were performed using the \iram\ antenna in two epochs (2020 and 2024) for T\,Dra, while one epoch (2020) was used for the other sources. 
 
Key findings include the detection of \hcoplus\ ($v$=0,  $J$=1--0 and $J$=3--2 transitions) in the C-rich X-ray emitting AGB star T\,Dra. The observed intensity of these lines doubled between 2020 and 2024, which is most likely linked to the variable X-ray emission in T\,Dra. 
Additional detections in T\,Dra include several species such as \sio, \hcn, \hctresn, \hnc, \sis, \sicdos, and \cch, highlighting its chemically rich envelope. In contrast, the other sources yielded very limited detections, with \sio\ observed in the O-rich stars, and \hcn\ in the other C-rich star.

Using radiative transfer modelling, we derived the physical properties of the envelopes for each object by fitting their CO emission lines (Table~\ref{tab:MADEX_models}). For T\,Dra, the best-fit model has a CO-depleted central region, probably related with photodissociation produced by the internal X-ray radiation. For the rest of the detected molecular species, a radiative transfer analysis was also performed in order to derive the average molecular abundances in the envelope (Table~\ref{tab:abundances}). In T\,Dra, the \hcoplus\ fractional abundance was found to be [1.5-3]\ex{-8}, exceeding the typical values in C-rich AGB CSEs. Furthermore, \hnc\, and \hctresn\, also exhibit elevated abundances ($\sim$1\ex{-7} and $\sim$11\ex{-5}, respectively) in T\,Dra compared with standard C-rich AGBs. These findings are supported by chemical kinetics models that incorporate the effects of internal X-ray and UV emission, thus marking a pioneering step in the study of X-ray induced chemistry in AGB circumstellar environments.   

Our chemical models reveal that X-rays have a profound effect on the molecular composition of the circumstellar envelope of T\,Dra. Enhanced abundances were observed not only for \hcoplus\ but also for \hnc\, and \hctresn, which are are found to be sensitive to X-rays from our models. Other molecules not detected in this study (e.g. \ndoshplus) were also found to be highly enhanced by X-rays, expanding the scope for future observational studies. In contrast, UV radiation had a negligible effect in optically thick envelopes such as those studied here, except under the extreme assumption of highly porous envelopes. 

In summary, the results presented here underline the crucial role of internal X-ray emission in shaping the chemistry of AGB CSEs, with implications for the abundances and distributions of both parent and daughter species. These findings provide a basis for further studies on the influence of high-energy radiation in circumstellar environments. Advancing our understanding requires high angular resolution interferometric observations. Such studies could provide detailed radial abundance profiles for molecules such as \hcoplus\, and offer deeper insights into the interplay between X-ray radiation and chemistry. Moreover, broader spectral surveys are needed to identify additional molecules influenced by X-ray and UV emission, and monitoring of molecular line variability and X-ray fluxes would further elucidate the dynamic chemical processes in these environments.

\begin{acknowledgements}
We thank the anonymous referee whose detailed review and valuable comments have helped us to improve the paper. This work is part of the I+D+i projects PID2019-105203GB-C22, PID2023-146056NB-C22, PID2023-147545NB-I00, PID2019-105203GB-C21, PID2023-146056NB-C21, and PID2020-117034RJ-I00 funded by Spanish MCIN/AEI/10.13039/501100011033 and by ``ERDF A way of making Europe". J.A.H. is supported by INTA grant PRE\_MDM\_05 and acknowledges CSIC grant iMOVE 23023. R.S. contribution to the research described here was carried out at the Jet Propulsion Laboratory, California Institute of Technology, under a contract with NASA, and funded in part by NASA via various ADAP awards. We thank the IRAM 30 m staff for their support during the observations. 

This work is based on observations carried out under projects numbers 040-20 and 132-23 with the IRAM 30m telescope. IRAM is supported by INSU/CNRS (France), MPG (Germany) and IGN (Spain).

This research has made use of the SIMBAD database, operated at CDS, Strasbourg, France. This work presents results from the European Space Agency (ESA) space mission Gaia. Gaia data are being processed by the Gaia Data Processing and Analysis Consortium (DPAC). Funding for the DPAC is provided by national institutions, in particular the institutions participating in the Gaia MultiLateral Agreement (MLA). The Gaia mission website is \url{https://www.cosmos.esa.int/gaia}. The Gaia archive website is \url{https://archives.esac.esa.int/gaia}. Some of the data presented in this paper were obtained from the Multimission Archive at the Space Telescope Science Institute (MAST). STScI is operated by the Association of Universities for Research in Astronomy, Inc., under NASA contract NAS5-26555. Support for MAST for non-HST data is provided by the NASA Office of Space Science via grant NAG5-7584 and by other grants and contracts. 

\end{acknowledgements}

\bibliographystyle{aa} 
\bibliography{aa53819-25.bib}

\begin{appendix}

\onecolumn

\section{Spectral line measurements}\label{line_table}

We present the compilation of the measured parameters of the detected spectral lines. In the case of T\,Dra, this includes the parameters of the \hcn\, vibrationally excited and \hcoplus\, lines measured on the two different epochs as they show variability.

\begin{table*}[h!]

\renewcommand{\arraystretch}{1.2}
\small
\centering

\caption{Spectral measurements for detected lines.} 
\label{tab:lines}

\begin{adjustbox}{max width=\textwidth}
\begin{threeparttable}[b]

\begin{tabular}{l >{\centering\arraybackslash}p{1.80cm} >{\centering\arraybackslash}p{1.50cm} >{\centering\arraybackslash}p{1.50cm} >{\centering\arraybackslash}p{1.50cm} >{\centering\arraybackslash}p{1.50cm} >{\centering\arraybackslash}p{1.50cm} >{\centering\arraybackslash}p{1.50cm} >{\centering\arraybackslash}p{1.50cm}}
\hline\hline 
Transition  & Rest frequency & $\int T_{MB}dv$ & Peak \tmb & rms & $V_{\rm sys}$ & \vexp &  FWHM & Obs. Epoch \\  
       & (MHz) & (K \kms) & (mK) & (mK) & (\kms) & (\kms) & (\kms) & (yr)\\ 
\hline 

\multicolumn{8}{c}{T Dra (C-rich with X-rays)} \\
\hline 

\siouno &  86846.99 &  0.10$\pm$0.03 & 6 & 2& -13$\pm$1 & 7$\pm$4  & 16$\pm$2 & 2020+2024\\
\hcnuno &  88631.60  & 0.80$\pm$0.04 & 42 & 2 &  -11.5$\pm$0.1 & 13$\pm$3 & 16.8$\pm$0.3 & 2020+2024\\
\hcnuno\, $v_{2}$=$2$ $^{(a)}$ & 89090.10 & $<$0.02 &  $<$8 & 3 & --  &  -- &  -- & 2020\\
\hcnuno\, $v_{2}$=$2$ $^{(b)}$ & 89090.10 & 0.46$\pm$0.02 & 58 & 2 & -8.1$\pm$0.1 & 0.7$\pm$0.2 & 1.1$\pm$0.1 & 2024\\
\hcouno$^{(a)}$ &   89188.53 &  0.12$\pm$0.05 & 4 & 3 & -17$\pm$3 & 10$\pm$3  & 16$\pm$6 & 2020\\
\hcouno$^{(b)}$ &   89188.53 &  0.13$\pm$0.04 & 7 & 2  & -17$\pm$1 & 9$\pm$4  & 15$\pm$3 & 2024 \\
\hncuno & 90663.56 &  0.10$\pm$0.02 & 6  & 2 & -11.7$\pm$0.8 & 6$\pm$3  & 15$\pm$2 & 2020+2024\\
\hctresnuno &   90978.99 &  0.08$\pm$0.02 & 6 & 2 & -13.3$\pm$0.7 & 10$\pm$5  & 14$\pm$2 & 2020+2024\\
$\rm N_{2}H^{+}$($J$=1-0) &  93173.40 &   $<$0.05 & $<$9 & 3 & -- & -- &  -- & 2020+2024 \\
\siodos & 260518.02 & 1.51$\pm$0.05 & 102 & 5 &-12.9$\pm$0.1 & 11$\pm$2  & 11.5$\pm$0.3 & 2020 \\
\sicdosdos & 261150.70 & 0.19$\pm$0.07 & 5 & 6 & -13$\pm$2 & 7$\pm$4  & 11$\pm$3 & 2020 \\
\sicdosuno & 261509.32 & 0.17$\pm$0.07 & 9 & 6 & -12$\pm$2 & 11$\pm$4  & 16$\pm$4 & 2020 \\
(\sicdostres +\cch)$^{(*)}$ & 261990.74  & 0.56$\pm$0.15 & 19 & 6 & -- & -- & -- & 2020 \\
\hctresndos &  263792.30 & 0.44$\pm$0.10 & 21 & 6 &-11.3$\pm$0.7 & 10$\pm$2  & 18$\pm$2 & 2020 \\
\hcndos\, $v_{2}$=$1$e $^{(a)}$ & 265852.71 & 0.50$\pm$0.02 & 225 & 6 & -12.8$\pm$0.2 & 1.5$\pm$0.9  & 3.0$\pm$3 & 2020 \\
\hcndos\, $v_{2}$=$1$e $^{(b)}$ & 265852.71 & 1.66$\pm$0.02 & 519 & 9 & -13.7$\pm$0.1 & 2$\pm$1  & 3.0$\pm$0.2 & 2024 \\
\hcndos & 265886.43  & 4.88$\pm$0.05 & 346 & 3 & -11.8$\pm$0.4 & 10$\pm$4 & 13$\pm$1 & 2020+2024 \\
\hcndos\, $v_{2}$=$1$f $^{(a)}$ & 267199.28 & 0.16$\pm$0.07 & 23 & 7 & -14.6$\pm$0.6 & 4$\pm$1  & 7$\pm$1 & 2020 \\ 
\hcndos\, $v_{2}$=$1$f $^{(b)}$ & 267199.28 &  0.17 $\pm$0.07 & 16 & 7  & -14.2$\pm$0.7 & 6$\pm$2  & 11$\pm$1 & 2024 \\
\hcodos$^{(a)}$ &  267557.63 & 0.26$\pm$0.07 & 12 & 5 &-15$\pm$1 & 9$\pm$5  & 16$\pm$3 & 2020 \\
\hcodos$^{(b)}$ &  267557.63 & 0.61$\pm$0.07 & 37 & 6 & -15$\pm$1 & 9$\pm$5  & 13.3$\pm$0.9 & 2024 \\
\hnctres & 271981.11 & 0.72$\pm$0.07 & 49 & 6 & -13.8$\pm$0.3 & 11$\pm$3  & 12.0$\pm$0.6 & 2024 \\
\sisquince & 272243.06 & 0.18$\pm$0.05 & 12 & 4 & -14$\pm$1 & 6$\pm$3  & 14$\pm$2 & 2024 \\

\hline

\multicolumn{8}{c}{EY Hya (O-rich with X-rays)} \\
\hline 

\siouno &  86846.99  & 0.09$\pm$0.03  &  7 & 3 & 24$\pm$1 & 5$\pm$3 & 9$\pm$3 & 2020 \\
\siodos & 260518.02  & 0.42$\pm$0.10  & 24 & 8 & 25.8$\pm$0.9 & 6$\pm$3 & 13$\pm$2 & 2020 \\
\hcouno & 89188.53 &     $<$0.05 & $<$9 & 3 & -- & -- &  -- & 2020 \\
\hcodos & 267557.63  &     $<$0.14 & $<$27 & 9 & -- & -- &  -- & 2020 \\
\hline 

\multicolumn{8}{c}{VY Uma (C-rich without X-rays)} \\
\hline 

\hcnuno &  88631.60   &   $<$0.02 & $<$8 & 3 & -- & -- &  -- & 2020 \\
\hcndos & 265886.43 &  0.43$\pm$0.05  & 22 & 6 & -4$\pm$3 &  5$\pm$2 & 9$\pm$1 & 2020 \\
\hcouno & 89188.53 &     $<$0.05 & $<$9 & 3 & -- & -- &  -- & 2020 \\
\hcodos & 267557.63  &     $<$0.09 & $<$18 & 6 & -- & -- &  -- & 2020 \\
\hline 

\multicolumn{8}{c}{V Eri (O-rich without X-rays)} \\
\hline 

\siouno & 86846.99  & 0.12$\pm$0.03 & 10  & 5 & -12.8$\pm$0.7  & 4$\pm$2 & 9$\pm$2& 2020 \\
\siodos & 260518.02 & 0.75$\pm$0.14 & 58 & 10 &-11.1$\pm$0.5   & 7$\pm$2 & 13$\pm$1& 2020\\
\hcouno & 89188.53 &     $<$0.08 & $<$15 & 5 & -- & -- &  -- & 2020 \\
\hcodos & 267557.63  &     $<$0.17 & $<$33 & 11 & -- & -- &  -- & 2020 \\
\hline 

\end{tabular} 
\begin{tablenotes}
\item \normalsize \textbf{Notes.} Column (1): Transition, Col. (2): Line rest frequency, Col (3): Integrated main beam temperature, Col (4):  Main beam temperature of the peak from the so-called Shell fit (see Sect.~\ref{spectra}), Col (5): Main beam rms of the noise, col (6): Systemic velocity, col (7): Expansion velocity estimated from a Shell fit, col (8): line FWHM estimated from a Gaussian fit. col (9): Observing date (year); for lines labelled as 2020+2024, showing no variability, parameters are derived from the time-averaged spectra. $^{(a)}$ and $^{(b)}$ indicate the lines with multi-epoch observations that have not been averaged. $^{(*)}$ In this case, the \sicdos\, line is blended with some \cch\, lines (see Sect.~\ref{MADEX_other} and Fig.~\ref{fig:T-DRA_spectra}).
\end{tablenotes}

\end{threeparttable}
\end{adjustbox}

\renewcommand{\arraystretch}{1.0}

\end{table*}

\newpage

\section{\iram\, Spectra divided in two epochs}\label{Two_epochs}

We present the rest of multi-epoch spectral lines detected in T\,Dra, which was the only source observed in two epochs, apart from \hcoplus\, and \hcn\, vibrationally excited lines (which were presented in Figs.~\ref{fig:HCO+_T-DRA_spectra} and \ref{fig:T-DRA_spectra_vib} respectively).

\begin{figure*}[h!]
     \centering
     \begin{subfigure}[b]{0.33\linewidth}
         \centering
         \includegraphics[width=\linewidth]{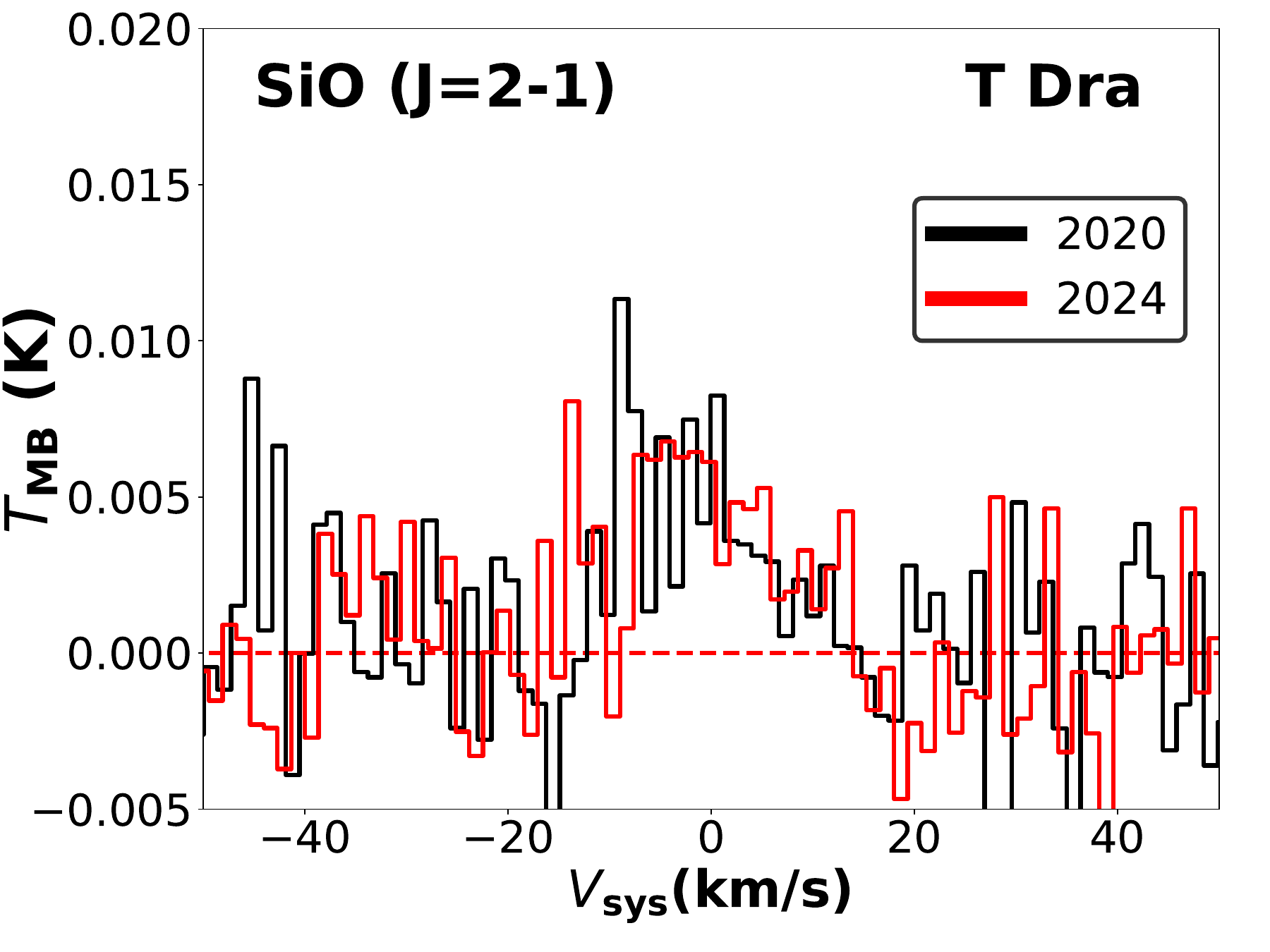}
     \end{subfigure}
     \begin{subfigure}[b]{0.33\linewidth}
         \centering
         \includegraphics[width=\linewidth]{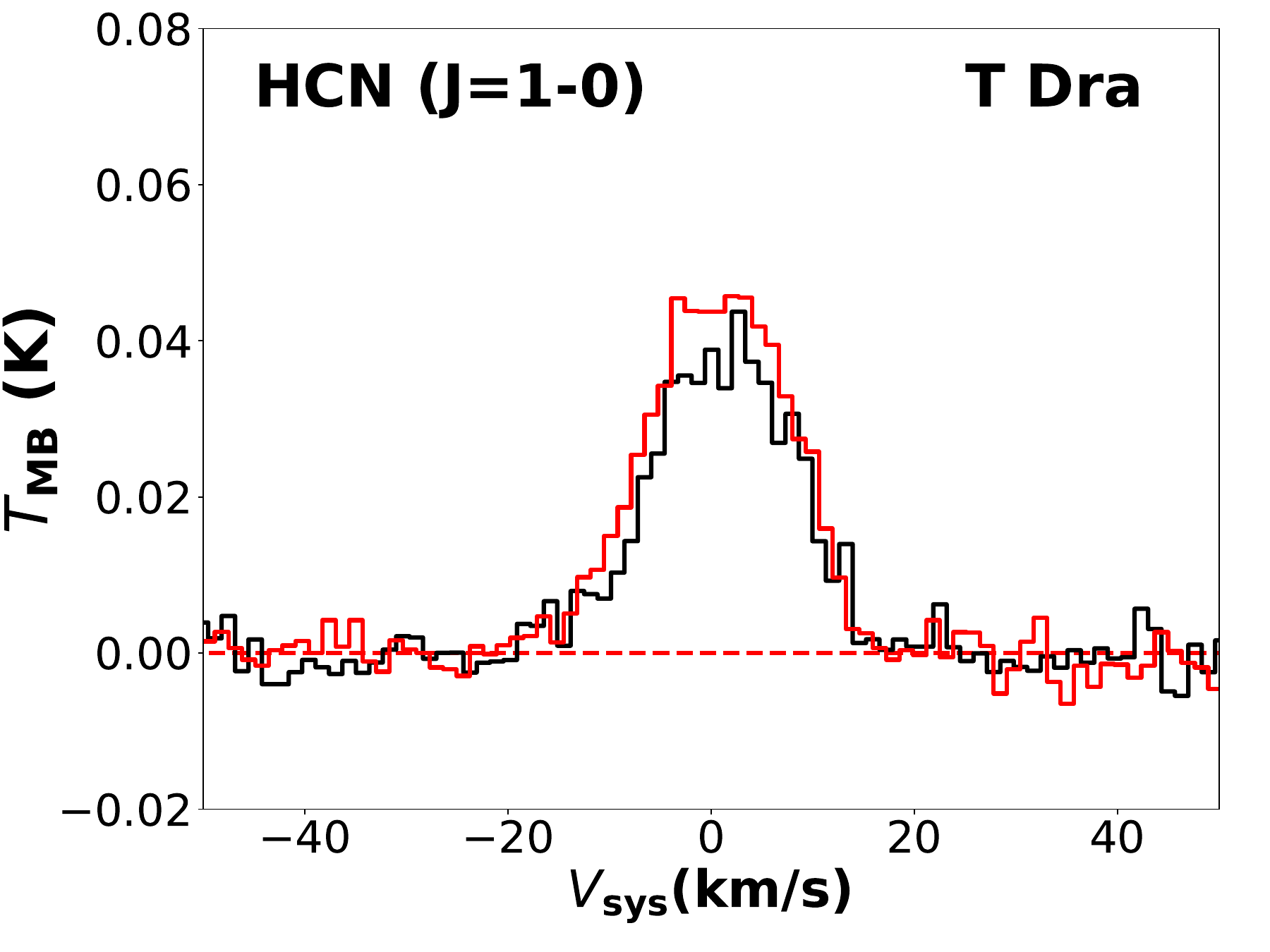}
     \end{subfigure}
     \begin{subfigure}[b]{0.33\linewidth}
         \centering
         \includegraphics[width=\linewidth]{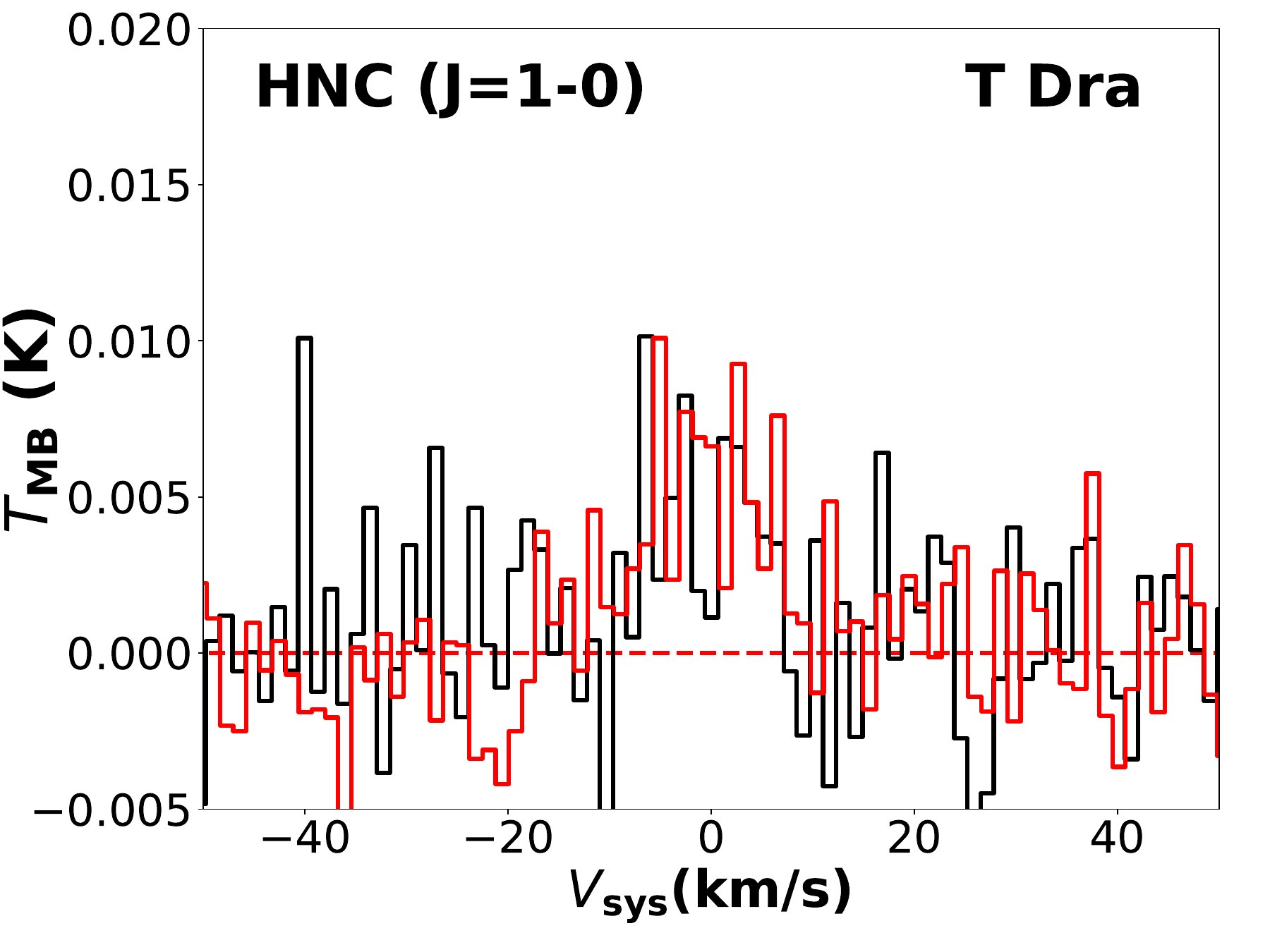}
     \end{subfigure}

     \begin{subfigure}[b]{0.33\linewidth}
         \centering
         \includegraphics[width=\linewidth]{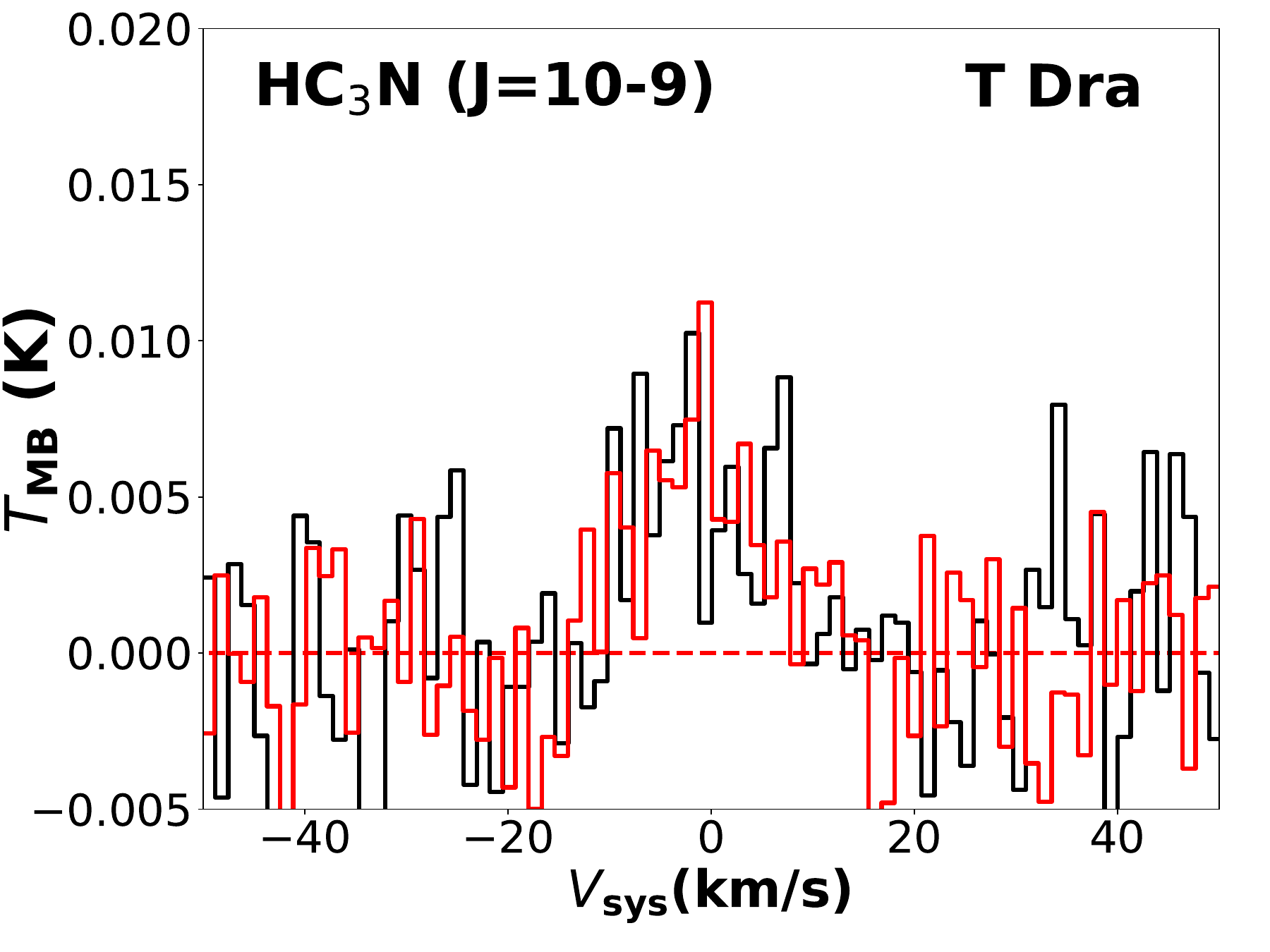}
     \end{subfigure}
     \begin{subfigure}[b]{0.33\linewidth}
         \centering
         \includegraphics[width=\linewidth]{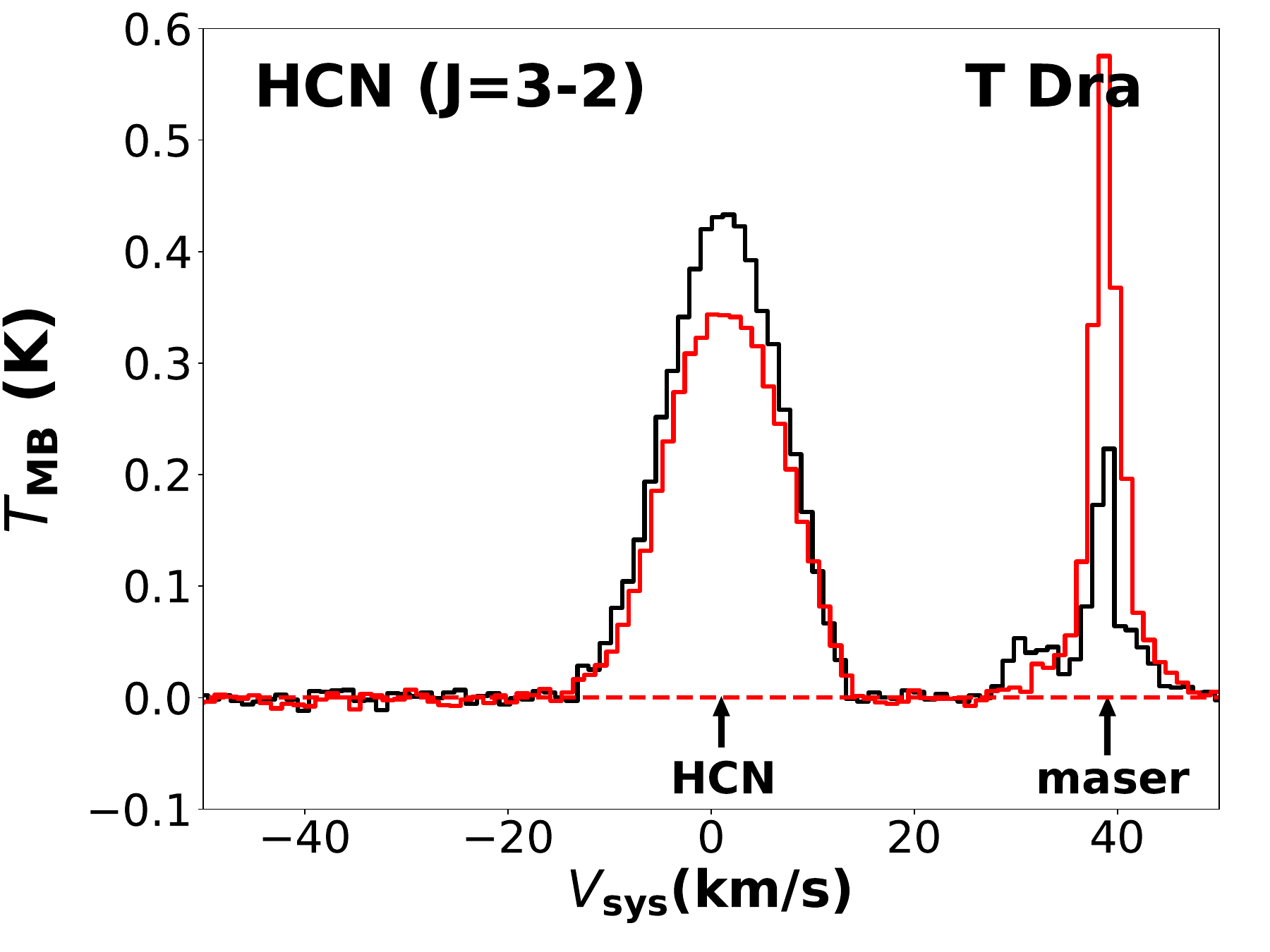}
     \end{subfigure}
     \hspace{0.33\linewidth}
        \caption{Multi-epoch spectra of the rest of lines in T\,Dra. Observed spectra are shown in black for 2020 observations and in red for 2024 observations (velocity resolution is $\delta v$=1.4\,\kms). No significant differences were found between the observed epochs for these lines considering flux calibration errors.}
        \label{fig:T-DRA_spectra_appendix_A1}
\end{figure*}

\begin{figure*}[h!]
     \centering
     \includegraphics[width=0.97\linewidth]{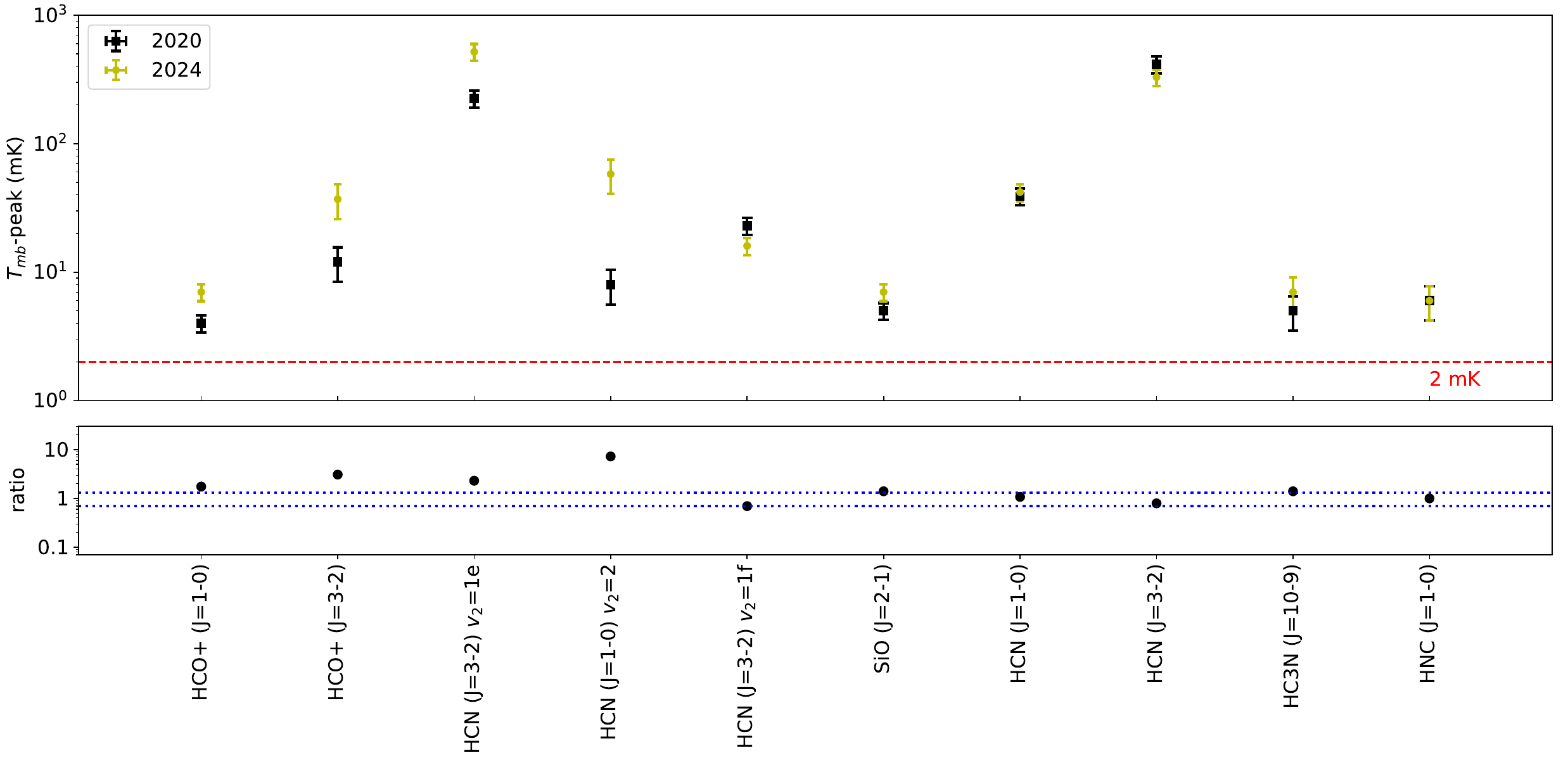}
     \caption{Comparison between the main beam peak temperatures for the lines observed for T\,Dra in the two observational campaigns. Upper: main beam peak temperatures of the lines, 2020 observations are shown in black and 2024 observations are shown in yellow. Bottom: Ratio between the 2024 and the 2020 main beam peak temperatures, blue dotted lines indicate the 1.3 and 0.7 range. It is shown that \hcoplus\, and \hcn\, maser lines show significant variability.}
     \label{fig:T-DRA_spectra_appendix_A2}
\end{figure*}

Fig.~\ref{fig:T-DRA_spectra_appendix_A1} shows the spectral lines divided in the two epochs and Fig.~\ref{fig:T-DRA_spectra_appendix_A2} shows the main beam peak temperature estimated from a Shell fit for both epochs and the ratio between them.

\newpage 

\section{Column densities from chemical kinetics modelling}\label{column_densities}

We present the tabulated values of the column densities estimates from our radiative transfer models and chemical kinetics models, which are shown in Fig.~\ref{fig:bar_plot}.

\begin{table*}[h!]

\renewcommand{\arraystretch}{1.2}
\small
\centering

\caption{Comparison between the empirically estimated molecular column densities derived from the radiative transfer analysis (Sect.~\ref{anal}) and those predicted by the chemical kinetics models (Sect.~\ref{chem}).} 
\label{tab:column_densities}

\begin{adjustbox}{max width=\textwidth}
\begin{threeparttable}[b]

\begin{tabular}{l >{\centering\arraybackslash}p{2.50cm} >{\centering\arraybackslash}p{2.50cm} >{\centering\arraybackslash}p{2.50cm} >{\centering\arraybackslash}p{2.50cm} >{\centering\arraybackslash}p{2.50cm}}
\hline\hline 
Molecule  & empirically estimated & model-x/uv & model-x & model-std & model-uv \\  
       & (cm$^{-2}$) & (cm$^{-2}$) & (cm$^{-2}$) & (cm$^{-2}$) & (cm$^{-2}$) \\  
\hline
\hline
\multicolumn{6}{c}{T Dra (C-rich with X-rays) --smooth model--} \\
\hline
CO  & 2.7$\times 10^{18}$ & 6.8$\times 10^{18}$ & 6.8$\times 10^{18}$ & 1.3$\times 10^{19}$ & 1.3$\times 10^{19}$  \\ 
HCN & 1.3$\times 10^{18}$ & 3.4$\times 10^{17}$ & 3.4$\times 10^{17}$ & 3.1$\times 10^{17}$ & 3.1$\times 10^{17}$ \\
H$\mathrm{C_{3}}$N &  1.7$\times 10^{17}$ & 2.6$\times 10^{15}$ & 2.6$\times 10^{15}$ & 2.5$\times 10^{15}$ & 2.5$\times 10^{15}$ \\
SiO & 3.5$\times 10^{16}$ & 3.0$\times 10^{15}$ & 3.0$\times 10^{15}$ & 1.9$\times 10^{15}$ & 1.9$\times 10^{15}$ \\
\sicdos  & 2.1$\times 10^{16}$ & 2.7$\times 10^{15}$ & 2.7$\times 10^{15}$ & 1.4$\times 10^{15}$ & 1.4$\times 10^{15}$ \\
\cch  & 8.2$\times 10^{15}$ & 2.8$\times 10^{15}$ & 2.8$\times 10^{15}$ & 7.4$\times 10^{15}$ & 7.4$\times 10^{15}$ \\
SiS  & 3.1$\times 10^{15}$ & 4.7$\times 10^{15}$ & 4.7$\times 10^{15}$ & 1.5$\times 10^{16}$ & 1.5$\times 10^{16}$ \\
HNC & 2.8$\times 10^{15}$ & 1.2$\times 10^{16}$ & 1.2$\times 10^{16}$ & 3.4$\times 10^{13}$ & 3.4$\times 10^{13}$ \\
\hcoplus (2024) & 4.9$\times 10^{14}$ & 1.3$\times 10^{15}$ & 1.3$\times 10^{15}$ & 5.6$\times 10^{11}$ & 5.6$\times 10^{11}$ \\
\hcoplus (2020) & 2.4$\times 10^{14}$   & --- & --- & --- & --- \\
$\rm N_{2}H^{+}$  & $<$4.8$\times 10^{14}$ & 5.1$\times 10^{13}$ & 5.1$\times 10^{13}$ & 3.2$\times 10^{9}$ & 3.2$\times 10^{9}$ \\

\hline
\multicolumn{6}{c}{T Dra (C-rich with X-rays) --porous model--} \\
\hline

CO & 2.7$\times 10^{18}$ & 7.6$\times 10^{18}$ & 6.7$\times 10^{18}$ & 1.3$\times 10^{19}$ & 1.3$\times 10^{19}$ \\
HCN & 1.3$\times 10^{18}$ & 1.1$\times 10^{18}$ & 3.5$\times 10^{17}$ & 3.1$\times 10^{17}$ & 1.1$\times 10^{18}$ \\
H$\mathrm{C_{3}}$N &  1.7$\times 10^{17}$ & 2.2$\times 10^{16}$ & 2.4$\times 10^{15}$ & 3.1$\times 10^{15}$ & 6.3$\times 10^{15}$ \\
SiO & 3.5$\times 10^{16}$ & 2.5$\times 10^{15}$ & 2.8$\times 10^{15}$ & 1.8$\times 10^{15}$ & 1.3$\times 10^{13}$ \\
\sicdos  & 2.1$\times 10^{16}$ & 8.3$\times 10^{15}$ & 2.7$\times 10^{15}$ & 1.7$\times 10^{15}$ & 7.6$\times 10^{15}$ \\
\cch  & 8.2$\times 10^{15}$ & 4.7$\times 10^{15}$ & 1.6$\times 10^{15}$ & 4.2$\times 10^{15}$ & 1.8$\times 10^{15}$ \\
SiS  & 3.1$\times 10^{15}$ & 1.7$\times 10^{14}$ & 4.5$\times 10^{15}$ & 1.5$\times 10^{16}$ & 3.2$\times 10^{14}$ \\
HNC & 2.8$\times 10^{15}$ & 4.4$\times 10^{15}$ & 9.9$\times 10^{15}$ & 4.1$\times 10^{13}$ & 7.9$\times 10^{14}$ \\
\hcoplus (2024) & 4.9$\times 10^{14}$ & 5.5$\times 10^{14}$ & 7.3$\times 10^{14}$ & 3.7$\times 10^{11}$ & 8.3$\times 10^{11}$ \\
\hcoplus (2020) & 2.4$\times 10^{14}$   & --- & --- & --- & --- \\
$\rm N_{2}H^{+}$  & $<$4.8$\times 10^{14}$ & 4.4$\times 10^{12}$ & 2.6$\times 10^{13}$ & 1.9$\times 10^{9}$ & 5.8$\times 10^{8}$ \\

\hline
\multicolumn{6}{c}{EY Hya (O-rich with X-rays) --smooth model--} \\
\hline

CO & 1.2$\times 10^{18}$ & 6.5$\times 10^{17}$ & 5.1$\times 10^{17}$ & 1.2$\times 10^{18}$ & 1.2$\times 10^{18}$ \\
SiO  & 1.19$\times 10^{15}$ & 3.0$\times 10^{14}$ & 5.2$\times 10^{14}$ & 9.7$\times 10^{14}$ & 4.5$\times 10^{14}$ \\
\hcoplus & $<$5.9$\times 10^{14}$ & 3.8$\times 10^{14}$ & 3.8$\times 10^{14}$ & 1.8$\times 10^{13}$ & 1.7$\times 10^{13}$ \\

\hline
\multicolumn{6}{c}{VY Uma (C-rich without X-rays) --smooth model--} \\
\hline

CO & 3.6$\times 10^{18}$ & -- & -- & 3.6$\times 10^{18}$ & 3.6$\times 10^{18}$ \\
HCN  & 6.1$\times 10^{15}$ & -- & -- & 8.4$\times 10^{16}$ & 6.2$\times 10^{16}$ \\
\hcoplus  &  $<$2.3$\times 10^{14}$ & -- & -- & 1.8$\times 10^{11}$ & 4.6$\times 10^{11}$ \\

\hline
\multicolumn{6}{c}{V Eri (O-rich without X-rays) --smooth model--} \\
\hline

CO & 1.1$\times 10^{18}$ & -- & -- & 1.0$\times 10^{18}$ & 1.0$\times 10^{18}$ \\
SiO  & 1.0$\times 10^{15}$ & -- & -- & 8.4$\times 10^{14}$ & 5.6$\times 10^{14}$ \\
\hcoplus  & $<$1.6$\times 10^{15}$ & -- & -- & 2.1$\times 10^{14}$ & 2.0$\times 10^{14}$ \\

\hline
\end{tabular} 
\begin{tablenotes}
\item \normalsize \textbf{Notes.} Column (1): Molecule, Col. (2): empirically estimated column densities from the radiative transfer modelling (see Sect.~\ref{anal}), Col (3): Column densities estimated for the chemical kinetics model-x/uv, Col (4): same for model-std, Col (5): same for model-uv, Col (6): same for model-x.).
\end{tablenotes}

\end{threeparttable}
\end{adjustbox}

\renewcommand{\arraystretch}{1.0}

\end{table*}

\end{appendix}

\end{document}